\documentclass[12pt]{article}
\usepackage{amsmath}
\usepackage{lineno}
\usepackage{graphicx}
\usepackage{tabularx}
\usepackage{natbib}
\bibpunct[,]{(}{)}{;}{a}{}{,}
\usepackage{multirow}
\usepackage{url} 
\usepackage{textcomp}
\usepackage{mathtools}
\usepackage{amsmath,amsthm,amsopn,amsfonts,pdfpages,dsfont}
\usepackage{caption}
\usepackage{url}

  \usepackage{stackengine}
  \usepackage{algorithm}
  \usepackage{algorithmicx}
    \usepackage{rotating}
    \usepackage{tikz}
    
    \newcommand{\X}{\mathbf{X}^{(t)}}
    \newcommand{\e}{\mathbb{E}}
    
    \newcommand{\real}{\mathbb{R}}

    \newtheorem{theorem}{Theorem}
    \newtheorem{assumption}{Assumption}
    
    \newtheorem{lemma}[theorem]{Lemma}
    \newtheorem{example}[theorem]{Example}
    
    \theoremstyle{definition}
    \newtheorem{definition}{Definition}
    
    \theoremstyle{remark}
    \newtheorem{remark}{Remark}

        \usepackage[normalem]{ulem}
        \newcommand\redsout{\bgroup\markoverwith{\textcolor{red}{\rule[0.5ex]{2pt}{0.4pt}}}\ULon}

        \algnewcommand\algorithmicinput{\textbf{Input:}}
        \algnewcommand\Input{\item[\algorithmicinput]}
        \algnewcommand\algorithmiciterate{\textbf{Iterate:}}
        \algnewcommand\Iterate{\item[\algorithmiciterate]}
        \algnewcommand\algorithmicinitialize{\textbf{Initialize:}}
        \algnewcommand\Initialize{\item[\algorithmicinitialize]}
        \algnewcommand\algorithmicoutput{\textbf{Output:}}
        \algnewcommand\Output{\item[\algorithmicoutput]}
        \algnewcommand\RETURN{\State \algorithmicreturn}%

        \newcommand{\mase}{\operatorname{MASE}}
        \newcommand{\rdpg}[1]{\operatorname{RDPG}\left(#1\right)}

              \newcommand{\cosie}[1]{\operatorname{COSIE}\left(#1\right)}
                \newcommand{\dirich}[1]{\operatorname{Dirichlet}\left(#1\right)}

                      \usepackage{thmtools}
                      \usepackage{float}
                      \usepackage[colorlinks=true,linkcolor=blue,citecolor=blue]{hyperref}
                      \usepackage{amssymb,subcaption}

                      \newcommand{\blind}{0}
                      
\usepackage{authblk}
\begin{document}

\def\spacingset#1{\renewcommand{\baselinestretch}%
{#1}\small\normalsize} \spacingset{1}


\if0\blind
{
  \title{\bf Multiple Network Embedding for Anomaly Detection in Time Series of Graphs}
  \author[1]{Guodong~Chen\thanks{\href{mailto:gchen35@jhu.edu}{gchen35@jhu.edu}}}
\author[2]{Jes\'{u}s~Arroyo}
\author[1]{Avanti~Athreya}
\author[3]{Joshua~Cape}
\author[4]{Joshua~T.~Vogelstein}
\author[5]{Youngser~Park}
\author[6]{Chris~White}
\author[6]{Jonathan~Larson}
\author[6]{Weiwei~Yang}
\author[1]{Carey~E.~Priebe\thanks{\href{mailto:cep@jhu.edu}{cep@jhu.edu}}}
\affil[1]{Department of Applied Mathematics and Statistics, Johns Hopkins University}
\affil[2]{Department of Statistics, Texas A\&M University}
\affil[3]{Department of Statistics, University of Wisconsin–Madison}
\affil[4]{Department of Biomedical Engineering, Kavli Neuroscience Discovery Institute, Johns Hopkins University}
\affil[5]{Center for Imaging Science, Johns Hopkins University}
\affil[6]{Microsoft AI and Research, Microsoft}
  \maketitle
} \fi

\if1\blind
{
  \bigskip
  \bigskip
  \bigskip
  \begin{center}
    {\LARGE\bf Multiple Network Embedding for Anomaly Detection in Time Series of Graphs}
\end{center}
  \medskip
} \fi

\bigskip
\begin{abstract}
This paper considers the graph signal processing problem of anomaly detection in time series of graphs. We examine two related, complementary inference tasks: the detection of anomalous graphs within a time series, and the detection of temporally anomalous vertices. We approach these tasks via the adaptation of statistically principled methods for joint graph inference, specifically 
\emph{multiple adjacency spectral embedding} (MASE). We demonstrate that our method is effective for our inference tasks. Moreover, we assess the performance of our method in terms of the underlying nature of detectable anomalies. We further provide the theoretical justification for our method and insight into its use. Applied to the Enron communication graph, a large-scale commercial search engine time series of graphs, and a larval \textit{Drosophila} connectome data, our approaches demonstrate their applicability and identify the anomalous vertices beyond just large degree change.
\end{abstract}

\noindent%
\vfill

\newpage
\spacingset{1.5} 
    \section{Introduction}
  \label{sec:Introduction}
  Given a time series of graphs $\mathcal{G}^{(t)} = (\mathcal{V},\mathcal{E}^{(t)}), t=1,2,\dots$ where the vertex set $\mathcal{V}=[n]=\{1, \dots, n\}$ is fixed and the edge sets $\mathcal{E}^{(t)}$ depend on time $t$, we consider two natural anomaly detection problems. The first problem involves detecting whether a particular graph $\mathcal{G}^{(t^{*})}$ is anomalous in the time series. The second problem involves detecting individual vertices anomalous in time. {They have shown applications in preventing detrimental events, such as computer network
    intrusion, social spam, and financial fraud. For example, a company's email history can be modeled as a sequence of graphs where agents are nodes, email exchanges are edges, and each month's emails form a graph. Graph anomaly detection can reveal changes in communication structure over time and potential fraud. Additionally, detecting anomalous vertices can help identify potentially fraudulent employees.} 
  
  Existing literature on anomaly detection in graphs (see, for instance, recent surveys \citep{ranshous2015anomaly, 10.1007/s10618-014-0365-y}) can roughly be categorized according to the characteristics of methods for modeling anomalies. Decomposition methods \citep{ide2004eigenspace, lakhina2004diagnosing} use eigenspace representations or tensor decomposition to extract features and monitor changes across time steps. Distance or (dis)similarity-based methods are also employed to monitor or identify changes \citep{koutra2013deltacon}. Probabilistic methods \citep{priebe2005scan,wang2013locality,heard2010} specify probability distributions to describe baseline ``typical'' behavior of features in networks (or networks themselves) and consider deviations from the baseline to be anomalies. {For instance, \cite{heard2010} employs a two-stage Bayesian procedure for detecting anomalies in dynamic graphs. Similarly, \cite{josephs2023bayesian} introduces a Gaussian process-based framework for anomaly detection with network inputs. Additionally, scan statistics \citep{priebe2005scan,wang2013locality} are often used in a moving time window analysis to compute local statistics across datasets for anomaly detection. The maximum value of these local statistics within the time window is called the scan statistic. There has also been some work on change point detection on graph data. For example, \cite{roy2017change} perform change point detection on a Markov random field, which is used for modeling the networks; \cite{bhattacharjee2020change} do change point detection in block models. \cite{wang2021optimal} achieve a near minimax rate of localization considering an independent sequence of inhomogeneous Bernoulli graphs. The latent position model for time series of graphs is considered in related works such as \cite{padilla2019change,marenco2022online}. Our approach differs by exploiting common structures in the nodes across the time series via joint graph embedding, which will be described next.  More recently, \cite{davis2023simple} proposed a dynamic network embedding method and introduced a hypothesis testing framework capable of testing for planted structures in networks. This framework bears similarities to our work.}
  
  Random graph inference has witnessed a host of developments and advancements in recent years \citep{JMLR:v18:16-480, JMLR:v18:17-448}. Much work has focused on single graph inference, while recently there has been increased interest in multiple graphs both with respect to modeling and performing statistical inference.
  Among recent developments \citep{xing2010state,durante2016locally,matias2017statistical,durante2017nonparametric,8215766,wang2019common,bhattacharyya2018spectral,pensky2019dynamic,macdonald2022latent}, multiple adjacency spectral embedding (\text{MASE}) \citep{arroyo2021inference} is a statistically principled multiple random graph embedding method for networks with latent space structure, absent dynamics or time dependency. This paper investigates a two-step procedure for detecting anomalies in time series of graphs that employs \text{MASE} for multiple network embedding. Notably, our approach benefits from simultaneous graph embedding to leverage common graph structure for parameter estimation, improving downstream discriminatory power for testing. Furthermore, as it hinges on probabilistic assumptions, our approach provides a statistically meaningful threshold for achieving a desired false positive rate of anomaly detection.
  
  This article is organized as follows. Section~\hyperref[sec:Preliminaries]{2} introduces notation. Section~\hyperref[sec:Method]{3} formulates two anomaly detection problems for graph-valued time series data, introduces \text{MASE} and our methodology. We present theoretical justifications in Section~\hyperref[sec:theory]{4} and its simulation results to assess the performance of \text{MASE} in anomaly detection in Section~\hyperref[sec:illus]{5}. For real data illustrations, we identify excessive activity in the Enron communication graph and a sub-region of a large-scale commercial search engine query-navigational graph in Section~\hyperref[sec:realdata]{6}. Section~\hyperref[sec:discussion]{7} concludes this paper with a discussion of outstanding issues and further summarizes our findings.
  
  

  \section{Setup}
  \label{sec:Preliminaries}
  \subsection{Notation and Preliminaries}
  This paper considers undirected, unweighted graphs without self-loops. Each graph is modeled via a random dot product graph (RDPG) \citep{JMLR:v18:17-448}, in which vertex connectivity is governed by latent space geometry. We begin by defining RDPGs as individual, static networks.


  \begin{definition} (\emph{Random Dot Product Graph}) 
  Let $X_{1}, X_{2}, \dots,  X_{n} \in\mathbb{R}^{d}$ be a collection of latent positions such that $0\leq X_{i}^{T} X_{j}\leq 1$ for each $i,j\in[n]$, and write $\mathbf{X} = [X_{1}|X_{2}|\cdots|X_{n}]^{T} \in \mathbb{R}^{n\times d}$. Suppose $\mathbf{A}$ is a symmetric hollow random adjacency matrix with
  \begin{equation}
  \mathbb{P}[\mathbf{A}]
  =\prod_{i< j}(X_{i}^{T}X_{j})^{\mathbf{A}_{ij}}(1-X_{i}^{T}X_{j})^{1-\mathbf{A}_{ij}}.
  \label{eq:RDPG-def}
  \end{equation}
  We then write $\mathbf{A}\sim \operatorname{RDPG}(\mathbf{X})$ and say that $\mathbf{A}$ is the adjacency matrix of a \emph{random dot product graph} with \emph{latent positions} given by the rows of $\mathbf{X}$ and positive semi-definite \emph{connectivity matrix} $\mathbf{P}=\mathbb{E}[\mathbf{A}]=\mathbf{X} \mathbf{X}^{T}$ with low rank structure $\text{rank}(\mathbf{P})\leq d$.
  \end{definition}

\begin{remark}(Identifiability)
	The random dot product graph (RDPG) model is a special case of a latent position model \citep{hoff2002latent}, which posits that edge existence between latent positions is determined by a symmetric link function. This model typically suffers from non-identifiability, as edge probabilities may be invariant under various transformations. For example, in the RDPG model, for any orthogonal transformation $\mathbf{W} \in \mathbb{R}^{d \times d}$, the relationship $\mathbf{X}\mathbf{X}^\top = (\mathbf{X}\mathbf{W})(\mathbf{X}\mathbf{W})^\top$ holds. This indicates that the latent positions $\mathbf{X}$ and $\mathbf{X}\mathbf{W}$ yield the same probability matrix as described in Equation~(\ref{eq:RDPG-def}). Therefore, the latent positions $\mathbf{X} \in \mathbb{R}^{n \times d}$ are identifiable only up to an orthogonal transformation.
	
\end{remark}

The latent state model and its indefinite extensions \citep{https://doi.org/10.1111/rssb.12509} can encapsulate low-rank independent-edge random graphs, such as stochastic block model (SBM) graphs \citep{HOLLAND1983109}. The Additive and Multiplicative Effects Network (AMEN) model by \cite{minhas2019inferential} bears similarities to the RDPG model. In these frameworks, the latent positions matrix $\mathbf{X}$ represents node behaviors and structures. For instance, an SBM with community structures arises when $\mathbf{X}$ contains distinct latent positions. This latent position matrix presents a natural, unobserved ``target'' that one might aim to estimate via the observed data $\mathbf{A}$.

  \subsection{Latent Position Models for Time Series of Graphs}
  \label{sec:model}
  Consider a series of graphs $\mathcal{G}^{(1)}, \ldots, \mathcal{G}^{(M)}$ observed across $M$ time points. Each shares the vertex set $\mathcal{V}$ but has distinct edge sets $\mathcal{E}^{(t)}$. This paper addresses graphs with matched vertices and known vertex correspondence, denoted as $\mathcal{V}=[n] = {1, 2, \dots, n}$, where $n$ represents the vertex count. The RDPG model for a time series of graphs is derived from $n$ \emph{individual vertex processes} $\{X_{i}^{(t)}\}_{t=1}^{M}$, where $X_{i}^{(t)} \in \mathbb{R}^{d}$ is the latent position for vertex $i$ at time $t$. Latent positions in graph $\mathcal{G}^{(t)}$ are assembled in the matrix $\mathbf{X}^{(t)}$ = $[X_{1}^{(t)}, \dots ,X_{n}^{(t)}]^{T} \in \mathbb{R}^{n\times d}$. We call the collection $\mathbf{X}^{(t)}, 1 \leq t \leq M,$ the {\em overall vertex process}.  
  
  Observing time series of graphs, it is natural to consider leveraging information from multiple graphs, which motivates us to assume some underlying structures in the overall vertex process. Note that any rank $d$ latent position matrix $\mathbf{X}^{(t)} \in \mathbb{R}^{n\times d}$ can be decomposed as $\mathbf{X}^{(t)} = \mathbf{V}^{(t)} \mathbf{S}^{(t)}$ via QR decomposition, where $\mathbf{V}^{(t)}\in \mathbb{R}^{n\times d}$ consists of orthonormal columns (we call $\mathbf{V}^{(t)}$ an orthonormal basis of the \emph{left singular subspace} of $\mathbf{X}^{(t)}$) and $\mathbf{S}^{(t)} \in \mathbb{R}^{d\times d}$. So it is intuitive to consider characterizing the underlying structure type in the overall vertex process by their subspaces. Next, we introduce three types of structures across time in the overall vertex process:
  \begin{enumerate}
  	\item The latent positions $\mathbf{X}^{(t)} \in \mathbb{R}^{n\times d}$ share the same left singular subspace, making $\mathbf{V}=\mathbf{V}^{(t)}\in \mathbb{R}^{n\times d}$ constant over time, while allowing unique matrices (up to an orthogonal transformation) $\mathbf{S}^{(t)}\in \mathbb{R}^{d\times d}$. This is represented as:
  	\begin{equation}
  	\label{eq:structure1}
  	\mathbf{X}^{(t)} = \mathbf{V} \mathbf{S}^{(t)} \mathbf{W}
  	\end{equation}
  	where $\mathbf{V}$ has $d$ orthonormal columns and $\mathbf{W} \in \mathcal{O}_{d}$ is an arbitrary orthogonal matrix. The subspace spanned by $\mathbf{V}$ is identical to the \emph{invariant subspace} of the connectivity matrices $\mathbf{P}^{(t)}$. A special case of this structure type is the multi-layer stochastic block model \citep{paul2020spectral, matias2017statistical} with positive definite connectivity matrices, wherein nodes' community structure remains fixed throughout time, but between and within community connectivity can vary via $\mathbf{S}^{(t)}$. Such a structure can also encapsulate other static node structures, like mixed memberships or hierarchical communities. This model aligns closely with \cite{draves2020bias} when $\mathbf{S}^{(t)}$ is diagonal.
  	
  	\item In practice, some of the graphs in the time series might present deviations from the shared invariant subspace assumption defined in Equation~(\ref{eq:structure1}), so we characterize  this behavior by allowing changes in $\mathbf{V}$ at some time points $\{t_1,\cdots,t_p\}$. Specifically, the latent positions $\mathbf{X}^{(t)} \in \mathbb{R}^{n\times d}$ share the same singular subspace for $t \in \{1,\cdots,M\}\setminus\{t_{1},\cdots,t_{p}\}$, while other $\mathbf{X}^{(t_{j})}$ are arbitrarily different i.e.,~
  	\begin{equation}
  	\label{eq:structure2}
  	\mathbf{X}^{(t)} = \begin{cases}	\mathbf{V}\mathbf{S}^{(t)}\mathbf{W}  ,& t \in [M]\setminus\{t_{1},\cdots,t_{p}\}  , \\ 			\mathbf{V}^{(t)}\mathbf{S}^{(t)}\mathbf{W} ,& t=t_j\text{, }j=1,\cdots,p.
  	\end{cases}
  	\end{equation}
  	This model can capture some deviations in the graphs at the node level, such as changes in community memberships for some vertices.
  	
  	\item More generally, when there is no shared structure in the vertices across time, all latent positions can be arbitrarily different, that is
  	\begin{equation}
  	\label{eq:structure3}
  	\mathbf{X}^{(t)}= \mathbf{V}^{(t)}\mathbf{S}^{(t)}\mathbf{W}.
  	\end{equation}  
  \end{enumerate}
  The model defined by Equation~(\ref{eq:structure2}) bridges the gap between a model with shared structure in the nodes via the common singular subspace in Equation~(\ref{eq:structure1}) and a model with arbitrarily different node structure in Equation~(\ref{eq:structure3}). Intuitively, statistical inferences about the time series should benefit from a shared structure in the vertices across time, and as such, one of our goals in this paper is to exploit a common structure when possible.

  We can characterize the evolution of the time process by observing the differences between adjacent time points $\mathbf{Y}^{(t)}=  \mathbf{X}^{(t)} - \mathbf{X}^{(t-1)}\mathbf{W}^{*}$, where $\mathbf{W}^{*}=\text{argmin}_{\mathbf{W}\in\mathcal{O}_{d}}\|\mathbf{X}^{(t)}-\mathbf{X}^{(t-1)}\mathbf{W}\|_F$, we call $\mathbf{Y}^{(t)}$ a latent position matrix \emph{difference} between two consecutive time points. When the left singular subspaces of $\mathbf{Y}^{(t)}$ and $\mathbf{X}^{(t-1)}$ are the same, we say that $\mathbf{Y}^{(t)}$ is a \emph{linearly dependent difference}; when the singular subspaces of $\mathbf{Y}^{(t)}$ and $\mathbf{X}^{(t-1)}$ are different, we call $\mathbf{Y}^{(t)}$ a \emph{linearly independent difference} (see Figure~\ref{fig:alpha0125} for a difference example, details in scenario $2$ in Section \hyperref[sec:datagen]{ 5.1}). 
  
  
  The models previously defined are constrained to have positive semidefinite connectivity matrices, but they can be extended to be able to generate  arbitrary low-rank connectivity matrices via the generalized random dot product model \citep{https://doi.org/10.1111/rssb.12509}. This model introduces an indefinite matrix $\mathbf{I}_{p,q}$ to express the connectivity matrix as $\mathbf{P}^{(t)}=\mathbf{X}^{(t)}\mathbf{I}_{p,q}\mathbf{X}^{(t)^{T}}$. Here  $\mathbf{I}_{p,q}\in\mathbb{R}^{d\times d}$ with $p+q=d$ is a diagonal matrix with its first $p$ diagonal entries equal to $1$ and the remaining $q$ entries equal to $-1$. For ease of exposition, we focus only on RDPG models (i.e.,~positive semidefinite GRDPGs).
\section{Methodology}
\label{sec:Method}

                                                                                                                                                                                                                                                                                                                                                                                                                                                                                   Defining our anomaly detection problem, we assume that in the absence of anomalies, the vertex process $\mathbf{X}^{(t)}, 1 \leq t \leq M,$ evolves with an unknown variability $\tau \geq 0$, such that $\|\mathbf{X}^{(t)}-\mathbf{X}^{(t-1)}\mathbf{W}^{*}\|\leq \tau$. The matrix norm $\|\cdot\|$ measures differences in latent positions at successive time points. The purpose of our inference is to detect a local (temporal) behavior change in a time series of graphs. In particular, we define a graph to be ``anomalous'' when a (potentially small, unspecified) collection of vertices change behavior at some time $t^{*}$ as compared to the recent past, while the remaining vertices continue with their normal behavior. As such, it is natural to consider a two-step procedure for anomaly detection: first perform spectral embedding and then assess changes in the estimated latent positions. 
                                                                                                                                                                                                                                                                                                                                                                                                                                                                                   
                                                                                                                                                                                                                                                                                                                                                                                                                                                                                   We consider the detection of anomalies in either the \emph{overall graph} or in \emph{individual vertices}. For the $i$-th vertex at time point $t^{*}$, \emph{individual vertex anomaly detection (VertexAD)} tests the null hypothesis $H_{0i}^{(t^{*})}$ (non-anomalous time for vertex $i$) against the alternative $H_{Ai}^{(t^{*})}$. The null assumes the $i$-th latent state remains within tolerance $\tau_{\text{vertex}}$ at $t^{*}$:
                                                                                                                                                                                                                                                                                                                                                                                                                                                                                     $$H_{0i}^{(t^{*})}: \|X_{i}^{(t^{*})} - X_{i}^{(t^{*}-1)}\|\leq \tau_{\text{vertex}},\quad \text{versus} \quad 
                                                                                                                                                                                                                                                                                                                                                                                                                                                                                   H_{Ai}^{(t^{*})}:  \|X_{i}^{(t^{*})} - X_{i}^{(t^{*}-1)}\|> \tau_{\text{vertex}},$$
                                                                                                                                                                                                                                                                                                                                                                                                                                                                                     where $X_{i}^{(t^{*})} = (\mathbf{X}^{(t)})_{i}$ and $X_{i}^{(t^{*}-1)} = (\mathbf{X}^{(t-1)}\mathbf{W}^{*})_{i}$.
                                                                                                                                                                                                                                                                                                                                                                                                                                                                                   Choosing $\tau_{\text{vertex}}>0$ allows us to consider some variability under the null. While letting $\tau_{\text{vertex}}=0$, this reduces to a classical two-sample test:
                                                                                                                                                                                                                                                                                                                                                                                                                                                                                     $$H_{0i}^{(t^{*})}: X_{i}^{(t^{*})} = X_{i}^{(t^{*}-1)},\quad \text{versus} \quad H_{Ai}^{(t^{*})}: X_{i}^{(t^{*})} \neq X_{i}^{(t^{*}-1)}.$$
                                                                                                                                                                                                                                                                                                                                                                                                                                                                                     
                                                                                                                                                                                                                                                                                                                                                                                                                                                                                     
                                                                  
                                                                
                                                                To perform VertexAD for the $i$-th vertex, we define the test statistic
                                                                \begin{equation}
                                                                y_{i}^{(t)}=\|\widehat{X}_{i}^{(t)} - \widehat{X}_{i}^{(t-1)} \|_{2},
                                                                \label{eq:indverstat}
                                                                \end{equation}
                                                                where $\widehat{X}_{i}^{(t)} $ is the latent position estimate of vertex $i$ at time $t$. Presumably, the latent position estimates are close to the true latent positions \citep{sussman2012consistent}, and this test statistic will be large if there exists a substantial change between the latent position for vertex $i$ between $t-1$ and $t$. An anomaly is detected at time $t$ if $y_{i}^{(t)}$ is large enough for the null hypothesis to be rejected at some specified level.
                                                                
                                                                Based on this formulation of VertexAD, we can analogously define \emph{graph anomaly detection (GraphAD)} as a test of the null hypothesis
                                                                $$H_{0}^{(t^{*})}: \|\mathbf{X}^{(t^{*})} - \mathbf{X}^{(t^{*}-1)} \mathbf{W}^{*}\|\leq \tau_{\text{graph}}, \quad \text{versus} \quad H_{A}^{(t^{*})}: 
                                                                  \|\mathbf{X}^{(t^{*})} - \mathbf{X}^{(t^{*}-1)}\mathbf{W}^{*}\|> \tau_{\text{graph}}. $$
                                                                  
                                                                  The corresponding test statistic is defined as
                                                                \begin{equation}
                                                                y^{(t)}=\|\widehat{\mathbf{X}}^{(t)} - \widehat{\mathbf{X}}^{(t-1)} \widehat{\mathbf{W}}^{*}\|_{F}.
                                                                \label{eq:ovegraphstat}
                                                                \end{equation}
                                                                Here, $\|\mathbf{X}\|_{F}$ denotes the Frobenius norm of a matrix $\mathbf{X}\in\mathbb{R}^{n\times d}$, which corresponds to the square root of the sum of the squares of elements of $\mathbf{X}$. Although other norms can gauge the changes in the time series, we use the Frobenius norm as test statistics since this norm is the one we developed the theory. Results are similar when applying norms such as the $\ell_2$ operator norm of a matrix, which is the maximum singular value of that matrix. 
                                                                
                                                                Now we introduce our method for obtaining latent position estimates: the multiple adjacency spectral embedding (MASE) of \cite{arroyo2021inference}. The MASE algorithm is to estimate the parameters of the common subspace independent edge (COSIE) model in which all the expected adjacency matrices of the graphs, denoted by $\mathbf{P}^{(t)}=\mathbb{E}[\mathbf{A}^{(t)}]$, share the same invariant subspace. That is, these matrices can be expressed as $\mathbf{P}^{(t)}=\widetilde{\mathbf{V}}\mathbf{R}^{{(t)}} \widetilde{\mathbf{V}}^{T}$ with
                                                                $\widetilde{\mathbf{V}}\in \mathbb{R}^{n\times \widetilde{d}}$ with orthonormal columns (we call $\widetilde{\mathbf{V}}$ \emph{common subspace}), but allow each individual matrix $\mathbf{R}^{{(t)}}\in \mathbb{R}^{\widetilde{d}\times \widetilde{d}}$ to be different. This framework accommodates the structure from Equation~(\ref{eq:structure1}). In executing the MASE approach, each adjacency matrix $\mathbf{A}^{(t)}$ undergoes separate spectral decomposition. We denote the corresponding $d$ leading eigenvectors  of $\mathbf{A}^{(t)}$ (corresponding to the $d$ leading eigenvalues in magnitude) as $\widehat{\mathbf{V}}^{(t)}\in\mathbb{R}^{n\times d}$ for each $t\in[M]$. The matrix $\widehat{\mathbf{U}}=\left( \widehat{\mathbf{V}}^{(1)} \ \cdots \ \widehat{\mathbf{V}}^{(M)}\right)$, which is of size $n\times\left(Md\right)$, concatenates these spectral embeddings. Defining $\widehat{\mathbf{V}}$ as the leading left singular vectors of $\widehat{\mathbf{U}}$, we then calculate $\widehat{\mathbf{R}}^{(t)} = \widehat{\mathbf{V}}^{T}\mathbf{A}^{(t)}\widehat{\mathbf{V}}$, resulting in the estimated latent positions $\widehat{\mathbf{X}}^{(t)}=\widehat{\mathbf{V}}\widehat{\mathbf{R}}^{(t)}$. Notably, MASE produces consistent parameter estimation for $\widetilde{\mathbf{V}}$ and $\mathbf{R}^{{(t)}}$ across graphs. Like RDPG, the individual matrix $\mathbf{R}^{(t)} \in \mathbb{R}^{\widetilde{d} \times \widetilde{d}}$ in the COSIE model has an unidentifiability issue in estimating $\mathbf{R}^{(t)}$. However, MASE avoids the cumbersome Procrustes alignments for this issue because it effectively aligns the multiple ASEs to a common spectral embedding in the joint SVD step. Therefore, our MASE-based method avoids alignment when computing the test statistics $y^{(t)}$ and $y_{i}^{(t)}$.

                                                                We discuss the motivation for \text{MASE} under the structure types described in Section \hyperref[sec:model]{ 2.2}. It is clear that Equation~(\ref{eq:structure1}) can be formulated as a COSIE model with $\widetilde{d}=d$. Consider cases where graphs may deviate from the shared structure represented by the singular subspace, $\mathbf{V}$. For example, consider latent positions of type~\eqref{eq:structure3}. Such deviations can still be represented in the common subspace, but this might require an increased rank dimension when using a COSIE model. Such representation pays the price of building up the model complexity with more parameters compared with~\eqref{eq:structure1}. Specifically, let $\mathbf{U}$ be a matrix $\mathbf{U}=[\mathbf{V}^{(1)} ,\mathbf{V}^{(2)},\cdots,\mathbf{V}^{(M)} ]\in\mathbb{R}^{n\times Md}$, and suppose that $d'=\text{rank}(\mathbf{U})$. This matrix can be decomposed via singular value decomposition as $\mathbf{U} = \mathbf{V}'\mathbf{W}$, where $\mathbf{V}'\in\mathbb{R}^{n\times d'}$ is a matrix with orthonormal columns and
                                        $\mathbf{W} = [\mathbf{W}^{(1)}, \cdots, \mathbf{W}^{(M)}]$ is a $d'\times (Md)$ matrix. Hence, the latent positions of type \eqref{eq:structure3} can be expressed as $\mathbf{X}^{(t)} = \mathbf{V}'\mathbf{S}^{(t)} \mathbf{W} $,
                                        which is similar to the shared singular subspace in~\eqref{eq:structure1}, but now the singular subspace has rank dimension $d'>d$. Therefore, it is natural to characterize the deviation from the shared singular subspace in~(\ref{eq:structure1}) via some distance between the subspaces spanned by the columns of $\mathbf{V}$ and $\mathbf{V}'$, or the difference between $d'$ and $d$. Intuitively, \text{MASE} is preferable when the graphs are close to~(\ref{eq:structure1}) as they are under the COSIE model with dimension $d$. 

\begin{algorithm}
 	\caption{Two-step anomaly detection with bootstrapped \text{p}-value}
 	\begin{algorithmic} 
 		\Input  A time series of graphs $\{\mathbf{A}^{(t)}\}_{t=1}^{M}$, embedding dimensions $d$, time span $s$.

 		\begin{enumerate}
   \item For each $t$ in $1:M-(s-1)$, obtain the latent position estimates $(\{\widehat{\mathbf{X}}^{(u)} \}_{u=t}^{t+s-1},d)=\mase(\mathbf{A}^{(t)}, \dots,\mathbf{A}^{(t+s-1)})$ within time span $s$, then calculate  $y^{(v)}$ and $y_{i}^{(v)}$ for vertex $i=1, \dots,n$ based on ~(\ref{eq:indverstat}) and~(\ref{eq:ovegraphstat}) for times $v=t+1, \dots, t+s-1$.
        \item Employ parametric bootstrap using $\widehat{\mathbf{X}}=\widehat{\mathbf{X}}^{(t)}$ to generate $B$ samples, $y_b^{(t)}$ under the null hypothesis $\X=\mathbf{X}^{(t-1)}$, and $y_{ib}^{(t)}$ assuming $X_{i}^{(t-1)}=X_{i}^{(t)}$. Determine empirical \text{p}-values at $t$ as $p^{(t)}= \frac{\sum_{b=1}^{B} I(y_b^{(t)}>y^{(t)})}{B}$ and for each vertex $i$ as $p_{i}^{(t)}= \frac{\sum_{b=1}^{B} I(y_{ib}^{(t)}>y_{i}^{(t)})}{B}$.

            \item {Order the empirical \text{p}-values $\{p^{(t)}\}_{t=1}^{M}$ in decreasing order as $p(i)$, $i=1,...,M$. Compute the BH-adjusted \text{p}-values $Q^{\text{BH}}_{\text{adj}}(p(i))=\min\biggl\{1, \min_{j\geq i} \{\frac{Mp(j)}{j} \}\biggr\}$. Similarly, for the decreasing order of \text{p}-values $\{p_{i}^{(t)}\}_{t=1}^{M}$ as $p(k)$, $k=1,...,nM$, compute: $Q^{\text{BH}}_{\text{adj}}(p(k))=\min\biggl\{1, \min_{j\geq k} \{\frac{nMp(j)}{j} \}\biggr\}$ }.

 		\end{enumerate}

 		\Output  Report adjusted empirical \text{p}-values $Q^{\text{BH}}_{\text{adj}}(p^{(t)})$ at $t$ for GraphAD and adjusted empirical \text{p}-values $Q^{\text{BH}}_{\text{adj}}(p_{i}^{(t)})$ at time point $t$ and vertex $i$ for VertexAD, $t\in [M]$, $i \in [n]$.
 		

 	\end{algorithmic}
  \label{alg:hypothesis testing fdr control}
 \end{algorithm}

For \text{MASE}, the distribution of the test statistic can be obtained via the semi-parametric bootstrap from \cite{tang2017semiparametric}. Specifically, for any $\widehat{\mathbf{X}}\in \mathbb{R}^{n\times d}$ we generate \text{i.i.d.} samples of $\mathbf{A}\sim \rdpg{\widehat{\mathbf{X}} }$ and obtain the corresponding test statistics under $\rdpg{\widehat{\mathbf{X}}}$. The two-step procedure for anomaly detection via reporting significant adjusted p-values based on semi-parametric bootstrap is as follows: first, perform joint spectral embedding with time span $s=2$ or $s=M$, i.e.~either jointly embed adjacent graphs or jointly embed all available graphs as the number of embedding graphs can affect downstream inference task; then the second step is to perform hypothesis testing for $H_{0}$ or $H_{0i}$ as summarised in Algorithm~\ref{alg:hypothesis testing fdr control}. {To control the false discovery rate (FDR) in detecting multiple anomalies within the time series of graphs, we apply a classical multiple-hypothesis testing procedure \citep{benjamini1995controlling}.}

Next, we introduce a second approach to GraphAD and VertexAD with a two-step procedure using control charts. Control charts \citep{shewhart1986statistical} are a tool for analyzing process changes over time and are utilized to provide quantitative evidence regarding whether process variation is in control or out of control. Our approach considers graphs in a time window of length $l$ ending just before $t^{*}$: $\mathcal{W}^{(t^{*},l)}:=\{t^{*}-l, \dots, t^{*}-1\} \subseteq \{1, \dots,M\} $. In our control charts the tolerance $\tau_{\text{vertex}}=\tau_{\text{vertex}}^{(t^{*},l)}$ is a measure of dispersion of the latent positions $\{\mathbf{X}^{(t)}:t \in \mathcal{W}^{(t^{*},l)}  \}$. There are four fundamental elements in control charts: estimated statistics, moving average mean, moving average measure of dispersion, and rule to claim out-of-control points. In step one, we do the same joint spectral embedding with time span $s=2$ or $s=M$ as in Algorithm~\ref{alg:hypothesis testing fdr control}, and calculate the corresponding estimated statistics in equations~(\ref{eq:indverstat}) and~(\ref{eq:ovegraphstat}). In step two, we test the null hypotheses $H_{0}^{(t^{*})}$ and $H_{0i}^{(t^{*})}$ sequentially for $t^{*}=l+1, \dots, M$ using control charts with time window length $l$. 
To determine the tolerances $\tau_{\text{graph}}$ and $\tau_{\text{vertex}}$ in $H_{0}^{(t^{*})}$ and in $H_{0i}^{(t^{*})}$, we jointly embed $l$ graphs over a time window $\mathcal{W}^{(t^{*},l)}$ and obtain the corresponding test statistics
$\widetilde{y}^{(t^{*}-l+1)}, \dots,\widetilde{y}^{(t^{*}-1)}$. Subsequently, the moving averages and adjusted moving range \citep{montgomery2007introduction} are determined as follows:
\begin{equation}
\label{eq:movingaverage}
    \bar{y}^{(t^{*})}= \frac{\sum_{v'=t^{*}-l+1}^{t^{*}-1} \widetilde{y}^{(v')} }{l-1} ,
\end{equation}
\begin{equation}
    \label{eq:movingrange}
    \bar{\sigma}^{(t^{*})}= \frac{1}{1.128(l-2)} \sum_{v'=t^{*}-l+2}^{t^{*}-1} |\widetilde{y}^{(v')} - \widetilde{y}^{(v'-1)} |  .
  \end{equation}
  Here the constant $1.128$ is the mean of \emph{relative range}, which is the ratio between sample range and standard deviation assuming two samples are drawn from the identical independently distributed normal distribution.

For VertexAD, we use the moving averages and UnWeighted AVErage of subgroup estimates based on subgroup Standard Deviations (``UWAVE-SD") \citep{wetherill1991statistical}
\begin{equation}
\label{eq:movingaveragever}
    \bar{y}_{i}^{(t^{*})}= \frac{\sum_{i=1}^{n} \sum_{v'=t^{*}-l+1}^{t^{*}-1} \widetilde{y}_{i}^{(v')} }{ n(l-1)} ,
\end{equation}
\begin{equation}
    \label{eq:movingsdver}
    \bar{\sigma}_{i}^{(t^{*})}= \frac{1}{c(n)(l-1)} \sum_{v'=t^{*}-l+1}^{t^{*}-1} \widehat{\sigma}(v') ,
\end{equation}
    where  $\widehat{\sigma}(v')$'s are the sample standard deviations of $\widetilde{y}_{i}^{(v')}$ over $n$ vertices at time $t$, $c(n)=\sqrt{2/(n - 1)}\exp(\log \gamma(n/2) - \log \gamma((n - 1)/2))$ and $\gamma(\cdot)$ is the Gamma function. 
The control chart plots include a central line (CL) representing the moving average, with the upper control line (UCL) determined by $\bar{y}^{(t^{*})}+ 3 \bar{\sigma}^{(t^{*})} $. Anomalies or out-of-control points are highlighted with triangles on the control charts. This procedure is further visualized in Figure~\ref{fig:idealconchart1} (details in Section \hyperref[sec:datagen]{ 5.1}) and encapsulated in Algorithm~\ref{alg:controlchart} with $s=2$. {The approach described in this section is implemented in the R\footnote{\url{https://github.com/gdchen94/TSG-Anomaly-Detection/}}.}
\begin{algorithm}[ht]
 	\caption{Two-step anomaly detection with control charts}
 	\begin{algorithmic} 
 		\Input A time series of graphs $\{\mathbf{A}^{(t)}\}_{t=1}^{M}$, embedding dimensions $d$, joint embedding method $\mase$, time span $s=2$, and time window length $l$.
 		
 		\begin{enumerate}
 		    \item \algorithmiciterate{
 		    for $t=1,\cdots,M-1$}
 		    \begin{enumerate}
		        \item At time $t$, obtain the latent position estimates $(\{\widehat{\mathbf{X}}^{(u)} \}_{u=t}^{t+1},d)=\mase(\mathbf{A}^{(t)}, \mathbf{A}^{(t+1)})$ within time span $s$, then calculate  $y^{(t+1)}$ and $y_{i}^{(t+1)}$ for vertex $i=1, \dots,n$ based on equations~(\ref{eq:indverstat}) and~(\ref{eq:ovegraphstat}).
		        \item If $t>l$, calculate $\bar{y}^{(t)}$ and  $\bar{\sigma}^{(t)}$, and $\bar{y}_{i}^{(t)}$ and  $\bar{\sigma}_{i}^{(t)}$ for vertex $i$ based on equations~(\ref{eq:movingaverage}),  (\ref{eq:movingrange}), (\ref{eq:movingaveragever}) and (\ref{eq:movingsdver}) using $\{\widetilde{y}^{(v')}\}_{v'=t-l+1}^{t-1}$ and $\{\widetilde{y}_{i}^{(v')}\}_{v'=t-l+1}^{t-1}$.
 		    \end{enumerate}

 		\end{enumerate}

 		\Output  Report anomalous graphs or vertices using $y^{(t)}$ and $y_{i}^{(t)}$ based on Shewhart's rule (see Appendix for details) from the control chart. 
		          \end{algorithmic}

		          \label{alg:controlchart}
		          \end{algorithm}
		          
		          \section{Theory}
		          \label{sec:theory}
		          
		          
		          Here, we focus on the two-sample hypothesis testing problem described in Section~\ref{sec:Method} for graph anomaly detection with $\tau_{\text{graph}}=0$. We prove that the two-sample hypothesis test is an asymptotic consistent level $\alpha$ test under mild assumptions. Our results are based upon known theoretical results on Multiple Adjacency Spectral Embedding of~\cite{arroyo2021inference}, which served as our estimates. For completeness, we briefly present them in the Appendix.

As we discussed in Section~\ref{sec:model}, consider any two latent positions $\mathbf{X}^{(t)}$ and $\mathbf{X}^{(t-1)}$, they can be represented as $\mathbf{X}^{(t)}=\mathbf{V}\mathbf{R}^{(t)}$ and $\mathbf{X}^{(t-1)}=\mathbf{V}\mathbf{R}^{(t-1)}$ where $\mathbf{V}$ is their shared common subspace and $\mathbf{R}^{(t)}$, $\mathbf{R}^{(t-1)}$ are their corresponding individual matrices. The Frobenius norm difference $\|\mathbf{X}^{(t)} - \mathbf{X}^{(t-1)}\|_{F}$ boils down to the Frobenius norm difference between $\mathbf{R}^{(t)}$ and $\mathbf{R}^{(t-1)}$. So the graph anomaly detection problem defined in {Section}~\ref{sec:Method} is equivalent to testing the hypothesis of equality of individual matrices $\mathbf{R}^{(t)}$ and $\mathbf{R}^{(t-1)}$. We thus present our two-sample semiparametric hypothesis testing results about the individual matrices in this section. Before stating our results, we define the semi-parametric graph hypothesis testing consistency as described in~\citep{tang2017semiparametric}, borrowing from the classical notion of consistency.

\begin{definition} (\emph{Consistent Asymptotic Level $\alpha$ Test}) 
	Let $\mathbf{R}^{(t)}, \mathbf{R}^{(t+1)}$ be two full rank symmetric matrices of size $d\times d$, a test statistics $T_{n}$ and associated rejection region $R_{n}$ to test the null hypothesis that
	$$ H_{0}^{n}: \mathbf{R}^{(t)} =  \mathbf{R}^{(t+1)},$$  
	against:
	 $$H_{a}^{n}: \mathbf{R}^{(t)} \neq  \mathbf{R}^{(t+1)}$$ 
	is a \emph{consistent asymptotically level $\alpha$ test} if for any $\eta >0$, there exists $n_{0}= n_{0}(\eta)$ such that
	\begin{enumerate}
		\item If $n>n_0$ and $H_{a}^{n}$ is true, then $\mathbb{P}(T_n\in R_n) > 1 - \eta$.
		\item If $n>n_0$ and $H_{0}^{n}$ is true, then $\mathbb{P}(T_n\in R_n) \leq \alpha + \eta$.
		
	\end{enumerate}
\label{def:consistency}
\end{definition}

We then have the following results.

\begin{theorem}
	For each fixed $n$, consider the hypothesis test
	
	$$H_{0}^{n}: \mathbf{R}^{(t)} =\mathbf{R}^{(t+1)} \quad \text{versus} \quad H_{a}^{n}:  \mathbf{R}^{(t)} \neq \mathbf{R}^{(t+1)} $$
where $\mathbf{R}^{(t)}$ and $\mathbf{R}^{(t+1)} \in \mathbb{R}^{d\times d}$ are score matrices defined in the COSIE model. Let the $\{\widehat{\mathbf{R}}^{(t)}\}_{t=1}^m$ be the multiple adjacency spectral embedding estimates for $\{\mathbf{R}^{(t)}\}_{t=1}^m$. Define the test statistics

 $$T_{n} = \frac{\| \widehat{\mathbf{R}}^{(t)} - \widehat{\mathbf{R}}^{(t-1)}  \|_{F}}{d^2}.$$ 
 
 Let $\alpha \in (0,1)$ be given and suppose the requirements in Theorem~\ref{thm:COSIE_CLT} in the Appendix holds. Then there exists a constant $C'$, for all $C>C'$, if the rejection region is $R:=\{r\in\mathbb{R}: r\geq C\}$, then there exists an $n_{1} = n_{1}(\alpha, C) \in \mathbb{N}$ such that for all $n\geq n_{1}$, the test procedure with $T_{n}$ and rejection region $R$ is at most level $\alpha$ test, i.e.,~for all $n\geq n_{1}$, if $\mathbf{R}^{(t)} =\mathbf{R}^{(t+1)}$, then
$$\mathbb{P}(T_{n}\in R) \leq \alpha.$$

Furthermore, consider the sequence of $(\mathbf{V};\mathbf{R}^{(1)},\cdots,\mathbf{R}^{(m)})$ for each $n\in \mathbb{N}$, and denote by $d_{n}$ the quantity
$$d_{n}:= \| \mathbf{R}^{(t)} - \mathbf{R}^{(t+1)} \|_{F}.$$

Suppose $d_{n}\neq 0 $ for infinity many $n$. Let $t_{1}=\min\{k>0:d_{k}>0\}$ and sequentially define $t_{n} = \min\{k>t_{n-1}:d_{k}>0\}$. Let $b_{n}=d_{t_n}$. If $\liminf b_{n} = \infty$, then this test procedure is consistent in the sense of Definition \ref{def:consistency} over this sequence of $(\mathbf{V};\mathbf{R}^{(1)},\cdots,\mathbf{R}^{(m)})$.
\label{thm:main}
\end{theorem}

To illustrate Theorem~\ref{thm:main}, consider the following example on the Erd\"os-R\'enyi (ER) model. In the ER model $G(n,p)$, the edges between any pair of vertices are connected via the same probability $p$ independently. Consider $m$ samples of $(\mathbf{A}^{(1)}, \ldots,\mathbf{A}^{(m)})\sim\cosie{\mathbf{V};R^{(1)}, \ldots, R^{(m)}}$ are a sample of adjacency matrices from the COSIE model with $\mathbf{V} = \frac{1}{\sqrt{n}}\mathbf{1}_{n}\in \real^{n}$ and $R^{{(t)}}=np_{t}$ such that each sample adjacency matrix $\mathbf{A}^{(t)}$ follows an ER model $G(n, p_t)$. Assuming $\{p_{t}\}_{t=1}^{m}$ are distinct constant probabilities, we observe that as $n\rightarrow \infty$ for any arbitrary adjacent pair of $t$, $t+1$, if $p_t \neq p_{t+1}$ then
$\|R^{(t)}- R^{(t+1)}\|_{F}=\sqrt{n(p_{t}^{2}-p_{t+1}^{2})}\rightarrow \infty$. Therefore, the sequence $b_n$ diverges as required by the conditions of the theorem.
Additionally, we examine assumptions for Theorem~\ref{thm:COSIE_V-Vhat} and technical assumptions~\ref{assump:Delocalization},~\ref{assump:variance_P} and~\ref{assump:score_lambdamin} for Theorem~\ref{thm:COSIE_CLT} in the Appendix. Simple calculations show that $\delta(\mathbf{P}^{(t)}) = np_{t+1}$, $\varepsilon = \sqrt{\frac{1}{mn} \sum_{t=1}^{m} \frac{1}{p_{t}  }}=O(\frac{1}{\sqrt{n}})$, $s^2(\mathbf{P}^{(t)})=n^2p_{t}(1-p_{t})$ and $|\lambda_{\min}( \mathbf{\Sigma}^{(t)})| = \Omega(\frac{1}{n})$. Theorem~\ref{thm:main} then implies that for $t' \in \{t, t+1\}$, $\|\widehat{R}^{(t')} - R^{(t')}\|_{F} = O_{P}(d^{2}), $ and thus the test procedure with test statistics $T_n=\frac{\| \widehat{R}^{(t)} - \widehat{R}^{(t-1)}  \|_{F}}{d^2}$ is consistent in the sense of Definition~\ref{def:consistency}. We end this section by noting that the results we obtained here can be improved by recent tighter concentration bounds in~\cite{zheng2022limit}.


\subsection{Detectability Analysis}
We performed a detectability analysis on our algorithms. Under the general assumptions of Theorem~\ref{thm:COSIE_CLT} introduced in the Appendix, assume we observe a time series of graphs that follow the COSIE model $(\mathbf{A}^{(1)}, \ldots, \mathbf{A}^{(m)}) \sim(\mathbf{V};\mathbf{R}^{(1)},\cdots,\mathbf{R}^{(m)})$. Consider our test statistic estimated via $\text{MASE}$, $y^{(t)} = \|\widehat{\mathbf{X}}^{(t)} - \widehat{\mathbf{X}}^{(t-1)} \widehat{\mathbf{W}}^{*}\|_{F}$. It will then be equal to
\[
y^{(t)} = \|\widehat{\mathbf{R}}^{(t)} - \widehat{\mathbf{R}}^{(t-1)}\|_{F} = \| \widehat{\mathbf{R}}^{(t)} - \mathbf{W}^\top \mathbf{R}^{(t)} \mathbf{W} + \mathbf{W}^\top \mathbf{R}^{(t)} \mathbf{W} - \widehat{\mathbf{R}}^{(t+1)} + \mathbf{W}^\top \mathbf{R}^{(t+1)} \mathbf{W} - \mathbf{W}^\top \mathbf{R}^{(t+1)} \mathbf{W}\|_F.
\]
Here $\mathbf{W} \in \mathcal{O}_{d}$ is an orthogonal matrix. With Lemma~\ref{lem:COSIE_R_decomp} introduced in the Appendix, there exist sequences of matrices $\mathbf{H}_{m}, \mathbf{W} \in \mathbb{R}^{d \times d}$ depending on $d$, $m$, and $n$
\[
\mathbf{W}\widehat{\mathbf{R}}^{(i)}\mathbf{W}^\top - \mathbf{R}^{(i)} + \mathbf{H}_{m} = \mathbf{V}^{\top}\mathbf{E}^{(i)}\mathbf{V},
\]
and $\mathbf{W} = \operatorname{\textnormal{arg inf}}_{\mathbf{W} \in \mathcal{O}_{d}} \|\widehat{\mathbf{V}} - \mathbf{V} \mathbf{W}\|_{F}$.

Thus, $y^{(t)}$ will be equal to
\[
\| -\mathbf{W}\mathbf{H}^{(t)}_m\mathbf{W}^\top + \mathbf{V}^\top \mathbf{E}^{(t)} \mathbf{V} + \mathbf{R}^{(t)} + \mathbf{W}\mathbf{H}^{(t+1)}_m\mathbf{W}^\top - \mathbf{V}^\top \mathbf{E}^{(t+1)} \mathbf{V} - \mathbf{R}^{(t+1)}\|_F.
\]

For a matrix $\mathbf{R} \in \mathbb{R}^{d \times d}$, we define $\operatorname{vec}(\mathbf{R}) \in \mathbb{R}^{d^2}$ as the vector of dimension $d^2$ that contains all elements of $\mathbf{R}$ in a column-major order. For $i, j \in [d]$,
\[
[\operatorname{vec}(\mathbf{R})]_{i + (j-1)d} := \mathbf{R}_{ij}.
\]

Let $\boldsymbol{\delta} = \operatorname{vec}(\mathbf{R}^{(t)} - \mathbf{R}^{(t+1)})$ and $\mathbf{e} = \operatorname{vec}(-\mathbf{W}\mathbf{H}^{(t)}_m\mathbf{W}^\top + \mathbf{W}\mathbf{H}^{(t+1)}_m\mathbf{W}^\top)$. We have
\[
\|\mathbf{e} + \boldsymbol{\delta}\|_{2} - \| \mathbf{V}^\top(-\mathbf{E}^{(t)} + \mathbf{E}^{(t+1)})\mathbf{V} \|_{F} \leq y^{(t)} \leq \|\mathbf{e} + \boldsymbol{\delta}\|_{2} + \| \mathbf{V}^\top(\mathbf{E}^{(t)} - \mathbf{E}^{(t+1)})\mathbf{V} \|_{F}.
\]
Taking expectation,
\[
\mathbb{E}[\|\mathbf{e} + \boldsymbol{\delta}\|_{2} - \| \mathbf{V}^\top(-\mathbf{E}^{(t)} + \mathbf{E}^{(t+1)})\mathbf{V} \|_{F}] \leq \mathbb{E}[y^{(t)}] \leq \mathbb{E}[\|\mathbf{e} + \boldsymbol{\delta}\|_{2} + \| \mathbf{V}^\top(\mathbf{E}^{(t)} - \mathbf{E}^{(t+1)})\mathbf{V} \|_{F}].
\]
Using Jensen's Inequality, we have
\[
\mathbb{E}[\|\mathbf{e} + \boldsymbol{\delta}\|_{2}] - \| \mathbb{E}[\mathbf{V}^\top(-\mathbf{E}^{(t)} + \mathbf{E}^{(t+1)})\mathbf{V}] \|_{F} \leq \mathbb{E}[y^{(t)}] \leq \mathbb{E}[\|\mathbf{e} + \boldsymbol{\delta}\|_{2}] + \sqrt{\mathbb{E} [\| \mathbf{V}^\top(\mathbf{E}^{(t)} - \mathbf{E}^{(t+1)})\mathbf{V} \|_{F}^2]}.
\]
Let $\mathbf{\sigma}_{\mathbf{P}}^{2} = \mathbb{E} [\| \mathbf{V}^\top(\mathbf{E}^{(t)} - \mathbf{E}^{(t+1)})\mathbf{V} \|_{F}^2]$. With Theorem~\ref{thm:COSIE_CLT} introduced in the Appendix, we have $\operatorname{vec}(\widehat{\mathbf{R}}^{(t)})$ and $\operatorname{vec}(\widehat{\mathbf{R}}^{(t+1)})$ have finite variance assuming $d$ is finite. Furthermore, their covariance $\operatorname{Cov}(\operatorname{vec}(\widehat{\mathbf{R}}^{(t)}), \operatorname{vec}(\widehat{\mathbf{R}}^{(t+1)}))$ will converge to zero, as $\lim\limits_{m \rightarrow \infty}\|\mathbf{H}^{(t)}_m \|_{F} = 0$ and $\mathbf{E}^{(t)}$ and $\mathbf{E}^{(t+1)}$ are mutually independent.

If $l \rightarrow \infty$, we have $\bar{y}$ will be approximately $\mathbb{E}[y^{(t)}]$ provided the variance of $y^{(t)}$ is bounded and the covariance $\operatorname{cov}(y^{(i)}, y^{(j)})$ goes to zero as $|i - j| \rightarrow \infty$. We thus have the change to be detectable if
\[
\|\mathbf{e} + \boldsymbol{\delta}\|_{2} - \mathbf{\sigma}_{\mathbf{P}} \geq \|\mathbf{e}\|_{2} + \mathbf{\sigma}_{\mathbf{P}}
\]
in the asymptotic regime. This feasibility condition implies
\begin{equation}
\|\boldsymbol{\delta}\|_{2}^{2} + 2 \|\mathbf{e}\|_{2} (\|\boldsymbol{\delta}\|_{2} \cos \theta - 2 \mathbf{\sigma}_{\mathbf{P}}) \geq 4 \mathbf{\sigma}_{\mathbf{P}}^2,
\label{inequality:feasibility}
\end{equation}
where $\theta$ is the angle between $\boldsymbol{\delta}$ and $\mathbf{e}$. This sufficient condition suggests that a larger $\boldsymbol{\delta}$ helps detection. Similarly, the detection will be better with a smaller bias term $\mathbf{e}$ in estimation or a smaller variance $\mathbf{\sigma}_{\mathbf{P}}$ stemming from sampling the connection probability matrices. Furthermore, if $\cos \theta \leq \frac{2\mathbf{\sigma}_{\mathbf{P}}}{\|\boldsymbol{\delta}\|_{2}}$, a small change $\boldsymbol{\delta}$ may be undetected.

We provide an illustrative Erd\"os-R\'enyi (ER) model to provide insights on this feasibility condition.

\begin{example}
	Consider $m$ samples of $(\mathbf{A}^{(1)}, \ldots, \mathbf{A}^{(m)}) \sim \cosie{\mathbf{V}; R^{(1)}, \ldots, R^{(m)}}$ are samples of adjacency matrices from the COSIE model with $\mathbf{V} = \frac{1}{\sqrt{n}} \mathbf{1}_{n} \in \mathbb{R}^{n}$ and $R^{{(t)}} = np_{t}$ such that each sample adjacency matrix $\mathbf{A}^{(t)}$ follows an ER model $G(n, p_t)$. We assume $p_t = p$ stays constant before $t_{c}$ and changes to $q = p + \Delta$ after $t_{c}$. In Appendix A, we show that the following bound: there exists a constant $C$, such that
	\[
	\mathbb{P}(\|e + \delta\|_2 - \mathbf{\sigma}_{\mathbf{P}} > \|e\|_2 + \mathbf{\sigma}_{\mathbf{P}}) \geq 1 - \frac{C \cdot d}{\sqrt{m}} \cdot \frac{2n\Delta + 4\sqrt{n}\sqrt{p^2 + (p + \Delta)^2}}{[n^2\Delta^2 - 4n[p^2 + (p + \Delta)^2]]}.
	\]
	holds asymptotically in $n$. 
\end{example}

This means that if we let $\frac{n\Delta\sqrt{m}}{d}$ go to infinity, then the change will be detected with high probability. In other words, the method detects changes up to an order of $dn^{-1}m^{-0.5}$.

\begin{remark}
	Comparing our feasibility condition \eqref{inequality:feasibility} with the result in \cite{marenco2022online}, we have an additional term $\mathbf{\sigma}_{\mathbf{P}}$, which is due to the difference between our control chart and CUSUM statistics. By switching to the CUSUM statistics, we can similarly overcome this variance in sampling from the connectivity probability matrix, but potentially at the cost of detection delay. Note the detectable change order coincides with the order in \cite{marenco2022online}. However, in \cite{marenco2022online}, they need to assume the latent position stays the same before and after the change, while here, we can take the flexibility of the COSIE model by allowing variable connectivity probability $q$.
\end{remark}

\section{Simulations of Time Series of Graphs}
\label{sec:illus}

In this section, we evaluate our methods' performance through simulations across various scenarios and explore the impact of hyper-parameter combinations on the outcomes. Regarding hyper-parameters, Algorithm~\ref{alg:hypothesis testing fdr control} utilizes a threshold for adjusted p-values set at level $\alpha=0.05$, and Algorithm~\ref{alg:controlchart} has a test statistic threshold defined as $\bar{y}^{(t)}+ 3 \bar{\sigma}^{(t)}$. The default time window length is $l=11$ unless stated otherwise. Since joint embedding methods' performance varies with the number of embedded graphs, we assess time spans $s=2$ or $s=M$ with both algorithms. As an example, for $M$ graphs and $s=2$, latent positions are estimated from adjacent graphs, we called it \text{MASE(2)}. When $s=M$, positions are derived by embedding all available graphs, we labeled it \text{MASE(12)} if $M=12$.

\subsection{Data Generation}
\label{sec:datagen}
\emph{Scenario 1:} Starting with an example, we sample one-dimensional ($d=1$) latent positions $X_1^{(1)}, \dots, X_n^{(1)} \in \mathbb{R}$ uniformly from $U(0.2, 0.8)$ for $n=100$. These positions shape the vector $X^{(1)}$, which generates the graph $\mathcal{G}^{(1)}$ with adjacency matrix $\mathbf{A}^{(1)}$: $\mathbf{A}^{(1)} \sim \rdpg{X^{(1)}}$. Analogously, we form graphs $\mathcal{G}^{(t)} \sim \rdpg{X^{(t)}}$ for $t=-9, \dots, 12$ (note that in order to show results of Algorithm~\ref{alg:controlchart} starting at $t=2$ with $l=11$, our graphs start at time $-9$). For all $t \notin \{6, 7\}$, the latent positions are the same: $X^{(t)} = X^{(1)}$. At time points $6$ and $7$, using $\delta_{x}$ and $\delta_{n}$ to govern the magnitude and count of altered vertices, we tweak their corresponding latent positions: $X^{(6)} = X^{(1)} + \delta_{x} \cdot \Delta$ and $X^{(7)} = X^{(1)} - \delta_{x} \cdot \Delta$, where $\Delta = (1_{\delta_n/2}^{T}, -1_{\delta_n/2}^{T}, 0_{n - \delta_n}^{T})^{T} \in \mathbb{R}^{n}$ affects the first $\delta_n = 20$ vertices with $\delta_{x}=0.12$. This results in a time series of $M = 22$ graphs, each comprising 100 vertices, with anomalies at times $6$ and $7$ encompassing 20 anomalous vertices each.

\begin{figure}[ht]
    \centering
    \begin{subfigure}[b]{.44\textwidth}
        \includegraphics[width=\textwidth]{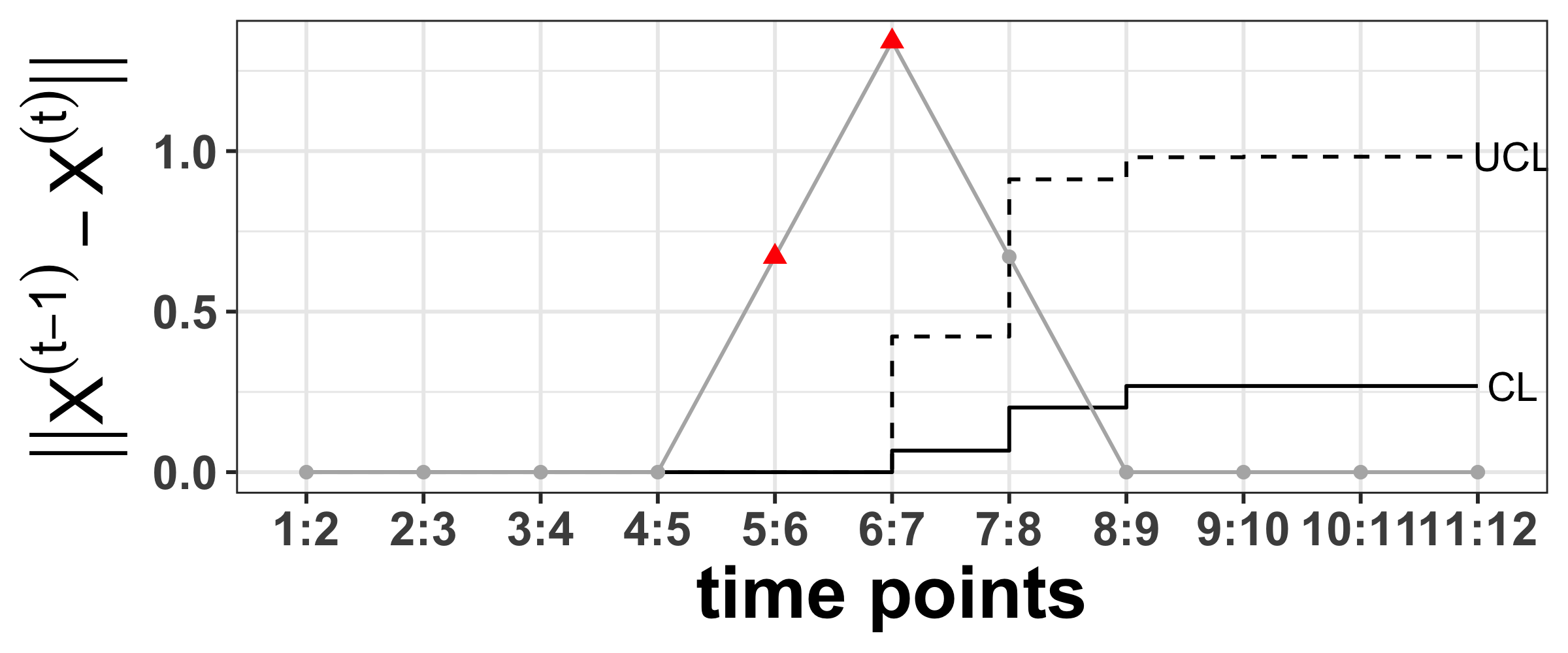}
        \caption{Illustrative control chart}
        \label{fig:idealconchart1}
    \end{subfigure}
    \begin{subfigure}[b]{.44\textwidth}
        \includegraphics[width=\textwidth]{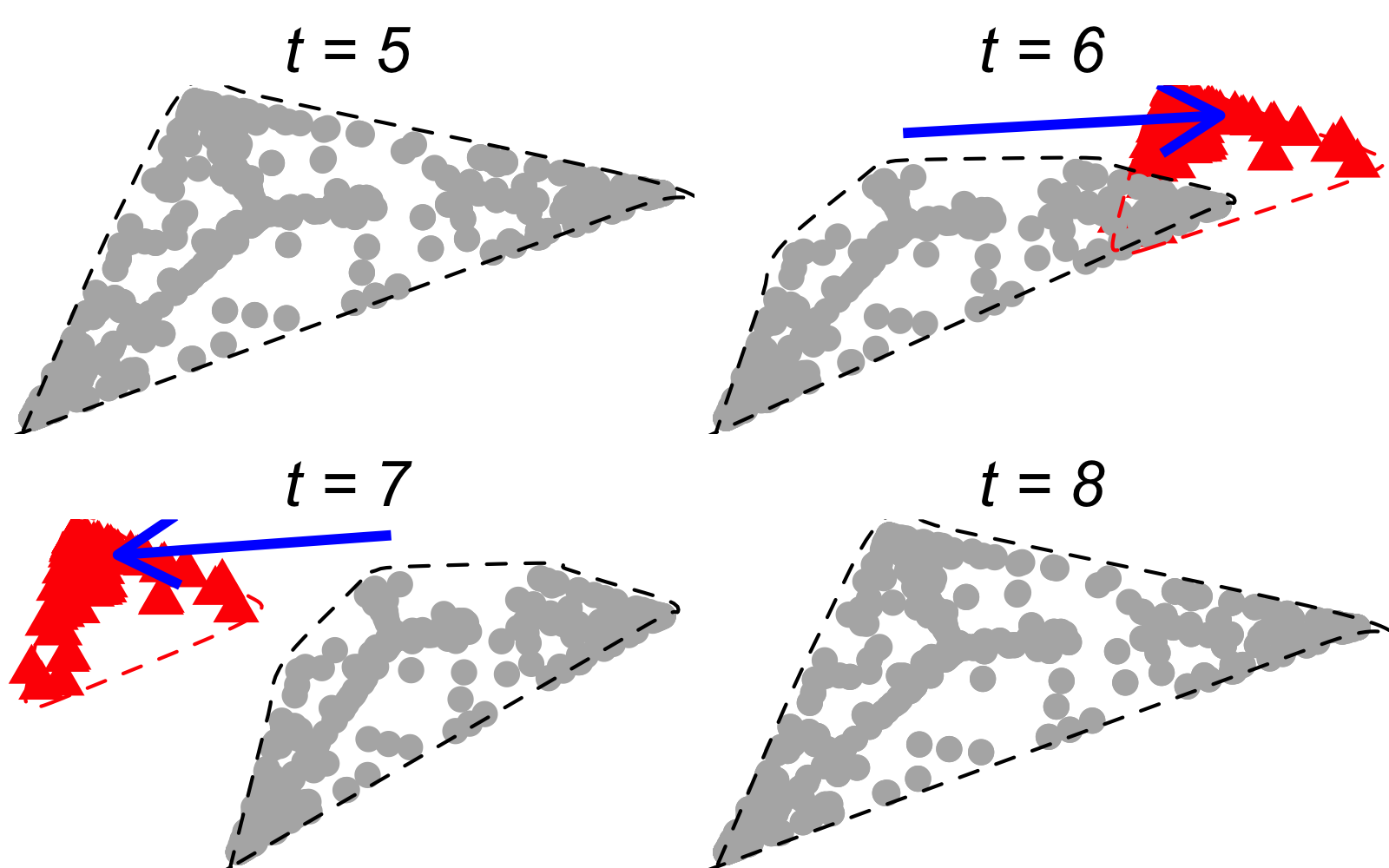}
        \caption{Latent position plot for time series of graphs, scenario $2$}
        \label{fig:alpha0125}
    \end{subfigure}
    \caption{For the control chart, dots are $\|X^{(t)}-X^{(t-1)}\|$, the solid line shows the mean, and the dashed line shows the mean plus three adjusted moving sample ranges. Triangles indicate anomalies. In the latent position plot, dots and triangles represent normal and anomalous latent positions at time points $5,6,7,8$, with anomalies at $6$ and $7$ affecting one community's connectivity.}
		            \end{figure}

The control chart presented in Figure~\ref{fig:idealconchart1} provides intuition for the test statistic $y^{(t)}$ and motivation for applying Algorithm~\ref{alg:controlchart} in GraphAD. Figure~\ref{fig:idealconchart1} plots $\|X^{(t-1)}-X^{(t)}\|$, $t=2,\dots,12$ from scenario $1$, with the central line (CL) being the average of $11$ of the $y^{(t)}$ and the dashed line (UCL) being the average plus three adjusted moving ranges. So, when $\widehat{X}^{(t)} $ is sufficiently close to $X^{(t)}$, $y^{(t)}$ at an anomalous time point will lie outside of UCL, which motivates us to investigate Algorithm~\ref{alg:controlchart} in this scenario.

\emph{Scenario 2:}
We utilize graphs derived from a mixed membership stochastic block model (MMSBM) \citep{airoldi2008mixed} with constant community structures over time. Consequently, the singular vectors of latent positions adhere to an invariant subspace $\mathbf{V} \in \mathbb{R}^{n\times 4}$ as per Equation~(\ref{eq:structure1}). For each time step $t=1,\cdots,12$, we generate $\mathcal{G}^{(t)}$ comprising $n=400$ vertices using the MMSBM. The block connectivity matrix is given by: $\mathbf{B}=(p-q)\mathbf{I}_{4}+q1_{4}1_{4}^\top$
with $p\sim U(0.5,1)$, $q\sim U(0,0.5)$. Each row of $\mathbf{Z}$, representing community memberships, generates from $\mathbf{Z}_{i}  \sim \dirich{\theta \cdot 1_{4}}$, where $\theta=0.125$. Anomalies are introduced at specific time points 6 and 7, perturbing only one MMSBM community's connectivity. The difference matrix $\mathbf{\Delta}$ primarily influences the first $\delta_{n}=100$ nearest vertices of a selected vertex. In order to assess the robustness of our methods under different parameters, we add a distribution for the perturbation $\mathbf{\Delta}_{i \cdot} \sim .6 \cdot \dirich{1_{4}}+.2$ in simulations.

\emph{Scenario 3:} Graphs mirror scenario $2$, but $\mathbf{Z}_{i}$ for graph $\mathcal{G}^{(t)}$ generates independently from $ \dirich{0.875 \cdot 1_{4}}$. As a result, the difference $\mathbf{\Delta}$ at anomalous time points influences not just community connectivity but also vertex memberships.

    \subsection{Results}
    \label{sec:result}
    
    \begin{figure}[!h]
    \centering
    \includegraphics[width=\linewidth]{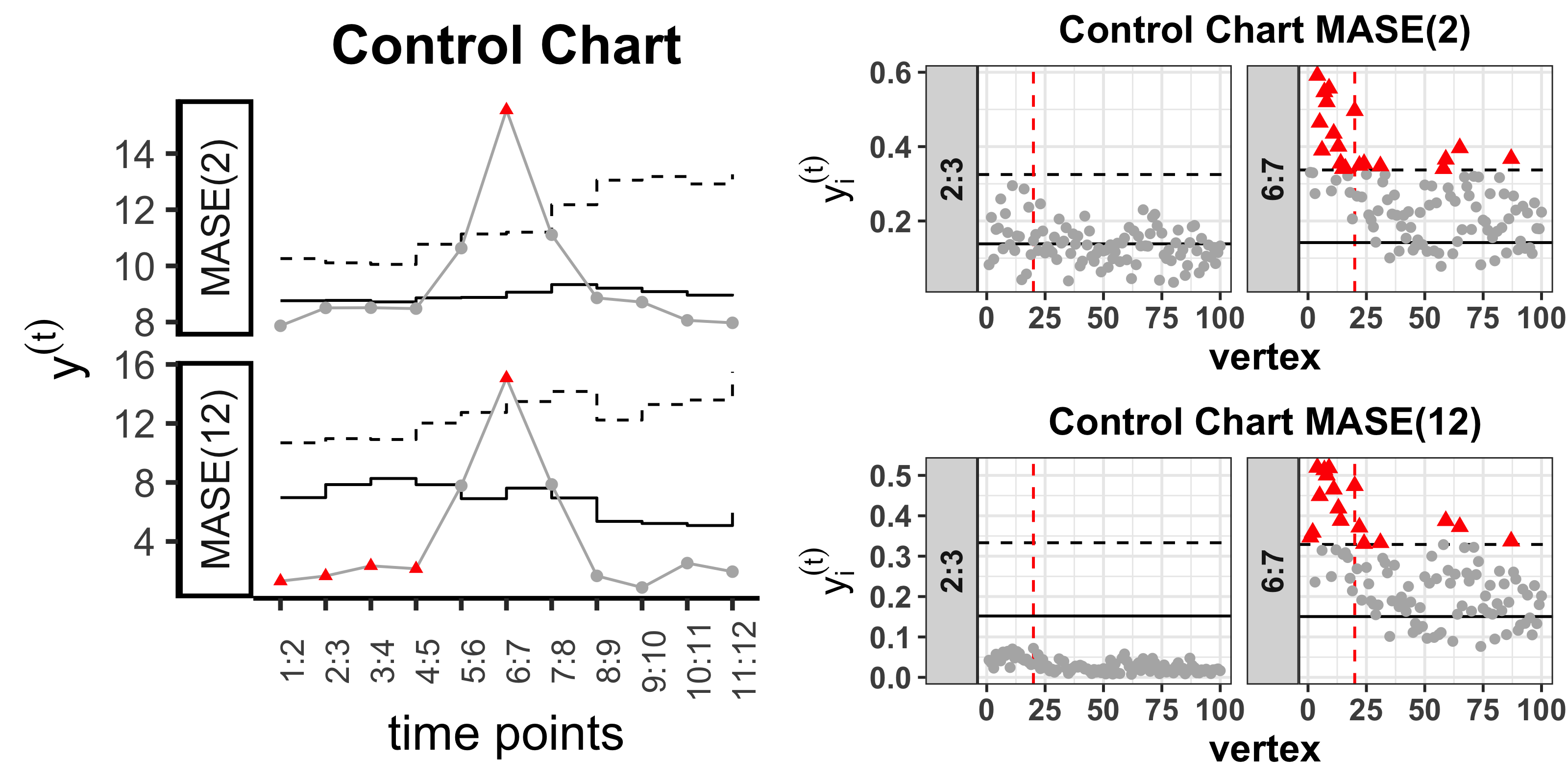}
    \caption{Control charts for time series of
      graphs with anomalies at time points $6$ and $7$ (scenario $1$). Left: GraphAD with the center line (CL) indicating the moving average of sample means and the dashed line (UCL) as the threshold for anomaly detection. Dots and triangles represent normal and anomalous graphs, respectively. Right: VertexAD with the same indicators, dots, and triangles denoting normal and anomalous vertices. True anomalous vertices are the first 20 at time points 6:7, marked by vertical dashed lines.}
    \label{fig:conchart1}
    \end{figure}

    With Figure~\ref{fig:idealconchart1} in mind, we first present results of our Algorithm~\ref{alg:controlchart} for scenario $1$, which should approximate Figure~\ref{fig:idealconchart1} but plotting $y^{(t)}$ instead of $\|X^{(t)}-X^{(t-1)}\|$ and adjusting the calculation of CL and UCL based on $y^{(t)}$. The control chart generated from Algorithm~\ref{alg:controlchart} for GraphAD is presented in the left panel of Figure~\ref{fig:conchart1}, and for VertexAD in the right panel, demonstrating the applicability of our approaches in scenario $1$.

    In Figure~\ref{fig:conchart1}'s left panel, Algorithm~\ref{alg:controlchart} with time spans $s=2$ and $s=12$ effectively detects anomalies at time points $6$ and $7$. The right panel uses the same time spans for VertexAD, highlighting its capability to detect the majority of anomalous vertices, with most of the first 20 vertices beyond the UCL. 
 All vertices at time $t$ have a common CL and UCL, as there's no inherent order among vertices like there is with time. We include only two-time points in Figure~\ref{fig:conchart1} for display purposes: one for normal adjacent time points $2:3$ and the other for anomalous time points $6:7$. Other normal cases are similar to the normal ones included here. Notably, charts based on \text{MASE} differ with varying time spans, indicating a substantial impact of time span $s$ on \text{MASE} for both GraphAD and VertexAD.

    
    \begin{figure}[ht]
    \centering
    \includegraphics[width=\linewidth]{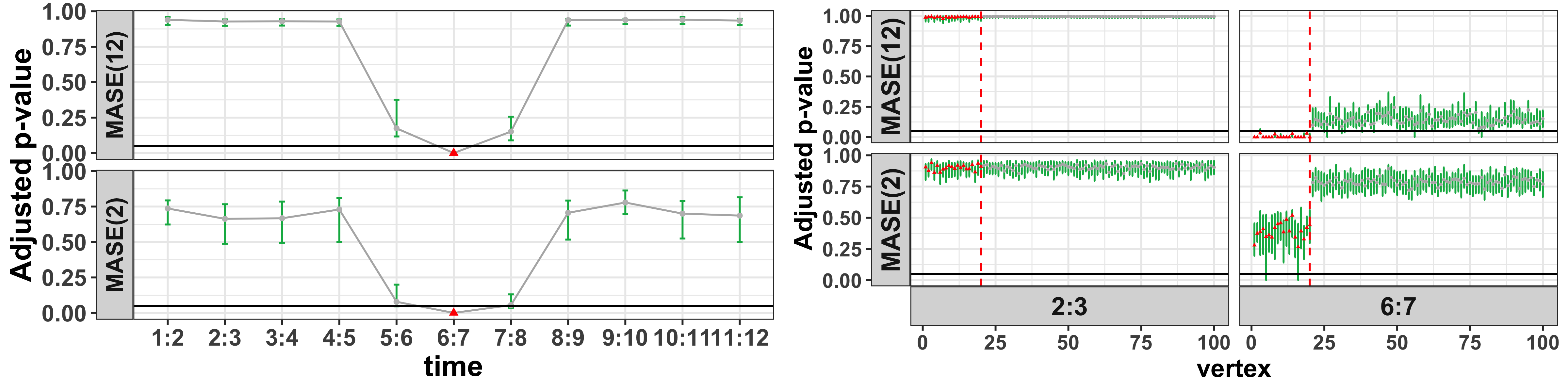}
    \caption{Hypothesis testing for time series of graphs from Figure~\ref{fig:conchart1}. Left: Error bars show confidence intervals for adjusted p-values, and horizontal lines mark the significant level of $0.05$. Right: Error bars indicate confidence intervals for vertex-specific adjusted p-values. True anomalous vertices are the first 20 at time points 6:7, marked by vertical dashed lines.}
    \label{fig:mco}
    \end{figure}
    
\begin{figure}
    \centering
    \begin{subfigure}[b]{.49\textwidth}
    \includegraphics[width=\textwidth]{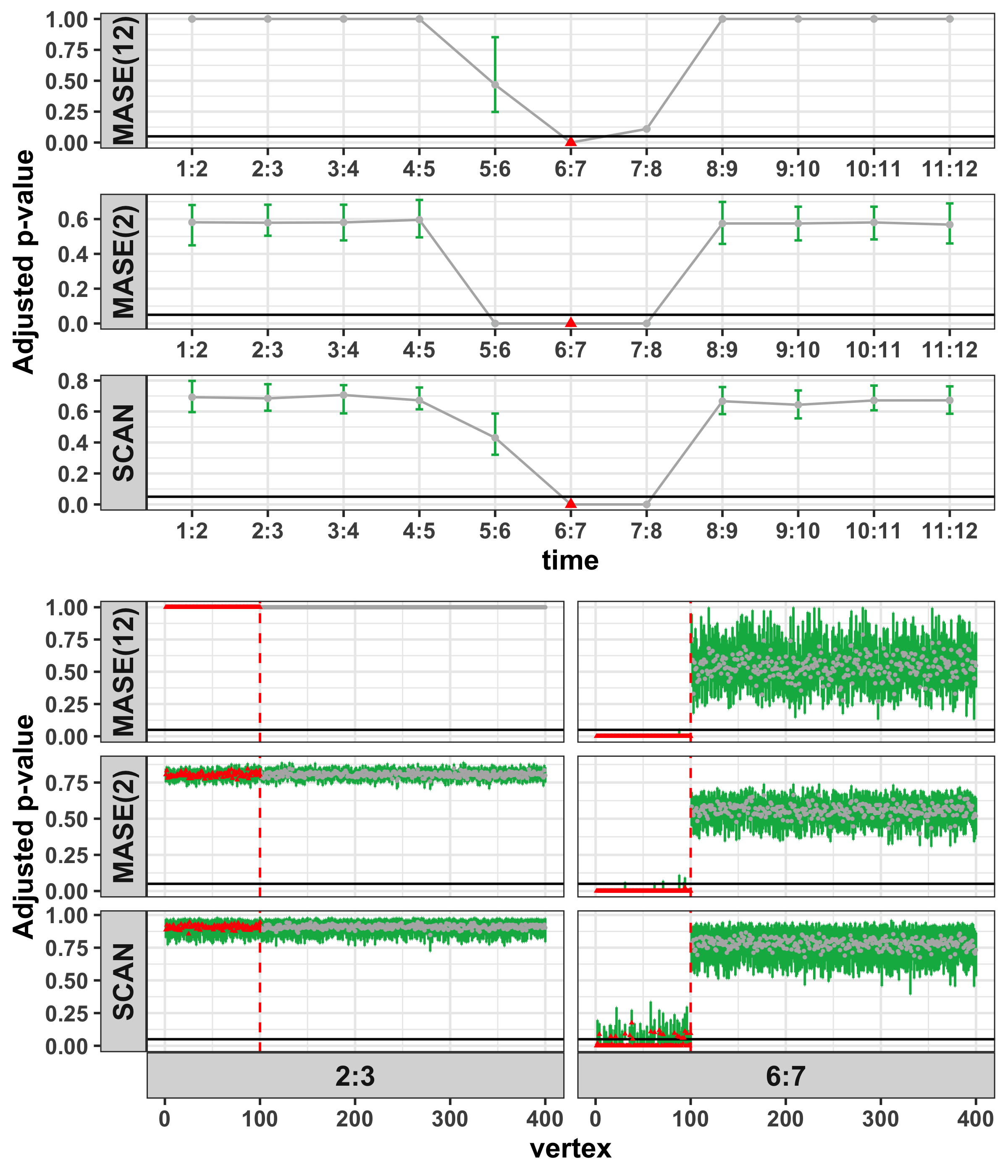}
    \caption{Scenario $2$} 
		            \label{fig:sdmultid4mar09}
		            \end{subfigure} 
		            \begin{subfigure}[b]{.49\textwidth}
		            \includegraphics[width=\textwidth]{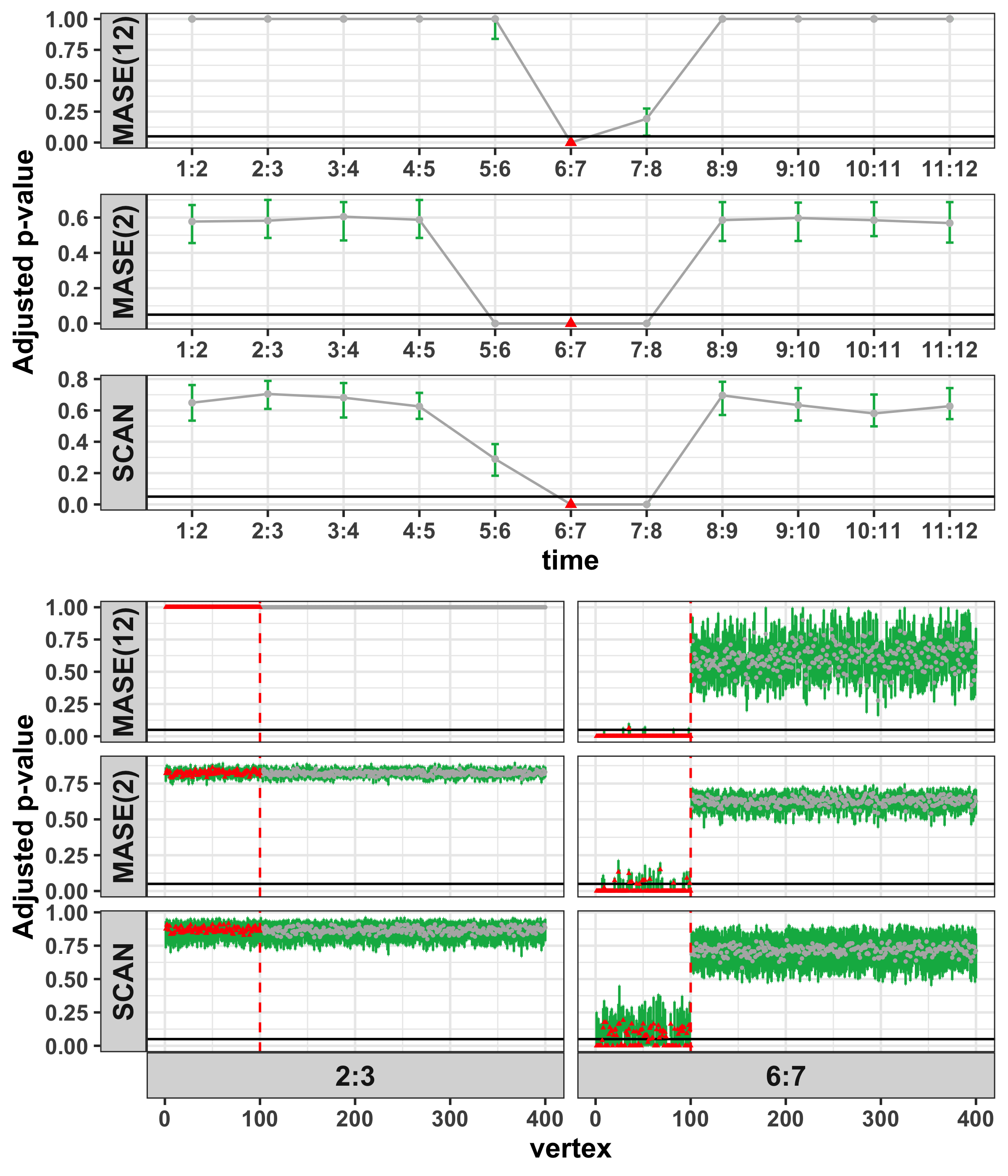}
		            \caption{Scenario $3$} 
		          \label{fig:ddmultid4mar09}
		          \end{subfigure}
		          \caption{Hypothesis testing for time series of graphs under scenarios $2$ and $3$. Error bars represent confidence intervals for adjusted p-values. Dots show the median adjusted p-values and horizontal lines mark the significant level of $0.05$. True anomalous vertices are the first 100 at time points 6:7, marked by vertical dashed lines.}
		          \label{fig:mmsbm4}
		          \end{figure}

    In the context of two-sample testing that considers $\tau=0$, we first generate $400$ samples of test statistic $y^{(t)} $ and $y_{i}^{(t)}$ under the null hypothesis to obtain the null distribution, then calculate adjusted p-values and claim the existence of an anomaly at statistically significant adjusted p-values with level $0.05$. Another $200$ Monte Carlo simulation is implemented to obtain the median estimate for adjusted \text{p}-values for \text{MASE} with different time spans $s$, and corresponding simultaneous confidence intervals are also supplied using the Bonferroni correction. The results of Algorithm~\ref{alg:hypothesis testing fdr control} are presented in Figures~\ref{fig:mco} and~\ref{fig:mmsbm4}. 
    {The results indicate that \text{MASE} performs effectively for GraphAD in scenario $1$ when using Algorithm~\ref{alg:hypothesis testing fdr control}. However, for VertexAD in the same scenario, \text{MASE(2)} fails to detect vertex anomalies at a significance level of $0.05$. In contrast, \text{MASE(12)} proves to be effective.}

		          {Figure~\ref{fig:sdmultid4mar09} and Figure~\ref{fig:ddmultid4mar09} address scenarios $2$ and $3$. We compare the scan statistics by \cite{wang2013locality} \text{SCAN} as a benchmark, which utilizes time-normalized edge count in subgraphs for anomaly detection. In these scenarios, both \text{MASE} and \text{SCAN} effectively detect GraphAD anomalies with significant adjusted \text{p}-values, as shown in the top panel of Figures~\ref{fig:sdmultid4mar09} and~\ref{fig:ddmultid4mar09}. The bottom panel of Figure~\ref{fig:sdmultid4mar09} demonstrates both \text{MASE(2)} and \text{MASE(12)} successfully identify anomalous vertices, underscoring \text{MASE}'s efficacy in the approximate \text{MMSBM} in scenario $2$. 
		            In the same panel of Figure~\ref{fig:ddmultid4mar09}, however, \text{MASE(2)} fails to detect a few vertex anomalies at time points $6:7$ in Scenario 3, at a significance level of $0.05$. This is due to the uniform distribution of membership vectors, which results in less clustered vertices, complicating the detection of shifted vertices. In contrast, \text{MASE(12)} successfully detects the anomalies, suggesting that a larger $s$, and consequently a larger sample size, can enhance detection power in this scenario. This effect might be due to \text{MASE(12)}'s ability to balance the bias-variance trade-off with a dimension smaller than the underlying common subspace's dimension, particularly when latent positions remain mostly constant. Finally, the bottom panel of both Figures~\ref{fig:sdmultid4mar09} and~\ref{fig:ddmultid4mar09} shows that \text{SCAN} fails to detect anomalous vertices in both scenarios at a significance level of $0.05$.}

\subsection{Comparison between \text{MASE} and baselines}
\label{subsec:compare}

In this section, we evaluate \text{MASE} against four literature baselines using simulated data derived from the scenarios described in Section \ref{sec:datagen}. We again generate $400$ samples of the test statistics $y^{(t)}$ and $y_{i}^{(t)}$ under the null hypothesis to obtain the null distribution. Similarly, we use adjusted p-values to claim the existence of an anomaly when they reach a statistically significant level of $0.05$. Additionally, we conduct $100$ Monte Carlo simulations to estimate the power and false positive rates. However, for \text{GP}, which is computation-intensive, we estimate its power and FPR using only $20$ Monte Carlo simulations, although it accesses the same number of graphs in training as the other methods. Graphs are generated using community membership preferences $\mathbf{Z}_{i}$, $i=1, \dots,n$ sourced from $\dirich{\theta \cdot 1_{4}}$, with fixed parameters $p=0.8$, $q=0.3$, and $\mathbf{\Delta}_{i \cdot}=0.3 \cdot 1_{4}\in \mathbb{R}^{4}$. The time series of graphs exhibit an invariant subspace $\mathbf{V}$ in Equation~(\ref{eq:structure1}), with perturbations in latent positions influenced by the parameter $\theta$; $\theta$ describes the extent of linear dependency between difference and subspace $\mathbf{V}$. For $\theta =0$, graphs are from SBM with invariant block structures, while for $\theta = 1$, they are from MMSBM with varying community memberships. {We benchmark our test statistics against several established methods: the scan statistics by \cite{wang2013locality}, the multiple graph embedding method \text{OMNI} by \cite{8215766}, the \text{GP} approach (where we use the Frobenius distance for comparability) by \cite{josephs2023bayesian}, and the direct comparison method that calculates the Frobenius norm difference between adjacency matrices in consecutive time pairs, $\|\mathbf{A}^{(t)}-\mathbf{A}^{(t+1)}\|_{F}$ (denoted as \text{DIST}). The \text{SCAN} method devises scan statistics to detect anomalies based on the time-normalized edge count in the subgraphs induced by vertices situated at a distance no greater than $k$.
\text{OMNI} performs scaled spectral embedding on the omnibus matrix, where multiple graphs on the same vertex set are jointly embedded into a single space with a distinct representation for each graph. \text{GP} encodes the networks in the kernel of a Gaussian process prior using their pairwise differences and discriminates anomalies based on the elbow method, specifically the largest jump among the sorted posteriors.} Our analysis focuses on comparing GraphAD and VertexAD performance using Algorithm~\ref{alg:hypothesis testing fdr control}. 
Both \text{MASE} and \text{OMNI} employ an embedding dimension $d=4$. Note that \text{GP} \citep{josephs2023bayesian} is only applicable to GraphAD. 

\begin{figure}
    \centering
    \includegraphics[width=\textwidth]{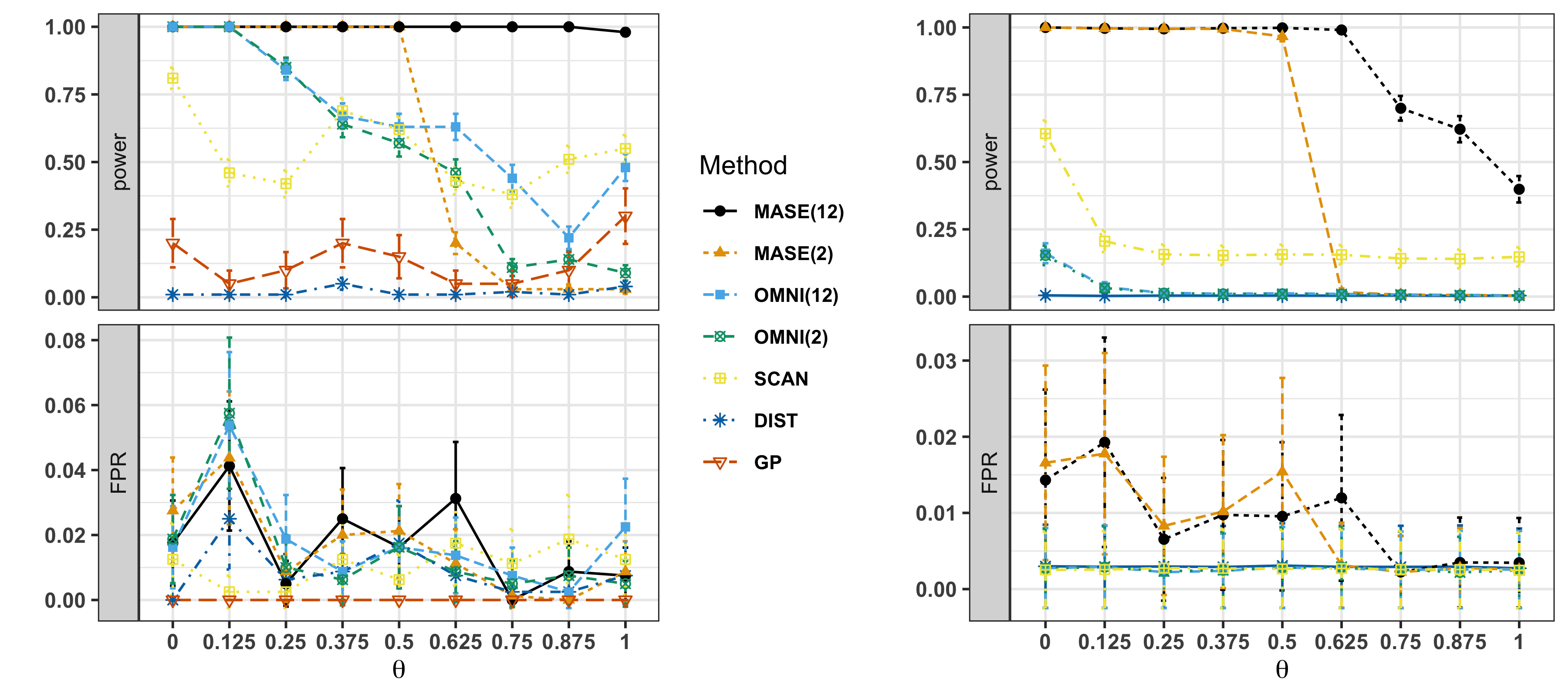}
    \caption{Measures of anomaly detection performance in time series of graphs (Scenario $2$, Section 5.1), generated with varying $\theta$ and fixed parameters $p=0.8$, $q=0.3$, and $\mathbf{\Delta}_{i \cdot}=0.3\cdot 1_{4}$. The left panel shows performance in GraphAD at a significance level of $0.05$, while the right panel illustrates performance in VertexAD. Error bars represent $\text{mean} \pm \text{standard error}$.}
    \label{fig:comparison}
\end{figure}

{Results for GraphAD at anomalous time points $6:7$ are depicted in the left panel of Figure~\ref{fig:comparison}, where we also display the corresponding false positive rate (bottom left). The false positive rates for all methods are effectively controlled to below the level of $0.05$ using adjusted p-values.} Our method demonstrates superior performance compared to other approaches when the model is correctly specified, i.e., when $\theta=0$. Additionally, it maintains a comparative advantage in different scenarios even when the model is misspecified, specifically when $\theta>0$. As $\theta$ ranges from $0$ to $1$, nodes are more inclined to associate with multiple communities, leading to ambiguous clustering patterns. This phenomenon results in the power for both methods approaching $0$ as the anomaly signal is less evident. For $\theta\approx 0$, \text{MASE} and \text{OMNI} outperform \text{SCAN}, largely due to their reliance on eigenvectors from adjacency matrices and the invariant subspace of connectivity matrices. The test statistics based on \text{MASE} and \text{OMNI} can benefit from this invariant subspace by embedding multiple graphs, which makes them more competitive compared with \text{SCAN}. As $\theta$ increases, \text{OMNI}'s efficiency diminishes due to variance dominating bias, especially given its greater parameter usage compared to \text{MASE}. {However, after $\theta$=0.5, \text{OMNI} surpasses \text{MASE(2)}, as it better captures graph differences with increasing $\theta$. As depicted in Figure~\ref{fig:comparison}, larger values of $s$ yield superior detection power, with \text{MASE(12)} nearing a power of 1 due to its balance of bias and variance, especially when the majority latent positions remain constant. We also observed that the power of \text{SCAN} statistics is more viable in the simulations from Figure~\ref{fig:comparison} as $\theta$ increases. There exist some theoretical results about the power of \text{SCAN} statistics \citep{wang2013locality}; however, it is beyond the scope of this paper. The direct comparison method \text{DIST} fails to detect any significant anomalies while the Gaussian process-based approach \text{GP} improves on it marginally.}

		          To evaluate \text{MASE}, \text{SCAN}, \text{OMNI}, and \text{DIST} within the context of VertexAD, we calculated the power and false-positive rates, aggregating across vertices and presenting the results in Figure~\ref{fig:comparison}. The analysis revealed that the false-positive rates are well-controlled at a significance level of $0.05$ for all the methods. Furthermore, \text{MASE(2)} performs well when the model is correctly specified with $\theta=0$ and demonstrates robust performance even when the model is misspecified ($\theta>0$). Similar to the observations in GraphAD, \text{MASE(12)} outperforms its competitors even under model misspecification. Additionally, we utilized the Area Under the Curve (AUC) metric and designed a rank-based metric for evaluating the detection of anomalous vertices at times $6:7$. Results are included in the Appendix Figure~\ref{fig:comparison_auc_mrr}. In these two metrics, the superiority of \text{MASE(12)} over other methods except \text{SCAN} persists. The comprehensive evaluation of vertex anomaly detection is achieved through the analysis of power, false positive rate, AUC, and rank-based metrics. When the model is correctly specified, the superiority of \text{MASE(12)} is pronounced; in scenarios where the model is misspecified, \text{MASE(12)} maintains its superiority at a significance level of $0.05$. It remains competitive in identifying anomalous vertices in general, though its efficacy diminishes slightly in pinpointing the most extreme anomalies.

		          \subsection{Algorithm Choice of Hyper-parameters}
		          
		          {Our algorithms depend on two critical parameters: the time span of multiple graph embedding $s$ and the time span used for constructing the in-process or null distribution $l$. As illustrated in Figure~\ref{fig:comparison}, a larger $s$ improves the detection power in graph-level anomaly detection when the network's eigenspace remains stable. However, a larger $s$ also results in a more significant variance when estimating the latent positions if the network's eigenspace changes regularly. Thus, the choice of $s$ should be based on the variability of the network's eigenspace and could be set as the number of time points where the eigenspace remains similar. As for $l$, we recommend selecting a value equal to the length of the stationary period in the time series of graphs, which allows for an accurate estimation of the null distribution. We make a last remark about the embedding dimension $d$. While the dimensions of the latent positions and common subspace are often unknown, they can be inferred by estimating the ranks of the adjacency matrices. The scree plot method \citep{zhu2006automatic}, which identifies an ``elbow" in the matrices' singular values, offers a straightforward and automated approach for dimension selection. In this work, we employed this method for dimension selection by applying it to the square root of the first $n$ leading eigenvalues, unless a specific dimension was explicitly stated otherwise. Our approaches require sequential multiple-graph embeddings; in practice, the dimensions for these embeddings can vary. To compare test statistics across different dimensions $d$ in real data analysis, we utilize the $\ell_2$ operator norm instead of the Frobenius norm. To assess the impact of dimension selection on our methods, we conducted a sensitivity analysis using the scree plot method on all $n$ leading eigenvalues for both simulation studies and real data analysis. The Bing data analysis was excluded from this part of our sensitivity study due to an out-of-memory error. The outcomes of this analysis, detailed in the Appendix, highlight our results' adaptability to dimension selection strategy changes. 

\section{Real Data Application}
\label{sec:realdata}
{In this section, we analyze the well-documented Enron email dataset and a dataset from a large-scale commercial search engine. For our analyses, we employ the competitors \text{OMNI} and \text{SCAN}, utilizing both adjusted p-values and control charts. 
}
\subsection{Enron Emails}

We analyze the Enron email dataset as used in \cite{priebe2005scan, wang2013locality}, which contains emails among $148$ users from $1998$ to $2002$. We make no distinction between sending and receiving the email, and we normalized the email counts as weights within the range $[0, 2]$. This process results in a monthly weighted undirected time series of graphs with matched vertices over $t=44$ periods. In Figure~\ref{fig:enronGraphADqcc}, we present the control chart results for parameters $s=2$ and $l=11$, corresponding to a time window of one year. Additional results for varied $s$ and $l$ are in the Appendix. Both \text{MASE(2)} and \text{MASE(12)} identified graph anomalies in August 2000, as well as May-June and August-September 2001. The \text{MASE(12)} detected additional anomalies in February and July of 2001. The August 2000 anomaly corresponds to Enron's stock price peak, while anomalies from May and June 2001 coincide with contentious exchanges during Enron's Quarterly Conference Call and the termination of Enron’s largest foreign investment in the Dabhol Power Company, India \citep{James.P.Galasyn.2002}. The August 2001 anomaly correlates with CEO Skilling's resignation \citep{James.P.Galasyn.2002}; September 2001's anomaly corresponds with Enron Chairman Lay's restructuring announcement and the controversial advice to employees to buy Enron shares \citep{James.P.Galasyn.2002}. In comparison, \text{OMNI(2)} detected fewer anomalies overall, but it identified an extra anomaly in February 2000. On the other hand, \text{SCAN} identified anomalies in October 2000 and during April-June 2001. 
\begin{figure}
    \centering
    \begin{subfigure}[b]{.59\textwidth}
    \includegraphics[width=\textwidth]{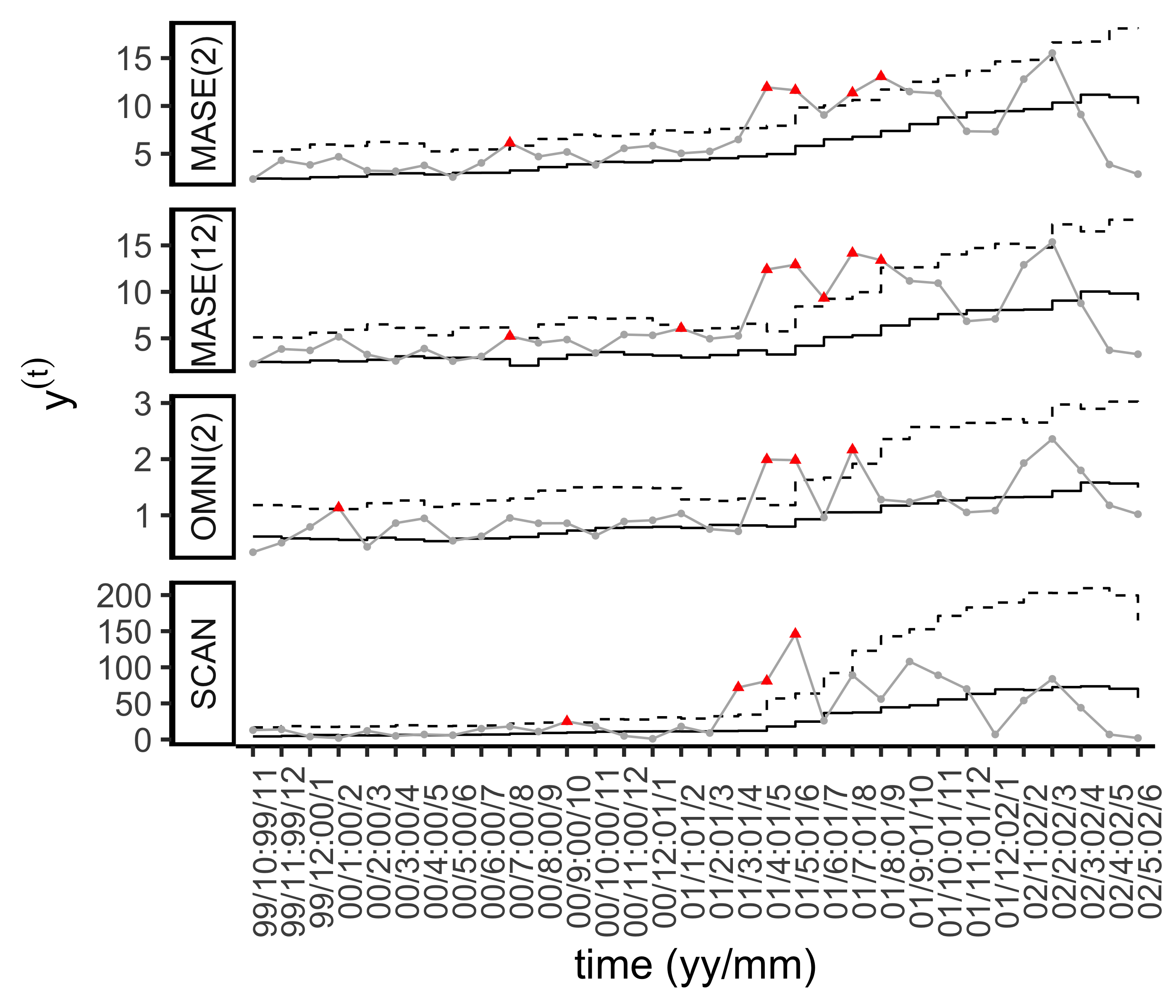}
    \caption{\label{fig:enronGraphADqcc}}
    \end{subfigure} 
    \begin{subfigure}[b]{.39\textwidth}
    \includegraphics[width=\textwidth]{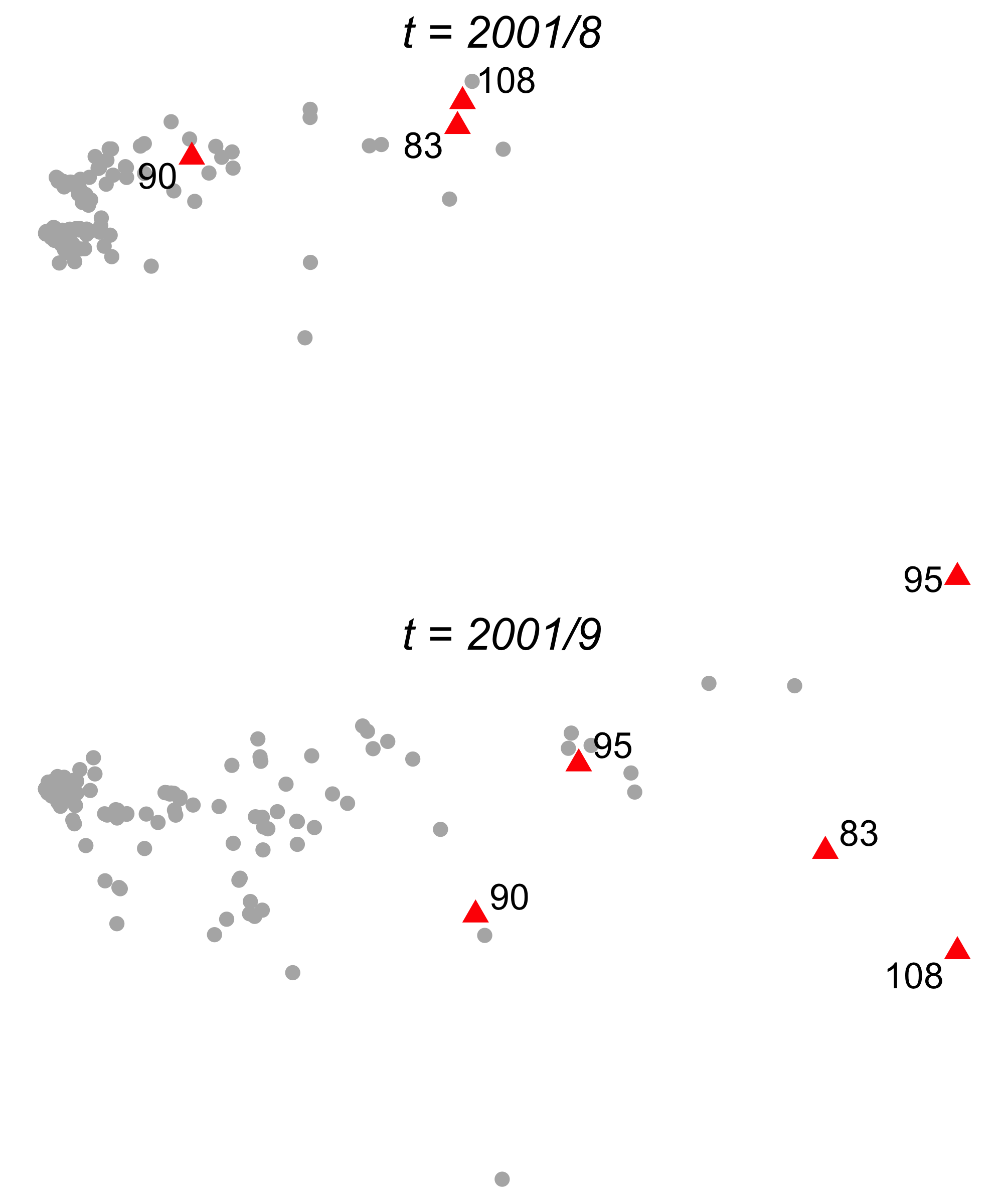}
    \caption{\label{fig:enronVertexADqcc}}
    \end{subfigure}
    \caption{Analysis of the Enron email graphs (Nov 1998 - Jun 2002). Left: GraphAD control charts with the moving averages (solid line) and three times the adjusted moving sample range (dashed line). Dots and triangles represent normal and anomalous graphs, respectively. The chart starts from Nov 1999 due to the training period. Right: Scatter plots of the first two MASE common latent dimensions for graphs between July and August 2001. Triangles represent vertex anomalies, while dots represent normal vertices.}

\end{figure}
We presented the latent positions for vertex anomalies detected by \text{MASE(2)} in August 2001 in Figure~\ref{fig:enronVertexADqcc} for illustration. The triangles\footnote{The correspondence between the vertex and the employees can be found at \url{https://www.cis.jhu.edu/~parky/Enron/employees}.} represent vertex anomalies detected by the control charts, while the dots represent normal vertices. These anomalies indicate significant changes in communication patterns. Among them, for example, $v_{95}$ is identified as the most extreme anomaly during this time period. Its shift from the center of the email community suggests a particularly notable change in the associated employee's email behavior. We included the detection results for adjusted \text{p}-values in the Appendix. 

\subsection{Large-scale Commercial Search Engine}
\label{sec:msr}
Using the Microsoft Bing (MSB) entity-transition dataset, we analyze monthly graphs from May 2018 to April 2019 ($M=12$). {The data in the Bing entity transition dataset was created to help develop better recommendation algorithms based on aggregated observed navigation patterns from users. What is contained in the data are transitions between different search phrases that represent transitions between those search phrases that are conducted in the same user session. To preserve anonymity, transitions are aggregated, and only transitions with enough occurrences are kept in the dataset. This provides us with relationship information between those terms purely based on the navigational behaviors of users. For example, if someone searches ``Lasagna" and then ``American recipes", we can infer a relationship between Lasagna and American recipes.} The graphs are undirected, unlooped, and positively weighted. Using the largest common component among the 12 graphs, each of the 12 graphs contains $|\mathcal{V}| = 33,793$ vertices. Given the uncertainty regarding the presence of anomalies, we introduced artificial ones at $t=6$ using a planted clique method. Details on the planted clique injection are provided in the Appendix. The results for a time window of $l=3$ are depicted in Figure~\ref{fig:MSRAD}. 

The left panel of Figure~\ref{fig:MSRAD} shows our method can detect the artificial anomaly involving planted clique of size $493$ in October for GraphAD. In addition, other than these artificial anomalies, our methods detected an anomaly in April 2019. That deviation is almost as evident as the artificial anomaly. Both \text{MASE(12)} and \text{SCAN} also detected an anomaly between August and September. We should emphasize that since we lack access to the ground truth here, it is difficult to confidently label a detected anomaly as a true positive. However, given that all methods, including \text{SCAN}, detected the April 2019 anomaly, we conjecture that it may indeed be a true anomaly. 
{The results obtained with adjusted p-values are similar to those with control charts and are included in the Appendix for reference.}

\begin{figure}[H]
\centering
\includegraphics[width=\linewidth]{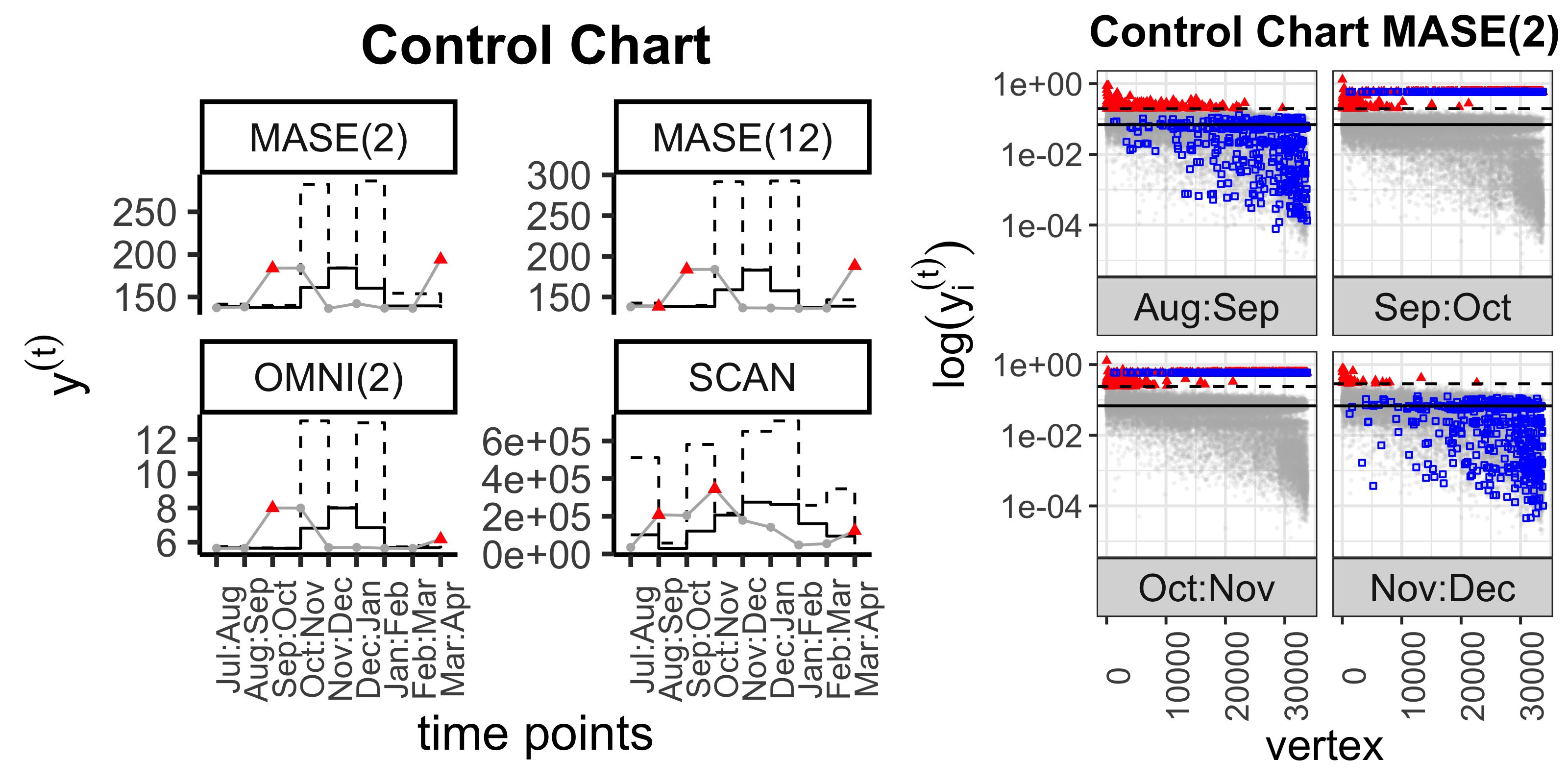}
\caption{Left: Control charts for GraphAD on the MSB time series, with an artificial anomaly inserted in October. The center solid line represents the moving average, and the dashed line represents the moving average plus three times the adjusted moving sample range. Dots and triangles represent normal and anomalous graphs. All methods except \text{SCAN} detect the artificial anomaly. Right: Control charts for VertexAD on the same time series with \text{MASE(2)}. The solid line represents the moving average; the dashed line represents the average plus three times EWAVE-SD. Dots represent normal vertices, triangles indicate detected anomalies, and squares are artificially anomalous vertices. The control chart starts in August due to the training period.}
\label{fig:MSRAD}
\end{figure}

The right panel of Figure~\ref{fig:MSRAD} displays the VertexAD results for \text{MASE} with span $s=2$. Our method detects all $473$ artificial anomalous vertices in October. Using these anomalies as benchmarks, we can explore other detected anomalies. Figure~\ref{fig:degchangeanomaly} in the Appendix aggregates vertices across months, illustrating degree changes between consecutive months. Detected anomalies identified by \text{MASE} are indicated with triangles. Artificially anomalous vertices are encircled by ellipses. This figure highlights \text{MASE}'s ability to detect anomalous vertices not limited to those with significant degree changes. {Latent positions for March and April 2019 are depicted in Figure~\ref{fig: msrVertexADqcc} (in the Supplementary Materials), showing anomalies with considerable shifts, such as vertex $v_{22020}$. }


We further carried out a sensitivity analysis to assess the robustness of the methods' performance in relation to different anomalies. To this end, we randomly selected vertices of varying sizes, forming a clique by adding edges within them at random time points. The results of this analysis are presented in Table~\ref{tab: sensitivity analysis adj-pval} (with adjusted \text{p}-values) and Table~\ref{tab: sensitivity analysis cc} (with control charts). These tables show that the \text{MASE(2)} method was able to detect the injected anomalies under different scenarios of artificially injected anomalies. However, it failed to detect the last anomaly in the final scenario.

\begin{table}
\centering
\begin{tabular}{rllrl}
\hline
Injected anomalous time & Injected clique size & Method & Detected anomalies\\ 
\hline
\multirow{4}{*}{\{6\}} & \multirow{4}{*}{\{100\}} & \text{MASE(2)} & \{5,\textbf{6},11,12\}   \\ 
&  & \text{MASE(12)} & \{\textbf{6},7,11,12\}   \\ 
&  & \text{OMNI(2)}  & \{\textbf{6},12\}  \\ 
&  & \text{SCAN}  & \{5,7,12\}   \\ 
\hline
\multirow{4}{*}{\{9\}} &\multirow{4}{*}{\{300\}} & \text{MASE(2)} & \{5,6,\textbf{9}\}    \\ 
&  & \text{MASE(12)} & \{5,6,\textbf{9}\}   \\ 
&   &\text{OMNI(2)}  & \{6,\textbf{9},10\}   \\ 
&   &\text{SCAN}  & \{5,7,12\}   \\ 
\hline
\multirow{4}{*}{\{12\}} &\multirow{4}{*}{\{1000\}} & \text{MASE(2)} & \{5,6,9,11,\textbf{12}\}  \\ 
&  & \text{MASE(12)} & \{5,6,9,11,\textbf{12}\}   \\ 
&   &\text{OMNI(2)}  & \{6,9,\textbf{12}\}  \\ 
&  & \text{SCAN}  & \{5,7,\textbf{12}\}  \\ 
\hline
\multirow{4}{*}{\{5,8,11\}} &\multirow{4}{*}{\{50,200,500\}} & \text{MASE(2)} & \{\textbf{5},6,\textbf{8},12\}  \\ 
&  & \text{MASE(12)} & \{\textbf{5},6,\textbf{8},12\}    \\ 
&  & \text{OMNI(2)}  & \{6,\textbf{8},9,12\}    \\ 
&  & \text{SCAN}  & \{\textbf{5},7,12\}   \\ 

\hline
\end{tabular}
\caption{Results of anomaly detection on Microsoft Bing (MSB) time series, where random anomalies were injected. The performance of three different methods using adjusted p-values, MASE, OMNI, and SCAN, are compared.}
\label{tab: sensitivity analysis adj-pval}

\end{table}

\begin{table}
\centering
\begin{tabular}{rllrl}
  \hline

Injected anomalous time & Injected clique size & Method & Detected anomalies\\ 
  \hline
\multirow{4}{*}{\{6\}} & \multirow{4}{*}{\{100\}} & \text{MASE(2)} & \{5,\textbf{6},12\}  \\ 
   &  & \text{MASE(12)} & \{\textbf{6},12\} \\ 
   &   &\text{OMNI(2)}  & \{\textbf{6},12\}  \\ 
   &   &\text{SCAN}  & \{5,12\}  \\ 
   \hline
  \multirow{4}{*}{\{9\}} &\multirow{4}{*}{\{300\}} & \text{MASE(2)} & \{5,\textbf{9}\}   \\ 
   &   &\text{MASE(12)} & \{\textbf{9}\} \\ 
   &   &\text{OMNI(2)}  & \{6,\textbf{9}\}    \\ 
   &   &\text{SCAN}  & \{5,12\}  \\ 
   \hline
   \multirow{4}{*}{\{12\}} &\multirow{4}{*}{\{1000\}} & \text{MASE(2)} & \{5,9,\textbf{12}\}  \\ 
   &   &\text{MASE(12)} & \{\textbf{12}\}   \\ 
   &   &\text{OMNI(2)}  & \{6,\textbf{12}\}   \\ 
   &   &\text{SCAN}  & \{5,\textbf{12}\}   \\  
   \hline
   \multirow{4}{*}{\{5,8,11\}} &\multirow{4}{*}{\{50,200,500\}} & \text{MASE(2)} & \{\textbf{5},\textbf{8}\}   \\ 
   &   &\text{MASE(12)} & \{\textbf{8}\}    \\ 
   &   &\text{OMNI(2)}  & \{6, \textbf{8}\}  \\ 
   &   &\text{SCAN}  & \{\textbf{5},12\}   \\ 
   
  \hline
\end{tabular}
\caption{Results of anomaly detection on Microsoft Bing (MSB) time series, where random anomalies were injected. The performance of three different methods using control charts, MASE, OMNI, and SCAN, are compared.}
\label{tab: sensitivity analysis cc}

\end{table}

\subsection{Biological Extinction Learning Networks}
\label{sec:neuro}
We utilized our method to analyze functional activity in biological learning networks using the Drosophila dataset, which contains known anomalies. Recently, the complete connectome of the larval brain has been mapped, allowing for the creation of biologically realistic models of these neural circuits based on their anatomical connections \citep{eschbach2020recurrent}. Investigations into how the learning-related circuits operate in the larval brain have involved associative learning simulations using connectome-constrained models. In these simulations, sequences of stimuli are delivered to the network, producing outputs that replicate responses observed in real animals (e.g., associating an odor with pain reduces its attractiveness).

We specifically trained the network models to perform extinction learning, a process wherein the association between a conditioned stimulus (CS, e.g., an odor) and reinforcement (pain or reward) diminishes after subsequent exposure to the CS without reinforcement. Network activity was simulated over 160 time points, representing a single extinction learning trial. In each trial, at $t=16$, a random odor (CS1) is presented to the neurons in the mushroom body. At $t=20$, a punishment or reward is administered. The odor is reintroduced at $t=80$ (CS2) and $t=140$ (CS3). Each stimulus lasted for 3 time points. Consequently, $t=16, 20, 80, 140$, and their subsequent 3 time points are identified as known anomalies. For further details about these network models, see \cite{eschbach2020recurrent}.

We performed 98 trials, each corresponding to an independent identical run of each parent network. This yielded 98 different time series, each consisting of 160 networks. Each network, directed and weighted with 140 nodes, is hollow, and all vertices are matched.

We applied our method with a default time window size of 12 on this dataset and compared it with the RDPG change point detection method by \cite{padilla2019change} (denoted as RDPG-CPD), using the default settings in its implementation in the R package \texttt{changepoints}, except the embedding dimension was estimated using the scree plot method. In our experiments, we measured the performance of an estimator $\hat{K}$ of the actual number of change points by the absolute error $\vert K - \hat{K} \vert$. For localization of change points, we used the one-sided Hausdorff distance defined as
\[
d(\hat{\mathcal{C}}| \mathcal{C}) \,=  \,  \underset{\eta \in \mathcal{C} }{\max }\,\underset{ x\in \hat{\mathcal{C}}   }{\min}\,\vert x - \eta\vert,
\]
where $\mathcal{C}$ is the true set of change points, and $\hat{\mathcal{C}}$ is the set of estimated change points. Similarly, the other one-sided Hausdorff distance metric is $d(\mathcal{C}|\hat{\mathcal{C}})$. We reported the symmetric Hausdorff distance by taking the maximum of these two one-sided distances for all methods across the 98 time series of graphs by plotting their density distribution at the bottom right of Figure~8. For $\vert K - \hat{K} \vert$, we reported the mean over 98 trials: 0.58 for RDPG-CPD, 9.48 for MASE(2), and 12.58 for MASE(12).

\begin{figure}[H]
	\centering
	\includegraphics[width=\linewidth]{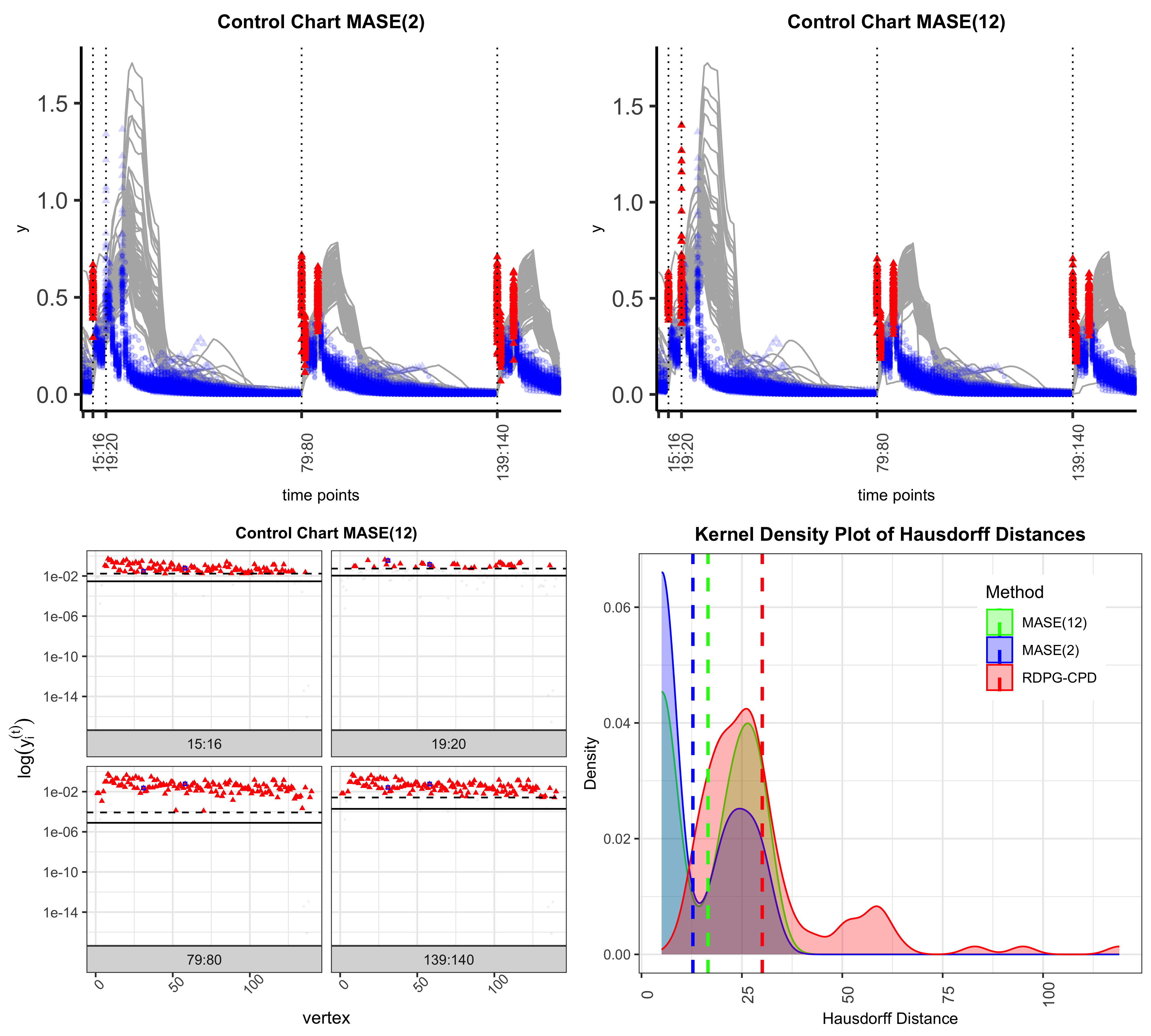}
	\caption{Analysis of the extinction learning networks. Top: GraphAD control charts with the moving averages plus three times the adjusted moving sample range (grey line). Dots and triangles represent normal and anomalous graphs, respectively. The red triangles denote time points detected as anomalous in at least 95 percent of the trials. The chart starts from $t=13$ due to the training period. Bottom left: typical vertex anomalies detected by MASE(12) at the four true anomalous time points for one trial. The blue squares are vertices 31 and 58, which were consistently identified as anomalous in at least 95 percent of the trials for the second CS. Bottom right: Kernel density for the Hausdorff distances for methods MASE(2), MASE(12), and RDPG-CPD.}
	\label{fig
	}
\end{figure}

Our results indicated that both MASE(2) and MASE(12) achieved smaller Hausdorff distances for these directed distances with a more significant number of detected points. This large number of detected points can potentially be because our methods are detecting anomalies. When the stimulus ends, it is also detected as an anomaly as the neural activity returns to normal (e.g., see Figure 4B, model output during an extinction learning trial in \cite{acharyya2024consistent}).

We presented our graph anomaly detection results in Figure 8. The triangles denote time points detected as anomalous, and those detected as anomalous in at least 95 percent of the trials are highlighted in red. MASE(2) and MASE(12) both successfully detected true anomalies corresponding to CS1, CS2, CS3, and CS4. Additionally, our method detected two more anomalies after CS3 and CS4, potentially corresponding to lasting stimulus ends and the network activities returning to normal. Note that the anomalies after CS2 are above the grey lines (the thresholds or the UCL), but since that time point is detected in less than 95 percent of the trials, it was not flagged red. However, according to the plot, if we decrease this 0.95 threshold for flagging anomalies across trials, it may still detect that return-to-normal anomaly for CS2.

We also identified typical vertex anomalies detected by MASE(12) at the four true anomalous time points plotted in Figure 8 (bottom left). Among the 140 vertices, only vertices 31 and 58 were consistently identified as anomalous in at least 95 percent of the trials for CS2, and only other vertices 8, 10, 13, 14, 20, 21, 23, 24, 26, 28, 36, 37, 38, and 62 were consistently identified as anomalous in at least 95 percent of the trials for CS1, CS3, and CS4, significantly narrowing down the potentially anomalous vertices for further investigation. However, determining which vertices in this setting are truly anomalous remains a significant neuroscientific question.

\section{Discussion}
\label{sec:discussion}

\subsection{Limitations and Extensions}
The simulations in this paper are done on graphs generated from conditionally independent RDPG models at each time step.  The next challenge is to design an algorithm to capture temporal structures (e.g.,~auto-regressive \citep{padilla2019change}) in time series of graphs. {Additionally, it would be interesting to explore other methods of controlling false positive rates, rather than the Benjamini and Hochberg procedure.} Finally, our theoretical results are only about graph-level hypothesis testing. The vertex anomaly detection results are based on empirical simulations; the theory describing relative strengths and limitations of this method in terms of the underlying nature of detectable vertex anomalies is of significant interest but technically challenging. 

\subsection{Conclusion}
We have proposed an algorithm that uses a test statistic obtained from a multiple adjacency spectral embedding methodology (\text{MASE}) for performing anomaly detection in time series of graphs. We have demonstrated, via simulation and theoretical results using a latent process model for time series of graphs, that our algorithms can be useful for two anomaly detection tasks. As demonstrated by the simulations in Section~\hyperref[subsec:compare]{5.3}, our proposed method outperforms competing methods when the model is correctly specified. Moreover, it continues to exhibit superior performance compared to competitors in certain scenarios, even under misspecification. We also assess our algorithms in real datasets and investigate the detected anomalies. Our approach outperforms alternative methods when the model is correctly specified. Moreover, it retains a competitive edge in various situations, even under conditions of model misspecification. While our adjusted p-values approach provides explicit FDR control at different $\alpha$ levels, our control chart approach does not. Interestingly, in our analysis of real-world datasets, control charts and methods for adjusting p-values both identify a similar number of graph anomalies. This suggests that the control chart might offer an implicit FDR control, by tuning the numbers of sigmas.

\bibliographystyle{apalike}

\bibliography{Bibliography-MM-MC}

\vspace{3mm}
\noindent {\bf Acknowledgements:}
This work was supported in part by DARPA programs D3M (FA8750-17-2-0112 ) and LogX (N6523620C8008-01) and MAA (FA8750-20-2-1001), and funding from Microsoft Research.
This material is based in part on research sponsored by the Air Force Research Laboratory and DARPA under agreement number FA8750-20-2-1001.
The U.S. Government is authorized to reproduce and distribute reprints for Governmental purposes notwithstanding any copyright notation thereon.
The views and conclusions contained herein are those of the authors and should not be interpreted as necessarily representing the official policies or endorsements, either expressed or implied, of the Air Force Research Laboratory and DARPA or the U.S. Government.

\noindent {\bf Conflicts of Interest:}
The authors declare no conflict of interest.

\begin{center}
{\large\bf SUPPLEMENTARY MATERIAL}
\end{center}

\begin{description}

\item[Code and Data:] The files containing R code and Enron data for reproducing all the plots can be found in \url{https://github.com/gdchen94/TSG-Anomaly-Detection/}. The commercial search engine data used in Section 6 are available from the authors upon reasonable request and with the permission of Microsoft Corporation.

\item [Appendix:] The supplemental files include an Appendix containing the following:
(i) Theoretical results for the multiple adjacency spectral embedding and proof of Theorem 1 and Example 2. 
(ii) Introduction of the Shewhart rule. 
(iii) Comparative analysis of VertexAD performance using AUC and rank-based metric. 
(iv) Sensitivity analysis for control charts applied to the Enron email dataset, with varying hyper-parameters $l$ and $s$. 
(v) Adjusted p-value results in the Enron email dataset, with varying hyper-parameters $l$ and $s$. 
(vi) Adjusted p-value plots in the MSB dataset. 
(vii) Analysis of the vertex anomaly detection results in the MSB dataset. 
(viii) Application of the planted clique method in the MSB dataset and analysis of the number of sigmas the data deviates from moving means in the MASE(12) test statistics, as shown in Figure 7. 
(ix) Sensitivity analysis of dimension selection method. 
(x) Asymptotic result via simulation study with increasing number of vertices.

\end{description}

\bigskip
\appendix





\section*{Appendix A: Proof}

\subsection*{Multiple adjacency spectral embedding results}
\label{subsec:mase-results}

In the existing literature, \cite{arroyo2021inference} established a non-asymptotic bound on the expected error for the common subspace estimate from \text{MASE}, and the bound decreases as the number of graphs grows. We first introduce some notations before introducing the theorem. For a given symmetric square matrix $\mathbf{M}\in \mathbb{R}^{r\times r}$, define $\lambda_{\min}(\mathbf{M})$ and $\lambda_{\max}(\mathbf{M})$ to be the minimum and maximum of the eigenvalues of the matrix $\mathbf{M}$. We denote $\delta(\mathbf{M}) = \max_{i=1,\cdots,r} \sum_{u=1}^{r} \mathbf{M}_{iu}$. Given two sequences of $\{a_{n}\}_{n=1}^{\infty}$ and $\{b_{n}\}_{n=1}^{\infty}$, we say $a_{n} \lesssim b_{n}$ if there exists constants $C>0$ and $n_{0}>0$ such that $a_{n} < C b_{n}$ for all $n\geq n_{0}$.

\begin{theorem}[Theorem~7 in \cite{arroyo2021inference}]
Let $\mathbf{R}^{(1)}, \dots, \mathbf{R}^{(m)}\in\real^{d\times d}$ be a collection of full rank symmetric matrices of size $d\times d$, and $\mathbf{V}\in\real^{n\times d}$ a matrix with orthonormal columns. Suppose $(\mathbf{A}^{(1)}, \ldots,\mathbf{A}^{(m)})\sim\cosie{\mathbf{V};\mathbf{R}^{(1)}, \ldots, \mathbf{R}^{(m)}}$ are a sample of adjacency matrices from the COSIE model. Define
\begin{equation}
\varepsilon = \sqrt{\frac{1}{m}\sum_{t=1}^m \frac{\delta(\mathbf{P}^{(t)})}{\lambda_{\min}^2(\mathbf{R}^{(t)})}}.\label{eq:AVPVerror}
\end{equation} Suppose the parameters satisfy $\min_{t\in[m]}\delta(\mathbf{P}^{(t)})=\omega(\log n)$ and $\varepsilon = o(1)$, then there exists $\mathbf{W}\in\mathcal{O}_d$ such that for $\widehat{\mathbf{V}}$ produced by \text{MASE} algorithm satisfy
	\begin{equation}
	\mathbb{E}\left[\min_{\mathbf{W}\in\mathcal{O}_d}\|\widehat{\mathbf{V}}-\mathbf{V}\mathbf{W}\|_F\right] \lesssim \sqrt{\frac{d}{m}}\varepsilon + \sqrt{d}\varepsilon^2. \label{eq:theorem-bound}
	\end{equation}
	\label{thm:COSIE_V-Vhat}.
\end{theorem}
 
 
 \cite{arroyo2021inference} also established the asymptotic distribution of the estimated score matrices as the number of vertices increased. This result depends on the following technical sparsity conditions on $\mathbf{P}^{(t)}$ and $\mathbf{V}$.

\begin{assumption}[Delocalization of $\mathbf{V}$]\label{assump:Delocalization} There exist constants $c_1, c_2>0$, and an orthogonal matrix $\mathbf{W}\in\mathcal{O}_d$ such that each entry of $\mathbf{V}\mathbf{W}$ satisfies
	$$\frac{c_1}{\sqrt{n}} < (\mathbf{V}\mathbf{W})_{kl} < \frac{c_2}{\sqrt{n}}, \quad\quad\forall k\in[n], l\in[d].$$
\end{assumption}

\begin{assumption}[Edge variance]\label{assump:variance_P} The sum of the variance of the edges satisfies for all $t\in[m]$

$$s^2(\mathbf{P}^{(t)}):=\sum_{s=1}^n\sum_{u=1}^n \mathbf{P}^{(t)}_{su}(1-\mathbf{P}^{(t)}_{su}) = \omega(1), \quad \text{and} $$

$$\sum_{s=1}^n\sum_{u=1}^n \mathbf{P}^{(t)}_{su}(1-\mathbf{P}^{(t)}_{su})  (1-\mathbf{P}^{(t)}_{su})(n \mathbf{V}_{sk}\mathbf{V}_{ul}+n \mathbf{V}_{uk}\mathbf{V}_{sl}) = \omega(1), \quad \text{for } k \neq l. $$

	. 
\end{assumption}

Given the matrices $\mathbf{V}$ and $\mathbf{P}^{(t)}=\mathbf{V}\mathbf{R}^{(t)}\mathbf{V}^\top $ we  define  $\mathbf{\Sigma}^{(t)}\in\mathbb{R}^{d\times d}$ as
\begin{equation}
	\mathbf{\Sigma}^{(t)}_{\frac{2k + l(l-1)}{2}, \frac{2k'+l'(l'-1)}{2}}  := \sum_{s=1}^{n-1}\sum_{u=s+1}^n\mathbf{P}^{(t)}_{su}(1-\mathbf{P}^{(t)}_{su})\left[(\mathbf{V}_{sk}\mathbf{V}_{ul} + \mathbf{V}_{uk}\mathbf{V}_{sl})(\mathbf{V}_{sk'}\mathbf{V}_{ul'} + \mathbf{V}_{sk'}\mathbf{V}_{ul'})\right].\label{eq:covarianceR}
\end{equation}

\begin{assumption}[Score matrix covariance]\label{assump:score_lambdamin} The magnitude of the smallest eigenvalue of $\mathbf{\Sigma}^{(t)}$ satisfies $|\lambda_{\min}(\mathbf{\Sigma}^{(t)})|=\omega(n^{-2})$.
\end{assumption}

With the above assumptions at hand, we are ready to introduce asymptotic normality results on individual entries on the matrix $\widehat{\mathbf{R}}^{(t)}$ in~\cite{arroyo2021inference}.

\begin{lemma}[Lemma~14 in \cite{arroyo2021inference}]\label{lem:COSIE_R_decomp} 
Under the general assumptions of Theorem~\ref{thm:COSIE_CLT}, there exist sequences of matrices $\mathbf{B}^{(i)}, \mathbf{N}^{(i)}, \mathbf{W} \in \mathbb{R}^{d \times d}$ depending on $d$, $m$, and $n$, such that
	\begin{equation}
	\mathbf{W}\widehat{\mathbf{R}}^{(i)}\mathbf{W}^\top - \mathbf{R}^{(i)}  - \mathbf{W} \mathbf{B}^{(i)} \mathbf{W}^{\top} - \mathbf{W} \mathbf{N}^{(i)} \mathbf{W}^{\top} =
	\mathbf{V}^{\top}\mathbf{E}^{(i)}\mathbf{V} ,\label{eq:R-decomposition}
	\end{equation}
	with $\mathbf{H}_{m}^{(i)} = -(\mathbf{W} \mathbf{B}^{(i)} \mathbf{W}^{\top} + \mathbf{W} \mathbf{N}^{(i)} \mathbf{W}^{\top})$ satisfying $\mathbb{E}[\|\mathbf{H}_m^{(i)}\|_F] = O_P(d/\sqrt{m})$, and $\mathbf{W} = \operatorname{\textnormal{arg inf}}_{\mathbf{W} \in \mathcal{O}_{d}}\|\widehat{\mathbf{V}} - \mathbf{V} \mathbf{W}\|_{F}$ is an orthogonal matrix.
\end{lemma}

\begin{theorem}[Theorem~11 in \cite{arroyo2021inference}]\label{thm:COSIE_CLT}
	Suppose $(\mathbf{A}{(1)}, \ldots,\mathbf{A}^{(m)})\sim\cosie{\mathbf{V};\mathbf{R}^{(1)}, \ldots, \mathbf{R}^{(m)}}$ are a sample of adjacency matrices from the COSIE model such that the parameters satisfy $\min_{t\in[m]}\delta(\mathbf{P}^{(t)})=\omega(\log n)$, $\varepsilon= O\left(\frac{1}{\max\sqrt{\delta(\mathbf{P}^{(t)})}}\right)$ as well as the delocalization requirements given in Assumption~\ref{assump:Delocalization}. Suppose that Assumption~\ref{assump:variance_P} and Assumption~\ref{assump:score_lambdamin}  hold, then there exists a sequence of matrices $\mathbf{W}\in\mathcal{O}_d$ such that
	$$\frac{1}{\sigma_{i,k,l}}(\widehat{\mathbf{R}}^{(t)} - \mathbf{W}^\top \mathbf{R}^{(t)} \mathbf{W}+\mathbf{H}_m^{(t)})_{kl}\overset{d}{\rightarrow}\mathcal{N}(0,1),$$
	where $\mathbf{H}^{(t)}_m$ is a random matrix satisfying $\e[\|\mathbf{H}^{(t)}_m\|_{F}]=O\left(\frac{d}{\sqrt{m}}\right)$ and $\sigma_{i,k,l}^2=O\left( \frac{d^2s^2(\mathbf{P}^{(t)})}{n^2}\right)=O(1)$.
\end{theorem}

\begin{proof}
[Proof of Theorem~\ref{thm:main}] 
Combine Portmanteau Lemma with CLT results Theorem~\ref{thm:COSIE_CLT}, then there exists a sequence of matrices $\mathbf{W}= \text{argmin}_{\mathbf{W}\in\mathcal{O}_d}\|\widehat{\mathbf{V}} - \mathbf{V}\mathbf{W}\|_F$ such that
\begin{equation}
\e[ \widehat{\mathbf{R}}^{(t)} - \mathbf{W}^\top \mathbf{R}^{(t)} \mathbf{W} + \mathbf{H}^{(t)}_m]_{k,l}^{2}\overset{}{\rightarrow}(\sigma_{i,k,l} )^{2},
\label{eq:convergence}
\end{equation}
where $\e[\|\mathbf{H}^{(t)}_m\|_F]=O\left(\frac{d}{\sqrt{m}}\right)$ and $\sigma_{i,k,l}^2=O\left( \frac{d^2s^2(\mathbf{P}^{(t)})}{n^2}\right)$.


Combine Markov inequality with the Equation (\ref{eq:convergence}), we have 
\begin{equation}
\| \widehat{\mathbf{R}}^{(t)} - {\mathbf{W}} ^\top \mathbf{R}^{(t)}\mathbf{W}  + \mathbf{H}_{m}^{(t)}\|_F = O_P\left( \frac{d^2\sqrt{s^2(\mathbf{P}^{(t)})}}{n} \right).
\label{eq:R-Rhat-bigO}
\end{equation}
Again using Markov inequality we have
\begin{equation}
\| \mathbf{H}_{m}^{(t)} \|_F = O_P\left( \frac{d}{\sqrt{m}} \right).
\label{eq:Hm-bigO}
\end{equation}

Because $s^2(\mathbf{P}^{(t)})=O(n^2)$, the right hand side of Equation~\eqref{eq:R-Rhat-bigO} is at most of an order $O_P(d^2)$. We thus have
\begin{equation*}
\| \widehat{\mathbf{R}}^{(t)} - {\mathbf{W}} ^\top \mathbf{R}^{(t)}\mathbf{W}  \|_{F} \leq \| \mathbf{H}_{m}^{(t)} \|_F  + \| \widehat{\mathbf{R}}^{(t)} - {\mathbf{W}} ^\top \mathbf{R}^{(t)}\mathbf{W}  + \mathbf{H}_{m}^{(t)}\|_F = O_P\left( \frac{d^2\sqrt{s^2(\mathbf{P}^{(t)})}}{n} +  \frac{d}{\sqrt{m}}   \right)
\end{equation*}
and so 
\begin{equation}
\| \widehat{\mathbf{R}}^{(t)} - {\mathbf{W}} ^\top \mathbf{R}^{(t)}\mathbf{W}  \|_{F}=O_P(d^2).
\label{eq: Rhat-diff-upper-3}
\end{equation}

Suppose the null hypothesis $H_{0}$ is true, then we have
\begin{align*}
&\| \widehat{\mathbf{R}}^{(t)} -  \widehat{\mathbf{R}}^{(t+1)} \|\\
\leq & \|  \widehat{\mathbf{R}}^{(t)} - {\mathbf{W}} ^\top \mathbf{R}^{(t)}\mathbf{W} \|_{F} + \| \widehat{\mathbf{R}}^{(t+1)} - {\mathbf{W}} ^\top \mathbf{R}^{(t+1)}\mathbf{W} \|_F, 
\label{eq:Rhat-diff-upper-1}
\end{align*}
where we have used the fact that under $H_{0}$, $\mathbf{R}^{(t)} =  \mathbf{R}^{(t+1)}$.

Let $\alpha$ be given, and let $\eta < \alpha/4$, by Theorem~\ref{thm:COSIE_CLT} and equality~(\ref{eq: Rhat-diff-upper-3}), we have for all $n$ sufficiently large, there exists a sequence of matrices $\mathbf{W}= \text{argmin}_{\mathbf{W}\in\mathcal{O}_d}\|\widehat{\mathbf{V}} - \mathbf{V}\mathbf{W}\|_F$ and a constant $C'$ such that with probability at least $1-\eta$,

$$\|\widehat{\mathbf{R}}^{(t)} - {\mathbf{W}} ^\top \mathbf{R}^{(t)}\mathbf{W}  \|_{F} \leq C(\mathbf{P}^{(t)}, m,d,n)+f(\mathbf{R}^{(t)}, \alpha, n)$$
$$\|\widehat{\mathbf{R}}^{(t+1)} - {\mathbf{W}} ^\top \mathbf{R}^{(t+1)}\mathbf{W} \|_{F} \leq C(\mathbf{P}^{(t+1)},m,d,n)+f(\mathbf{R}^{(t+1)}, \alpha, n)$$
where $C(\mathbf{P}^{(t)}, m,d,n) = \frac{C'}{2}( \frac{d^2\sqrt{s^2(\mathbf{P}^{(t)})}}{n} +  \frac{d}{\sqrt{m}} ) \leq C' d^{2}/2 $ and  $f(\mathbf{R}^{(t)}, \alpha, n)\rightarrow 0$ as $n\rightarrow \infty$ for fixed $\alpha$, $C'$ and $\mathbf{R}^{(t)}$ satisfying edge variance assumption~\ref{assump:variance_P} for all $t\in[m]$.

Therefore, for all $n>n_{0}$, with probability at least $1-\alpha$ we have

$$ \frac{\| \widehat{\mathbf{V}}  \widehat{\mathbf{R}}^{(t)} - \widehat{\mathbf{V}} \widehat{\mathbf{R}}^{(t+1)} \|_{F}}{d^2}\leq C' + r(\alpha, n).$$
where $r(\alpha,n) \rightarrow 0$ as $n \rightarrow \infty$ for a fixed $\alpha$.

We can thus take $n_{1}=n_{1}(\alpha, C) = \inf\{n \geq n_{0}(\alpha):r(\alpha,n) \leq C-C' \} < \infty$. Then for all $n>n_{1}$ and when null hypothesis is true, we conclude
$\mathbb{P}(T\in R) < \alpha$.

We next prove consistency, denote
$D(\mathbf{R}^{(t)}, \mathbf{R}^{(t+1)}) = \|\mathbf{R}^{(t)} - \mathbf{R}^{(t+1)}  \|_{F} $. 

With triangle inequality, we then have 

\begin{equation}
\| \widehat{\mathbf{R}}^{(t)} -  \widehat{\mathbf{R}}^{(t+1)} \|_F \\
\geq D(\mathbf{R}^{(t)}, \mathbf{R}^{(t+1)})-\| \widehat{\mathbf{R}}^{(t)} - {\mathbf{W}} ^\top \mathbf{R}^{(t)}\mathbf{W} \|_{F} - \| \widehat{\mathbf{R}}^{(t+1)} - {\mathbf{W}} ^\top \mathbf{R}^{(t+1)}\mathbf{W} \|_F
\label{eq:Rhat-lower-bound}
\end{equation}

Therefore, for all $n$,
\begin{align*}
\mathbb{P}(T \notin R) \leq& \mathbb{P}(\| \widehat{\mathbf{R}}^{(t)} -  \widehat{\mathbf{R}}^{(t+1)} \|_F/d^2 \leq C)\\
= & \mathbb{P}\bigg( \|\widehat{\mathbf{R}}^{(t)} - {\mathbf{W}} ^\top \mathbf{R}^{(t)}\mathbf{W} \|_{F} +  \| \widehat{\mathbf{R}}^{(t+1)} - {\mathbf{W}} ^\top \mathbf{R}^{(t+1)}\mathbf{W} \|_F + Cd^2 \geq D(\mathbf{R}^{(t)}, \mathbf{R}^{(t+1)})\bigg) .
\end{align*}

Now let $\beta>0$ given, according to Equation~\eqref{eq: Rhat-diff-upper-3}, we deduce that there exists a constant $M_{1}(\beta)$ and a positive integer $n_{0}= n_{0}(\alpha, \beta)$ so that, for all $n\geq n_{0}(\alpha, \beta)$,
$$\mathbb{P}\bigg( \| \widehat{\mathbf{R}}^{(t)} - {\mathbf{W}} ^\top \mathbf{R}^{(t)}\mathbf{W} \|_{F} + \frac{Cd^2}{2}\geq M_{1}/2\bigg) \leq \beta/2 $$
$$\mathbb{P}\bigg( \| \widehat{\mathbf{R}}^{(t+1)} - {\mathbf{W}} ^\top \mathbf{R}^{(t+1)}\mathbf{W} \|_{F} + \frac{Cd^2}{2}\geq M_{1}/2\bigg) \leq \beta/2 $$
If $b_{n}\rightarrow\infty$, there exists some $n_{2}(\alpha,\beta, C)$ such that, for all $n\geq n_{2}$, either $D(\mathbf{R}^{(t)}, \mathbf{R}^{(t+1)})=0$ or $n\geq n_{2}$, either $D(\mathbf{R}^{(t)}, \mathbf{R}^{(t+1)})\geq M_{1}$. Hence, for all $n\geq n_{2}$, if $D(\mathbf{R}^{(t)}, \mathbf{R}^{(t+1)}) \neq0 $, then $\mathbb{P}(T \notin R) \leq \beta$, i.e., our test statistics $T$ lies within the rejection region $R$ with probability at least $1-\beta$, as required.

\end{proof}

\subsection*{Proof of the Bound in Example 2}
\begin{proof}
	
	Here we assume $\Delta>0$. By direct calculation we have $\mathbf{\sigma}_{\mathbf{P}} = \sqrt{n}\sqrt{p^2 + (p + \Delta)^2}$ and $\delta = -n\Delta$. 
	\[
	\|e + \delta\|_2 - \mathbf{\sigma}_{\mathbf{P}} > \|e\|_2 + \mathbf{\sigma}_{\mathbf{P}},
	\]
	is equivalent to
	\[
	\|e\|_2^2 + 2\|e\|_2\delta + \delta^2 > \|e\|_2^2 + 4\mathbf{\sigma}_{\mathbf{P}}\|e\|_2 + 4\mathbf{\sigma}_{\mathbf{P}}^2.
	\]
	Plug in
	\[
	\delta = -n\Delta, \mathbf{\sigma}_{\mathbf{P}} = \sqrt{n}\sqrt{p^2 + (p + \Delta)^2}
	\]
	\begin{equation*} \label{eq1}
	\begin{split}
	&\Leftrightarrow n^2\Delta^2 - 4n[p^2 + (p + \Delta)^2] > \|e\|_2 (2n\Delta + 4\sqrt{n}\sqrt{p^2 + (p + \Delta)^2}) \\
	&\Leftrightarrow \frac{n^2\Delta^2 - 4n[p^2 + (p + \Delta)^2]}{2n\Delta + 4\sqrt{n}\sqrt{p^2 + (p + \Delta)^2}} > \|e\|_2.
	\end{split}
	\end{equation*}
	By Markov Inequality,
	\[
	\mathbb{P}(|e| \geq t) \leq \frac{\mathbb{E}(|e|)}{t}.
	\]
	Plug in 
	\[
	t = \frac{n^2\Delta^2 - 4n[p^2 + (p + \Delta)^2]}{2n\Delta + 4\sqrt{n}\sqrt{p^2 + (p + \Delta)^2}}
	\]
	and by Theorem~\ref{thm:COSIE_CLT} in the Appendix, we thus have finished the proof with
	\[
	\mathbb{P}(|e| \geq t) \leq \frac{C \cdot d}{\sqrt{m}}\left(\frac{n^2\Delta^2 - 4n[p^2 + (p + \Delta)^2]}{2n\Delta + 4\sqrt{n}\sqrt{p^2 + (p + \Delta)^2}}\right).
	\]
	
\end{proof}
\newpage
\section*{Appendix B: Shewhart rule}
Shewhart rule declares a process being out-of-control if a point plots outside the 3-sigma control limits. A warning is declared if more than a default of 7 consecutive points lies on one side of the center line, which represents the process mean. We do not show the warnings in this work.

\newpage
\section*{Appendix C: Additional simulation results}

\subsection*{Comparative analysis of VertexAD performance using AUC and rank-based metric}

\begin{figure}
    \centering
    \includegraphics[width=\textwidth]{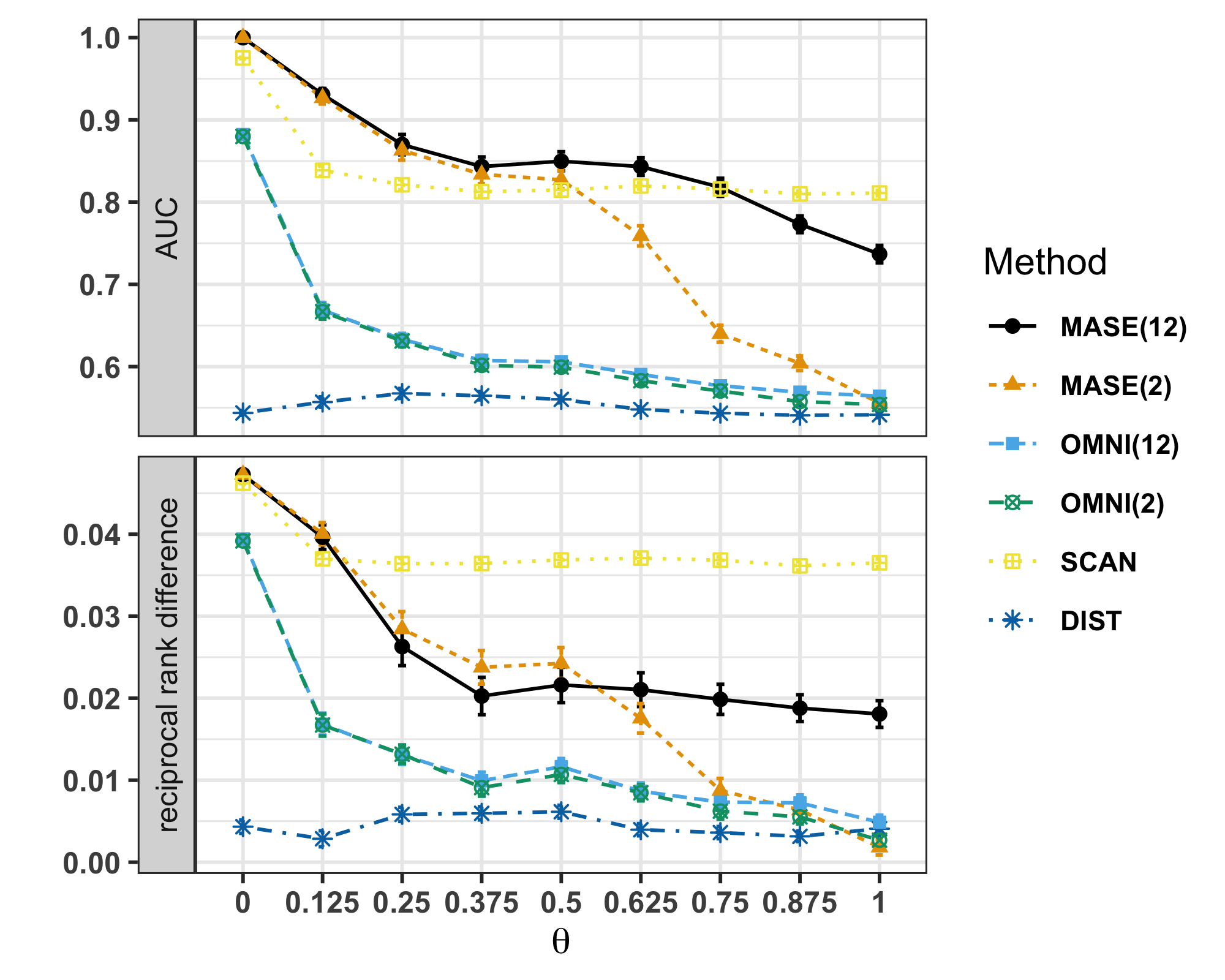}
    \caption{Measures of vertex anomaly detection performance in time series of graphs (Scenario $2$, Section 5.1), generated with varying $\theta$ and fixed parameters $p=0.8$, $q=0.3$, and $\mathbf{\Delta}_{i \cdot}=0.3\cdot 1_{4}$. The top panel shows performance in AUC at a significance level of $0.05$, while the bottom panel illustrates performance in the rank-based metric. Error bars represent $\text{mean} \pm \text{standard error}$.}
    \label{fig:comparison_auc_mrr}
\end{figure}

We computed the Area Under the Curve (AUC) for detecting anomalous vertices at times $6:7$ across $100$ Monte Carlo simulations to evaluate \text{MASE}, \text{SCAN}, \text{OMNI}, and \text{DIST} within a comparative simulation scenario (see Section~\ref{subsec:compare}). Results in Figure~\ref{fig:comparison_auc_mrr} show \text{MASE} surpassing \text{SCAN} for $\theta\approx 0$ due to a shared subspace assumption. However, as $\theta$ nears $1$, \text{SCAN} becomes superior, indicating its robustness against subspace variations. Larger $s$ values boost vertex anomaly detection performance consistent with graph anomaly detection. To assess VertexAD's efficiency in detecting top anomalies, we employ a rank-based metric. 
Specifically, we determine the reciprocal ranks $\text{RR}_{i}^{(t)}$ at time $t$ as $\text{RR}_{i}^{(t)}=1/r(y_{i}^{(t)})$, where $r(y_{i}^{(t)})$ is decreasing order of $y_{i}^{(t)}$ in $\{y_{i}^{(t)}\}_{i=1}^{n}$. We then compute the discrepancy between the average reciprocal ranks of anomalous and non-anomalous vertices at the anomalous time $t^{*}$. This reciprocal rank difference serves as a VertexAD metric, with larger values indicating a pronounced distinction between vertex types. This measure 
is particularly sensitive to the most extreme anomalies. We present the results of these differences in the bottom right of Figure~\ref{fig:comparison_auc_mrr}. \text{MASE} test statistics perform comparative with the \text{SCAN} statistics when $\theta\approx 0$, while \text{SCAN} statistic is consistent dominating the other two as $\theta$ increases to $1$. The superiority of \text{SCAN} in detecting the most extreme vertices is expected since \text{SCAN} uses a local graph scan statistic concerning each vertex. Figure~\ref{fig:comparison_auc_mrr} also shows that the performance of \text{MASE} in VertexAD climbs well above that of \text{OMNI} at all $\theta$. The direct comparison method \text{DIST} fails to detect any significant vertex anomalies. 

\newpage
\section*{Appendix D: Additional real data results}

\subsection*{Enron email network results}
In the context of the Enron dataset, we conducted an in-depth investigation into the influence of hyper-parameters $s$ and $l$ on GraphAD. The results, visualized in control charts for \text{MASE(2)}, \text{MASE(12)}, \text{OMNI(2)}, \text{OMNI(12)}, and \text{SCAN} across varying values of $l=10,11,12$ and $s=2,12$, are depicted in Figures~\ref{fig:enronGraphAD_cc_l10}, \ref{fig:enronGraphAD_cc_l11}, and \ref{fig:enronGraphAD_cc_l12}. Correspondingly, the results based on adjusted p-values for the combinations as mentioned earlier can be found in Figures~\ref{fig:enronGraphAD_pval_l10}, \ref{fig:enronGraphAD_pval_l11}, and~\ref{fig:enronGraphAD_pval_l12}. A noticeable observation is the robustness of our control chart approach to variations in $s$ and $l$. As an illustration, both \text{MASE(2)} and \text{MASE(12)} consistently identified similar anomalies for different $l$, though with $l=10,12$, \text{MASE(2)} did not detect the anomaly in February 2000. On a parallel note, our adjusted p-values-based approach also manifests robustness across different $s$ and $l$, with all techniques identifying same anomalies irrespective of the $l$ values.
Furthermore, our control charts tend to identify fewer graph anomalies than our adjusted p-values methodology. While our adjusted p-values procedure offers explicit FDR control at varying $\alpha$ thresholds, our control chart method does not. This observation might suggest that the control chart implicitly controls FDR, dependent on the fine-tuning of sigma values.

\subsubsection*{Control charts in the Enron email dataset with varying in-process parameter $l$}

\begin{figure}[H]
\centering

\includegraphics[width=\linewidth]{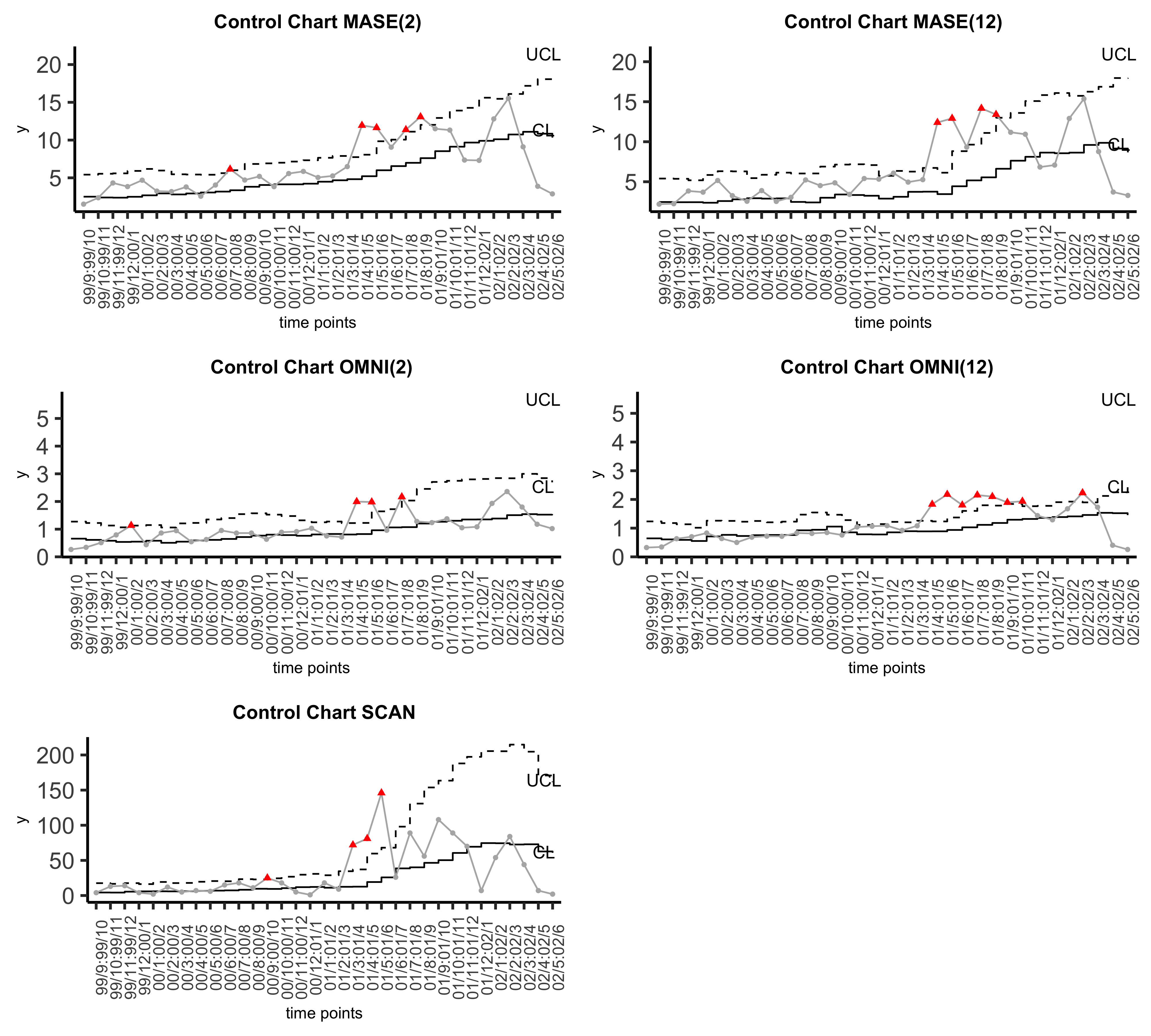}
\caption{GraphAD control charts with the moving averages (solid line), and three times the adjusted moving sample range (dashed line). Dots and triangles represent normal and anomalous graphs. The chart starts from Oct 1999 due to the training period $l=10$. }
\label{fig:enronGraphAD_cc_l10}
\end{figure}

\begin{figure}[H]
\centering
\includegraphics[width=\linewidth]{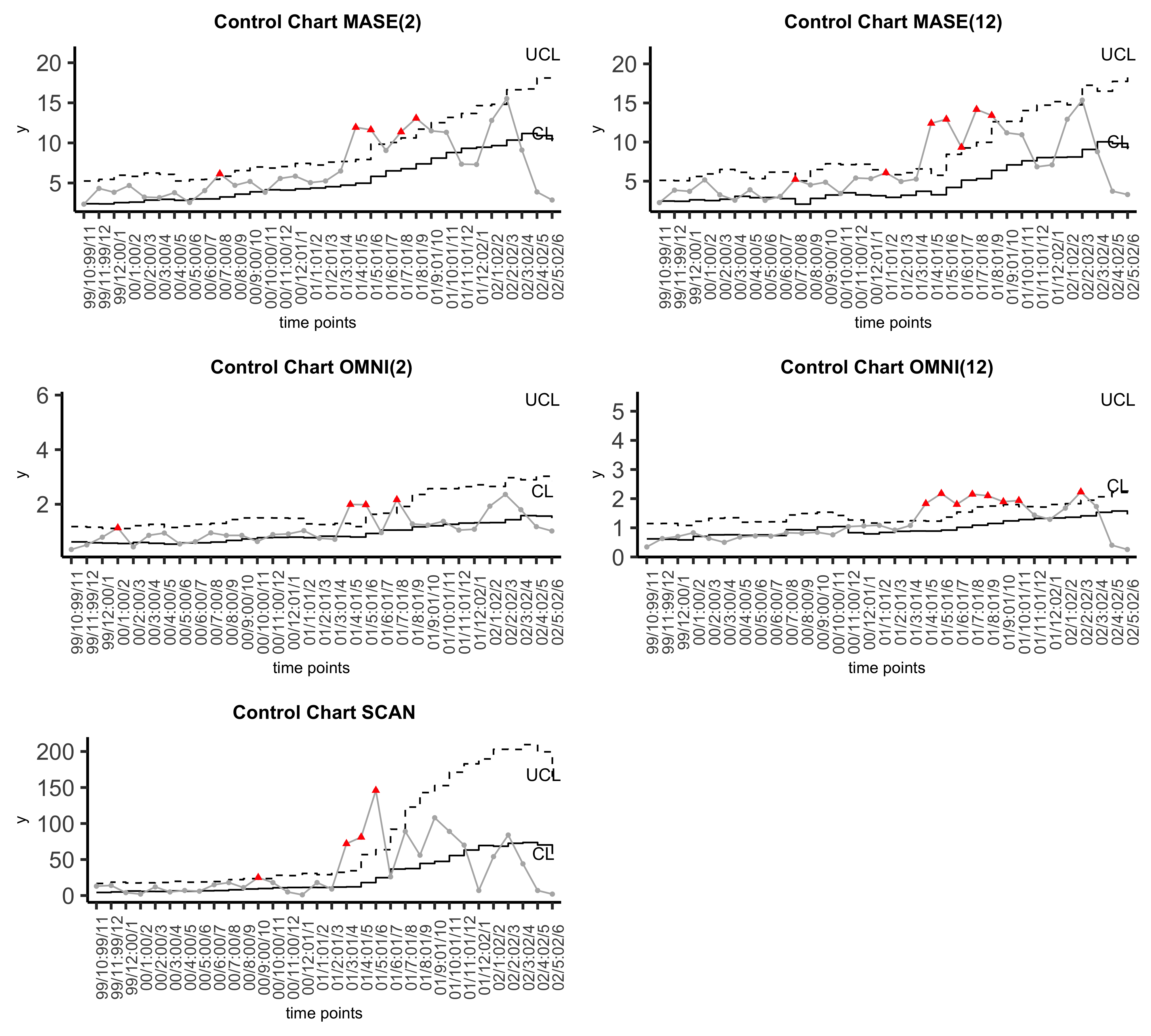}
\caption{GraphAD control charts with the moving averages (solid line), and three times the adjusted moving sample range (dashed line). Dots and triangles represent normal and anomalous graphs. The chart starts from Nov 1999 due to the training period $l=11$. }
\label{fig:enronGraphAD_cc_l11}
\end{figure}

\begin{figure}[H]
\centering
\includegraphics[width=\linewidth]{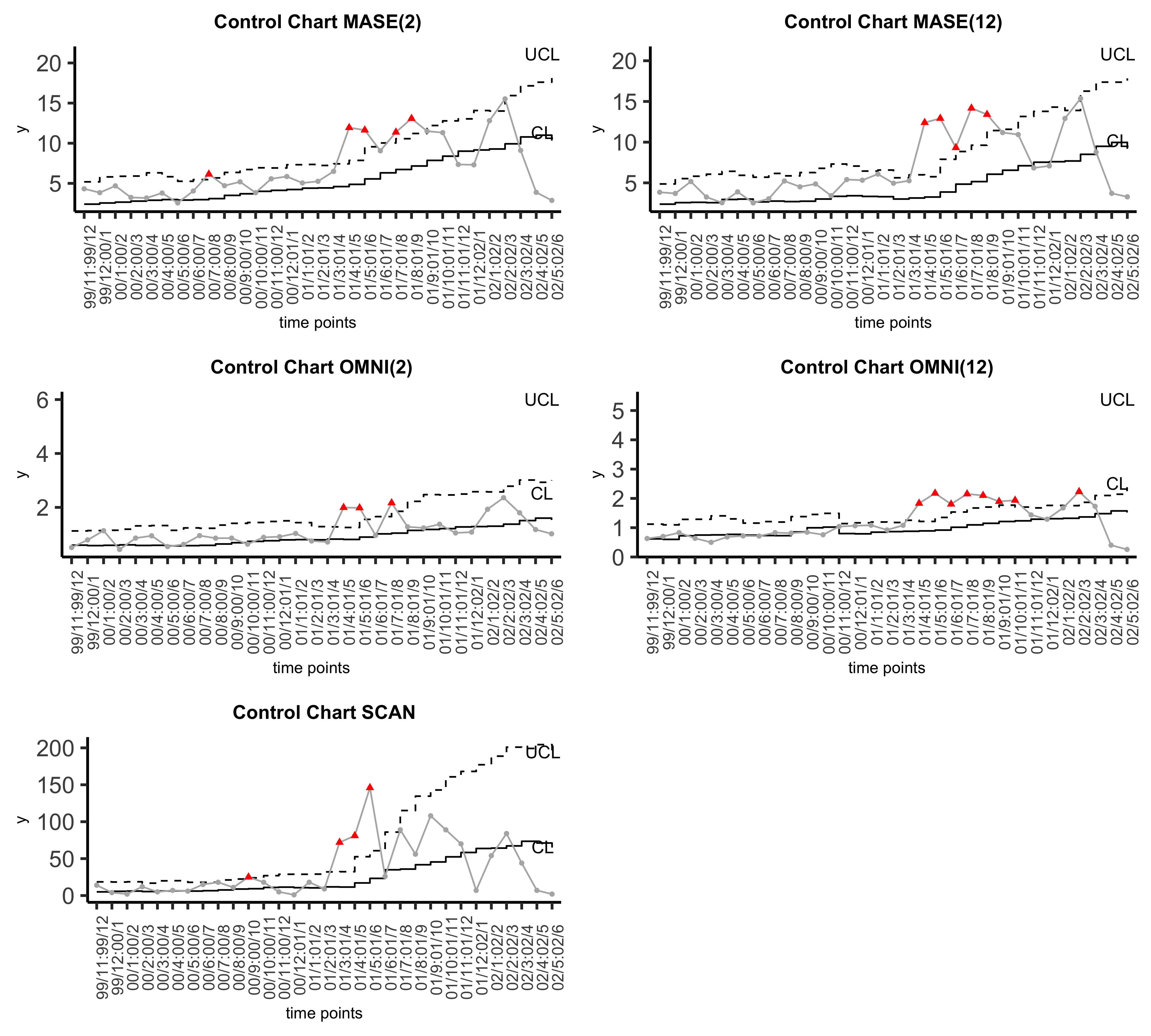}
\caption{GraphAD control charts with the moving averages (solid line), and three times the adjusted moving sample range (dashed line). Dots and triangles represent normal and anomalous graphs. The chart starts from Dec 1999 due to the training period $l=12$. }
\label{fig:enronGraphAD_cc_l12}
\end{figure}

\subsubsection*{Adjusted p-values plots in the Enron email dataset}

\begin{figure}[H]
\centering
\includegraphics[width=\linewidth]{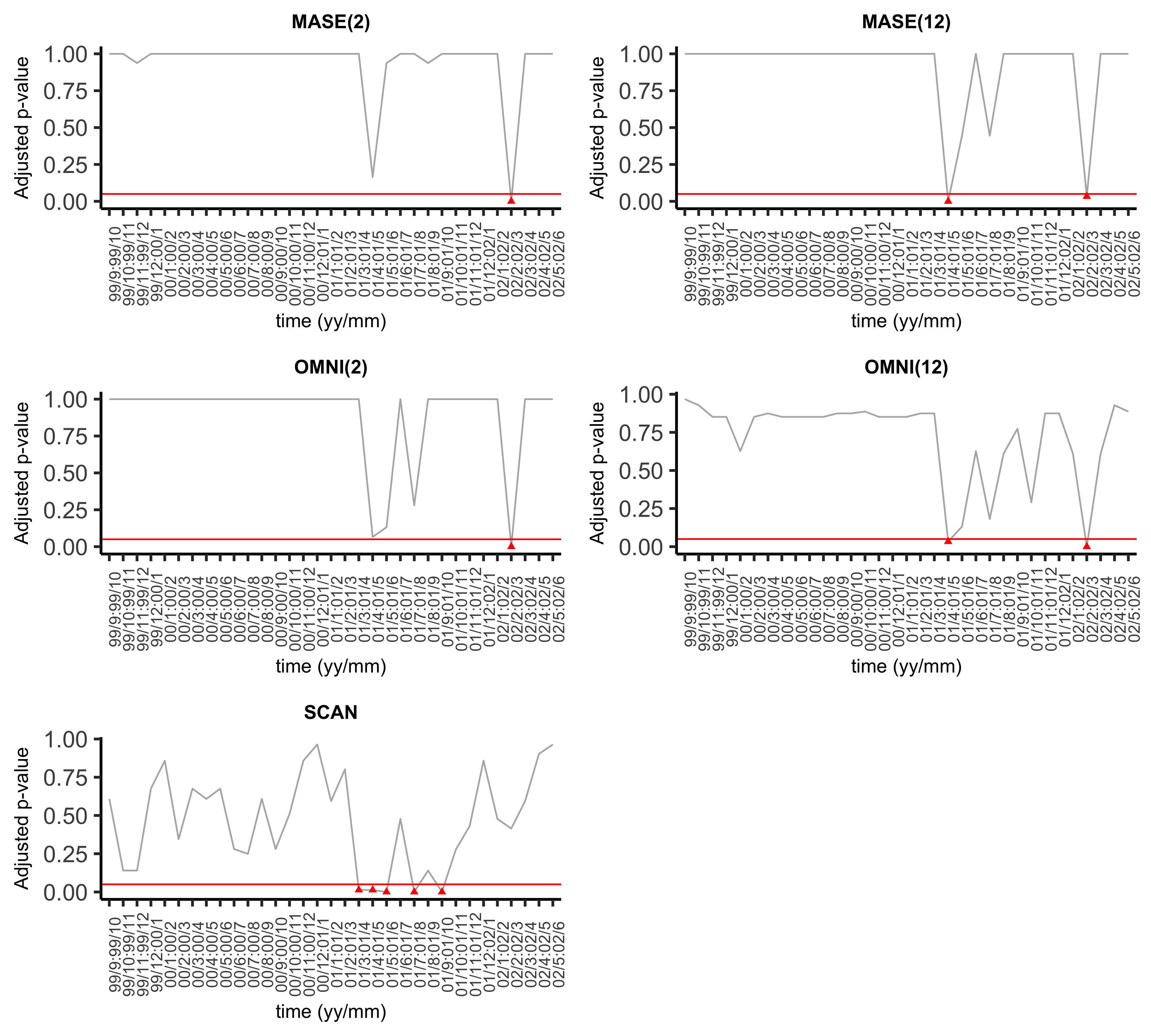}
\caption{Hypothesis testing for time series of
Enron email graphs using $l=10$. Horizontal lines mark the significant level of $0.05$.}
\label{fig:enronGraphAD_pval_l10}
\end{figure}

\begin{figure}[H]
\centering
\includegraphics[width=\linewidth]{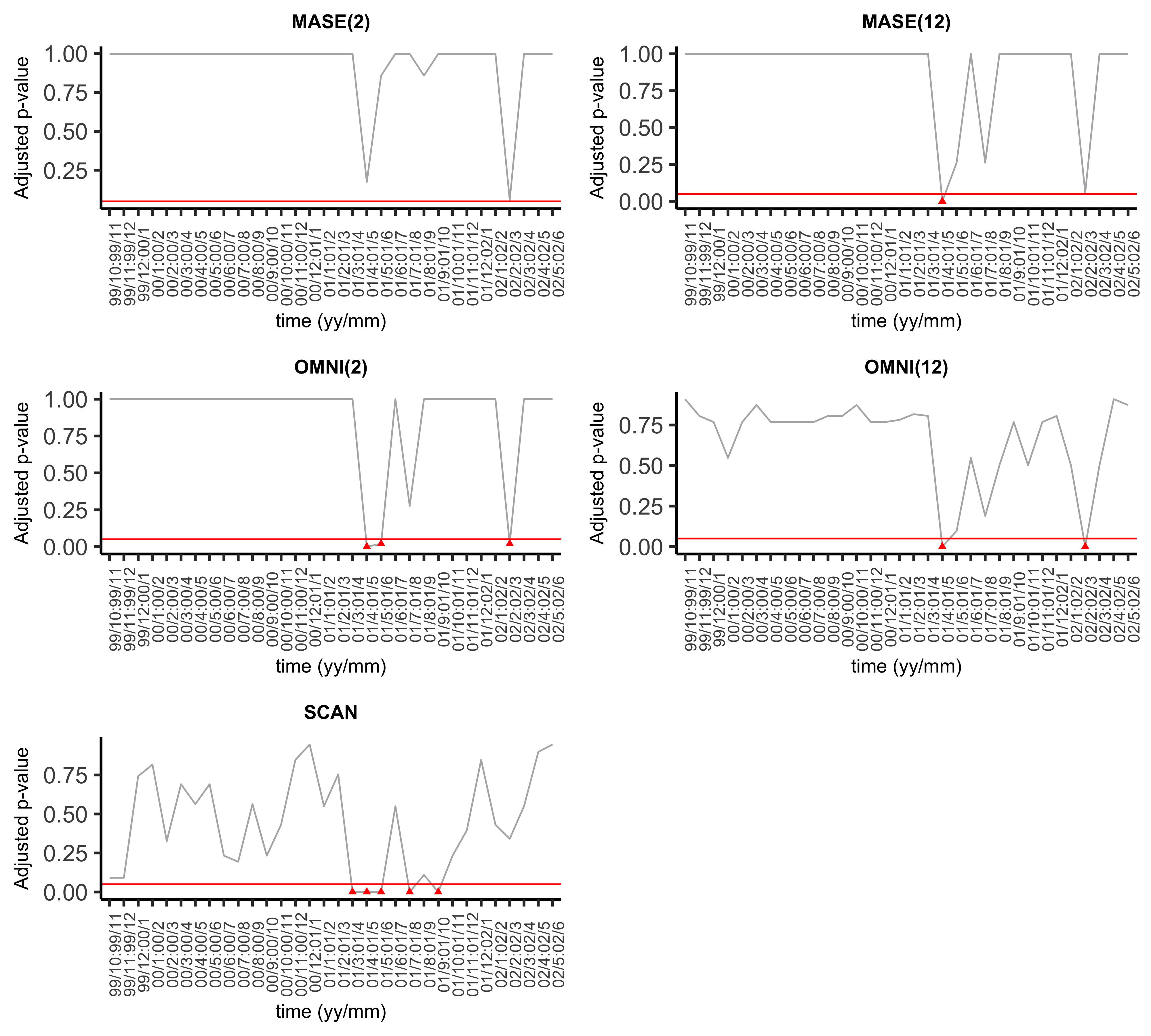}
\caption{Hypothesis testing for time series of
Enron email graphs using $l=11$. Horizontal lines mark the significant level of $0.05$.}
\label{fig:enronGraphAD_pval_l11}
\end{figure}

\begin{figure}[H]
\centering
\includegraphics[width=\linewidth]{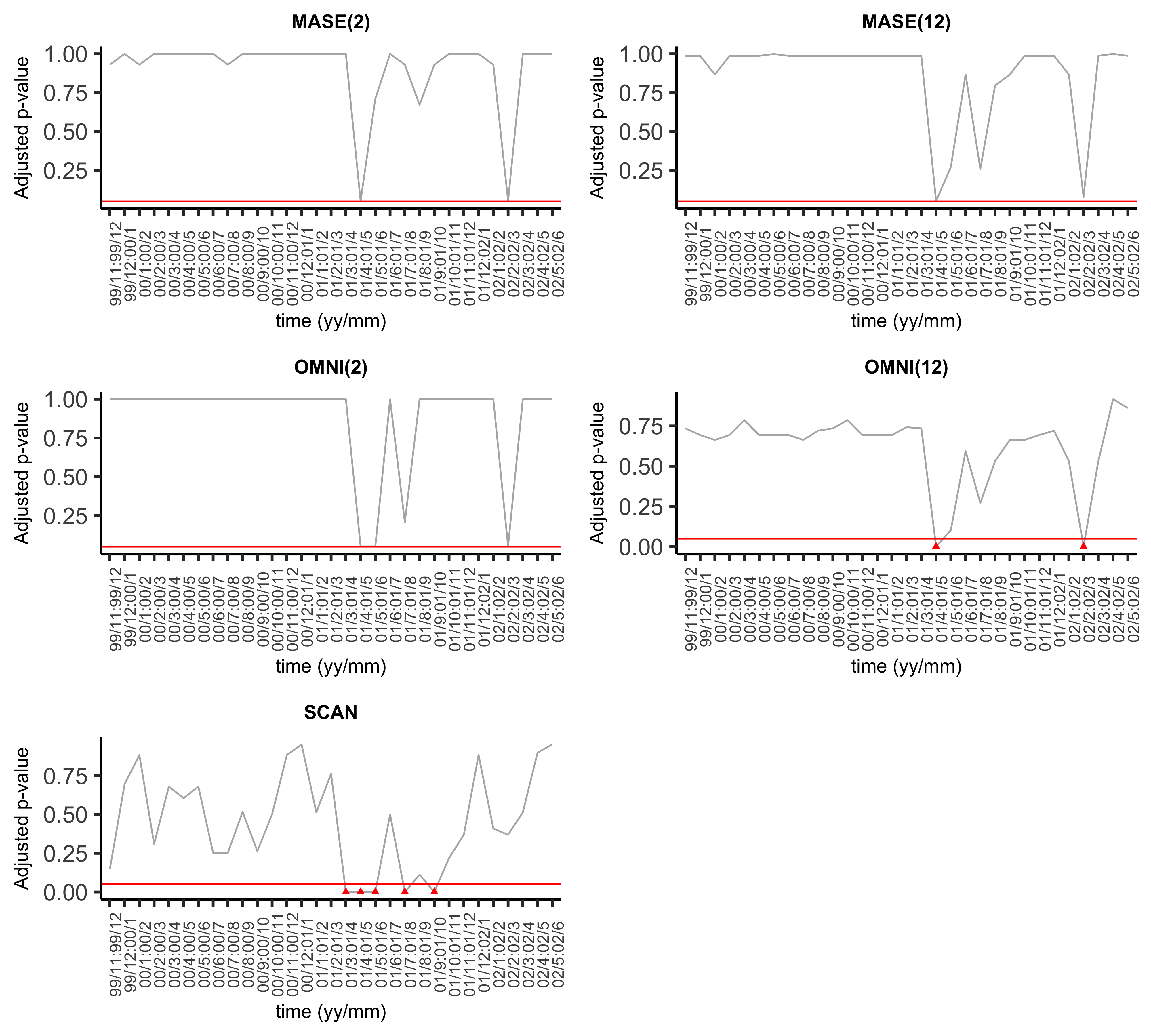}
\caption{Hypothesis testing for time series of
Enron email graphs using $l=12$. Horizontal lines mark the significant level of $0.05$.}
\label{fig:enronGraphAD_pval_l12}
\end{figure}

\subsection*{Microsoft Bing entity-transition results}

The results obtained from the adjusted p-values align closely with those from the control charts. Note that the anomalies detected using adjusted p-values encompass those identified with control charts and are included in Figure~\ref{fig:MSRGraphADFDR}. We present the detailed VertexAD results for \text{MASE(2)} in the right panel of Fig.~\ref{fig:MSRVertexAD(2)}. Generally speaking, our method detects all $473$ artificially anomalous vertices in Sep:Oct and Oct:Nov. For \text{MASE(2)}, $2,947$ vertices are claimed to be anomalous across time. Among them, $244$ and $295$ vertices are at the first two adjacent time pairs, while $562$ and $615$ vertices are detected where the artificial anomaly is injected (in Sep:Oct and Oct:Nov). There are $35$, $383$, and $415$ vertices detected as anomalies for the next three time pairs, and $514$ and $917$ anomalous vertices for the last two time pairs. {To investigate this, we have plotted the latent positions for March 2019 and April 2019 in Figure~\ref{fig: msrVertexADqcc}. The dots represent undetected vertices in April 2019, while the triangles represent the anomalies detected that appear to have undergone significant changes, such as vertex $v_{22020}$, for instance. }Furthermore, by treating the artificial anomalies as references, we can investigate other anomalous vertices that are detected by our methods. For example, in Fig.~\ref{fig:degchangeanomaly}, we assemble vertices across months and plot the histogram of changes of degrees between adjacent months for each vertex. The triangles are the detected anomalous vertices for \text{MASE}. For reference, the triangles circled by ellipses in Fig.~\ref{fig:degchangeanomaly} are degree changes for artificially anomalous vertices. 
Fig.~\ref{fig:degchangeanomaly} shows \text{MASE} detects vertices that change degree -- such vertices are likely to be anomalous. However, degree changes of artificially anomalous vertices are comparatively small, but we are still able to detect them. This demonstrates that our approaches can detect anomalous vertices beyond just vertices with large degree change. 

\subsubsection*{Adjusted p-values plots in the commercial search engine dataset}
\begin{figure}[H]
\centering
\includegraphics[width=\linewidth]{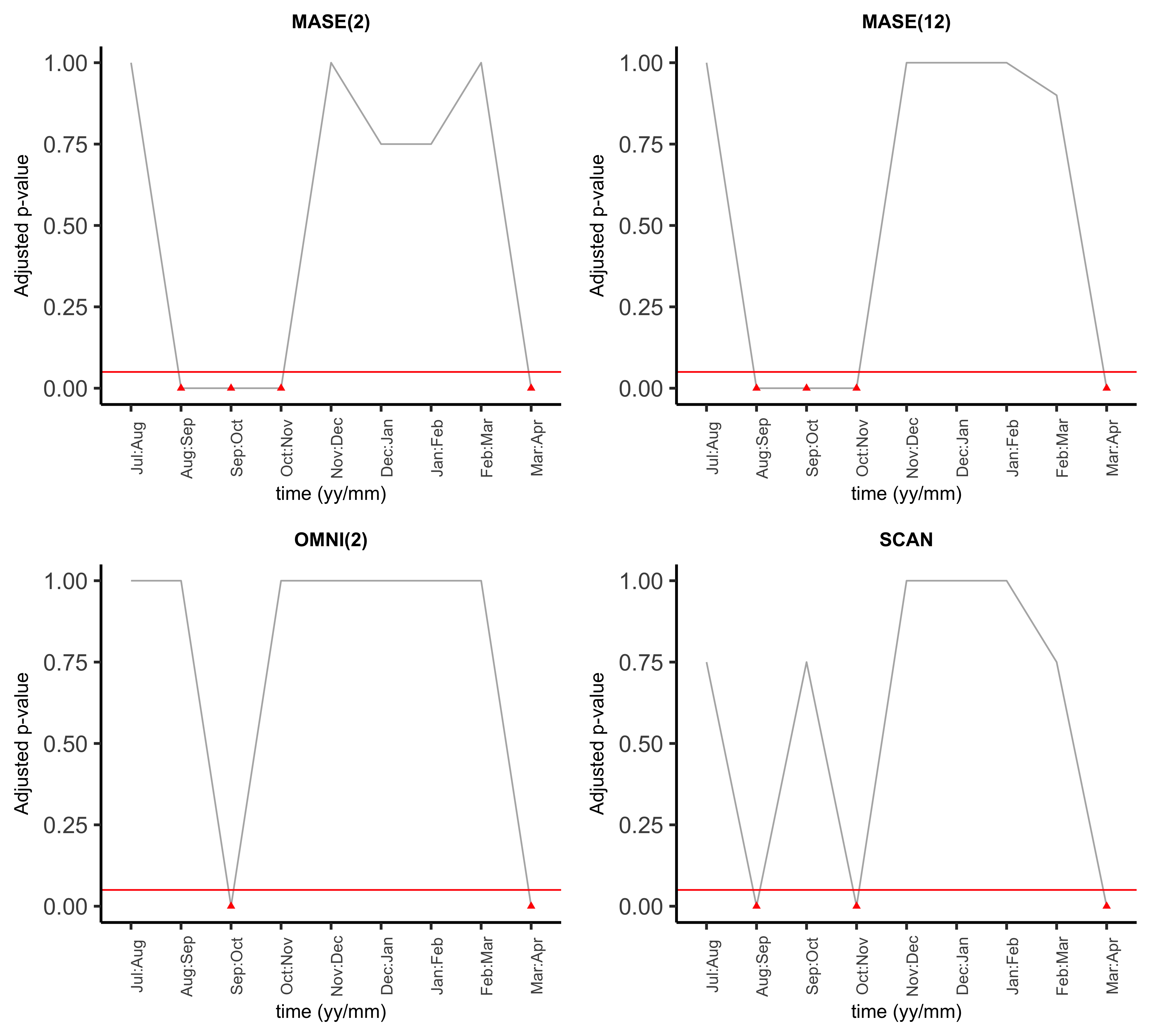}
\caption{Adjusted \text{p}-values on the MSB time series of graphs for which an artificial anomaly is inserted in October. \text{MASE} detects the artificial anomaly while \text{Scan} misses the anomaly in the middle of September and October. Because of the period of determining the null distribution, the chart starts in the middle of July and August.}
\label{fig:MSRGraphADFDR}
\end{figure}

\begin{figure}[H]
\centering

\includegraphics[width=\linewidth]{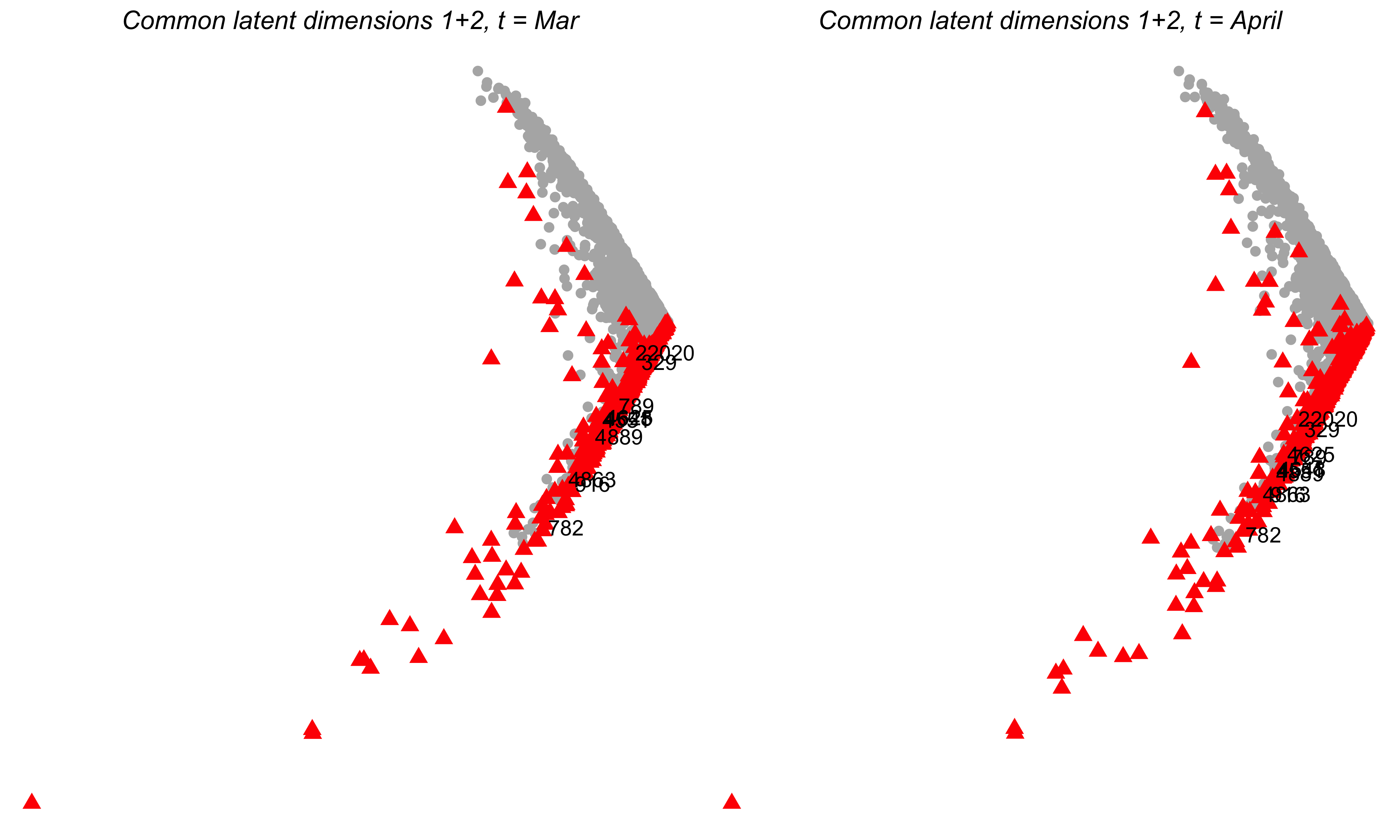}
\caption{Scatter plots of first two \text{MASE} common latent dimensions of MSB graphs from March 2019 to April 2019. The triangles represent vertex anomalies detected by the \text{MASE(2)} method using control charts, while the dots represent normal vertices. The top $10$ anomalous vertices are explicitly annotated.}
\label{fig: msrVertexADqcc}
\end{figure}

\begin{figure}[H]
\centering

\includegraphics[width=\linewidth]{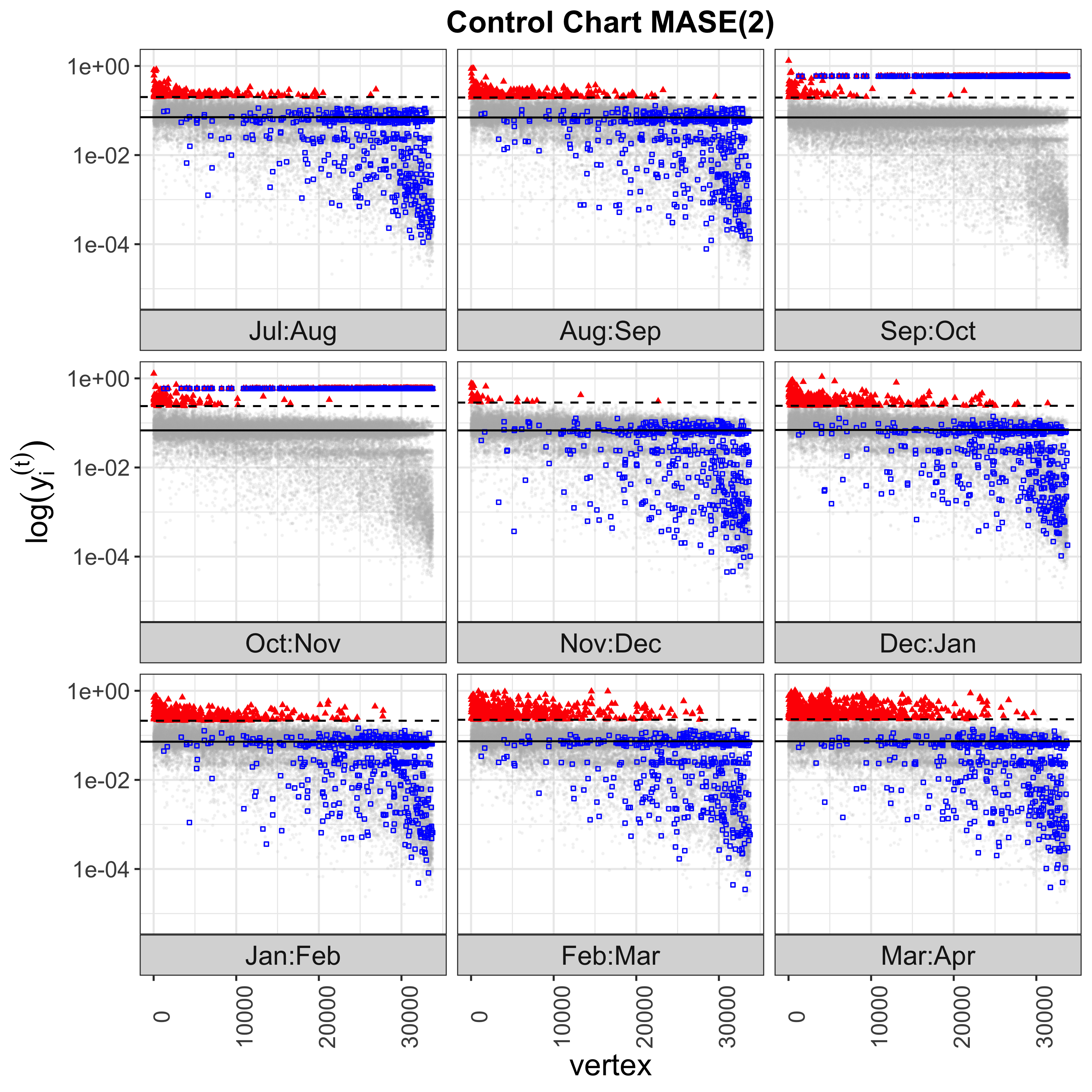}
\caption{Control charts for VertexAD on the MSB time series of graphs with a subgroup of artificially anomalous vertices inserted in October with $s=2$. Center solid line (CL) represents moving average of the sample means ${\bar{y}_{i}}^{(t)}$, dashed line (UCL) represents ${\bar{y}_{i}}^{(t)} + 3{\bar{\sigma}_{i}}^{(t)}$, where ${\bar{\sigma}_{i}}^{(t)}$ is EWAVE-SD. Dots are $y_{i}^{(t)}$ at times when the latent positions are claimed as normal, and the triangles are those $y_{i}^{(t)}$ which lie outside of UCL and are claimed as anomalous. Squares are artificially anomalous vertices. Because of the training period to construct the control charts, the control chart starts in the middle of July and August.}
\label{fig:MSRVertexAD(2)}
\end{figure}

\begin{figure}[H]
    \centering
    \includegraphics[width=\linewidth]{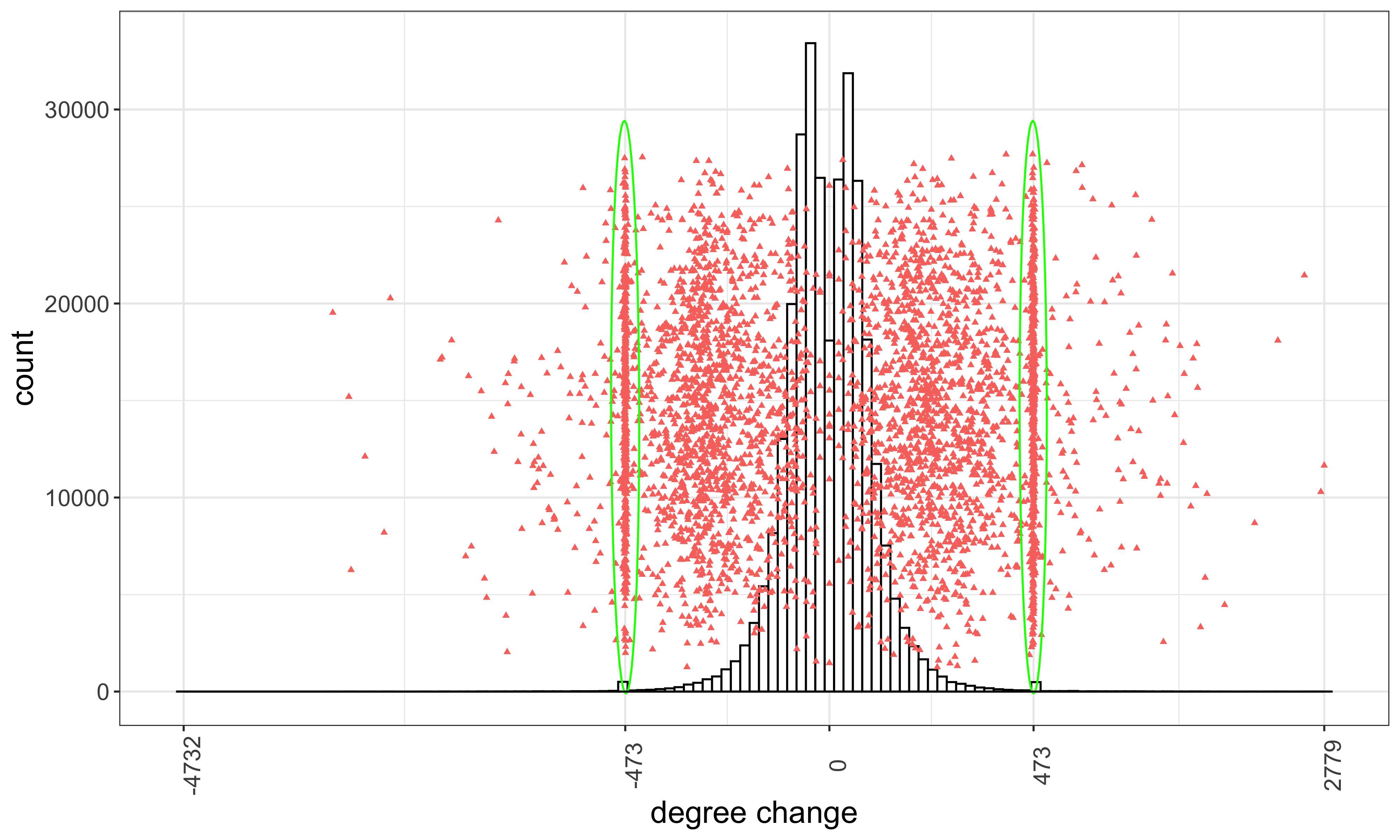}
 \caption{Histogram of degree changes for vertices across time points in the MSB time series of graphs with artificially injected anomalies. The triangles are both perturbed vertices (circled in ellipse) and the vertices detected by \text{MASE} (the degree change scale is square-root transformed, and the dots are vertically jittered for display purposes). This figure demonstrates that \text{MASE} can detect anomalous vertices beyond just vertices with large degree change. }
 \label{fig:degchangeanomaly}
\end{figure}

\begin{figure}[H]
\centering

\includegraphics[width=\linewidth]{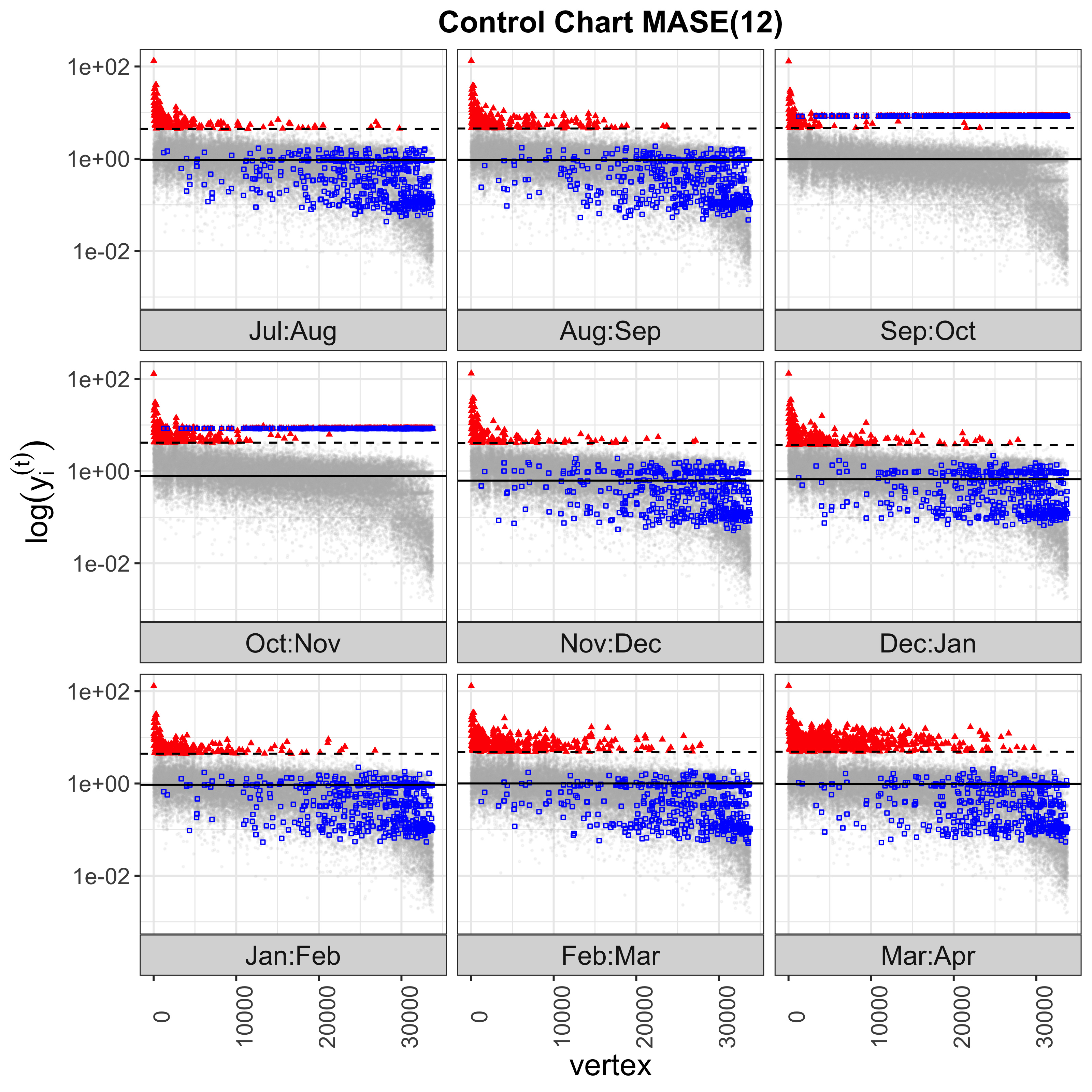}
\caption{Control charts for VertexAD on the MSB time series of graphs with a subgroup of artificially anomalous vertices inserted in October with $s=12$. Center solid line (CL) represents moving average of the sample means ${\bar{y}_{i}}^{(t)}$, dashed line (UCL) represents ${\bar{y}_{i}}^{(t)} + 3{\bar{\sigma}_{i}}^{(t)}$, where ${\bar{\sigma}_{i}}^{(t)}$ is EWAVE-SD. Dots are $y_{i}^{(t)}$ at times when the latent positions are claimed as normal, and the triangles are those $y_{i}^{(t)}$ which lie outside of UCL and are claimed as anomalous. Squares are artificially anomalous vertices. Because of the training period to construct the control charts, the control chart starts in the middle of July and August.}
\label{fig:MSRVertexAD(12)}
\end{figure}

\subsubsection*{The planted clique method in the MSB dataset}
\label{sec:msb appendix}
In the absence of ground truth for the existence of anomalies, we design an anomaly-insertion strategy,
creating artificially anomalous vertices for the graph at time point $t = 6$ (October).
So long as our methods detect these artificial anomalies, other detected anomalies may have merit.
Thus, our final result is that anomalies detected at the same level
in the original, unperturbed data, are ``real''.
Our approach to the creation of artificial anomalies involves a planted clique \citep{10.5555/1540612}, as follows. We perform \text{ASE} with $d = 20$ for the $6$-th graph (October),
and then apply Gaussian mixture modeling (GMM) to cluster the latent position estimates.
This results in nine clusters of vertices;
we add an edge with a weight equal to $1$ for each pair of vertices in the smallest cluster
($n^{*} = 473$) to create a complete subgraph.
Finally, we normalize edge weights for the entire (perturbed) time series of graphs
so that the normalized weights lie in the interval $[0, 2]$.

\begin{table}[h!]
\centering
\begin{tabular}{|l|r|r|r|r|r|r|r|r|r|}
\hline
Month & Aug & Sep & Oct & Nov & Dec & Jan & Feb & Mar & Apr \\
\hline
No. of Sigmas & -0.8 & 13.4 & 82.3 & 0.6 & -321.5 & -0.5 & -4.9 & -0.9 & 211.0 \\
\hline
\end{tabular}
\caption{Number of sigmas by which the data deviates from the moving means in the \text{MASE(12)} test statistics, as illustrated in Figure~7.}

\label{tab:sigmarage}
\end{table}

\newpage
\section*{Appendix E: Sensitivity results on dimension selection}
In the main section of the manuscript, we employed the scree plot method \citep{zhu2006automatic} for dimension selection by applying it to the square root of the first $n$ leading eigenvalues. To assess the impact of dimension selection on our methods, we conducted a sensitivity analysis using the scree plot method on all $n$ leading eigenvalues. This analysis was performed for both simulation studies (see Figures~\ref{fig:conchart1_full},~\ref{fig:mmsbm4_full}, and~\ref{fig:mco_full}) and real data analysis (see Figure~\ref{fig:enronGraphADqcc_full} and~\ref{fig:enronVertexADqcc_full}). Additional figures are provided for the Enron dataset (see Figures~\ref{fig:enronGraphAD_cc_l10_full},~\ref{fig:enronGraphAD_cc_l11_full},~\ref{fig:enronGraphAD_cc_l12_full}, and~\ref{fig:enronGraphAD_pval_l10_full}). These figures vary the parameter $l$ to assess sensitivity in both $l$ and dimension selection methods simultaneously for both control charts and p-value based approaches. The Bing data analysis was excluded from this part of our sensitivity study due to an out-of-memory error. The outcomes of this analysis, shown in the figures mentioned above, are consistent with the results in the main section. This highlighted our results' adaptability to dimension selection strategy changes.

\begin{figure}[h]
\centering
\includegraphics[width=\linewidth]{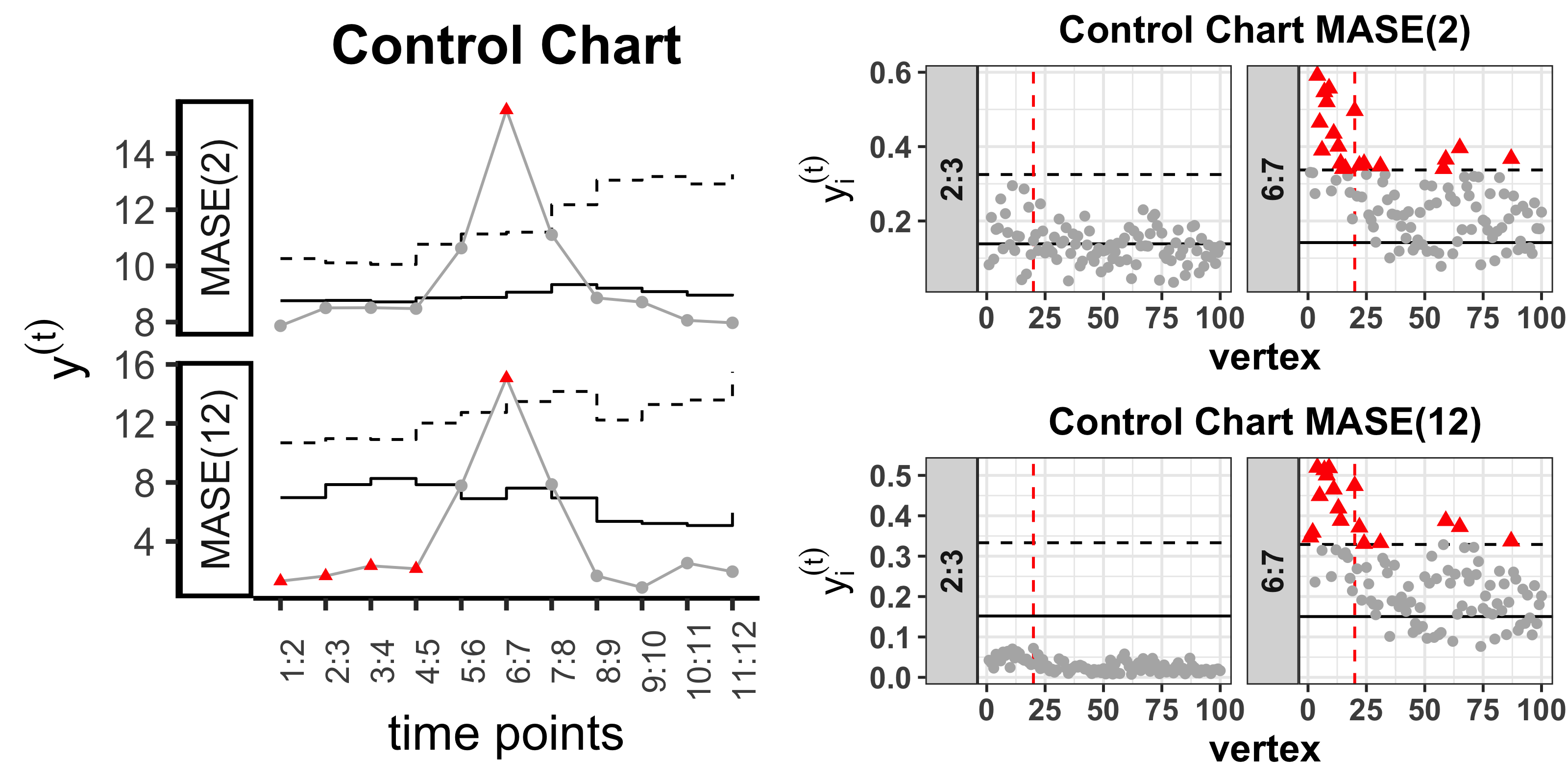}
\caption{Control charts for time series of
graphs with anomalies at time points $6$ and $7$ (scenario $1$). This figure is similar to Figure~\ref{fig:conchart1} but differs in its approach to determining the optimal number of dimensions: here, the elbow is selected based on an analysis of all $n$ eigenvalues.}
\label{fig:conchart1_full}
\end{figure}

\begin{figure}[h]
\centering
\includegraphics[width=\linewidth]{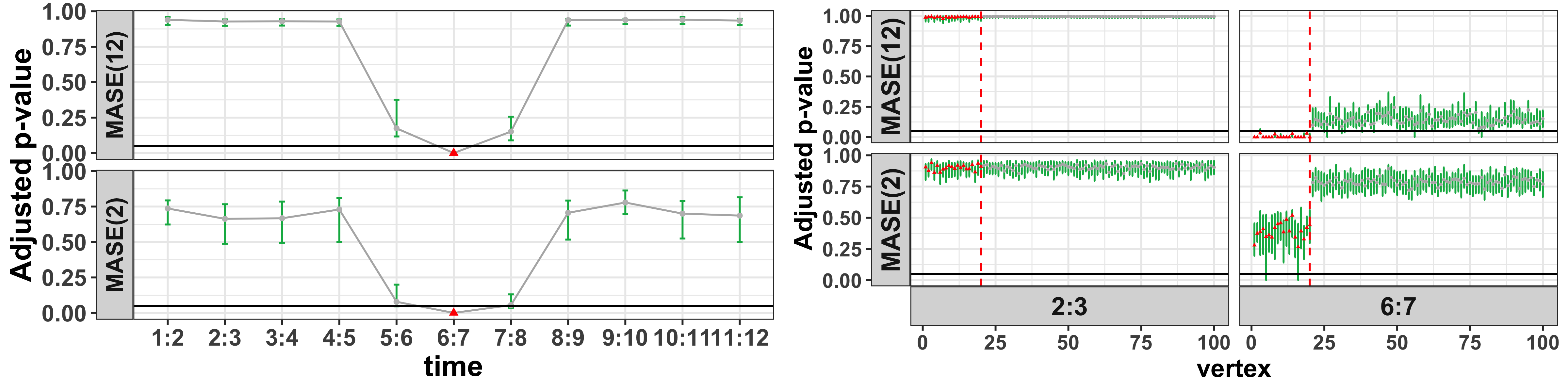}

\caption{Hypothesis testing for time series of graphs from Figure~\ref{fig:conchart1}. This figure is similar to Figure~\ref{fig:mco} but differs in its approach to determining the optimal number of dimensions: here, the elbow is selected based on an analysis of all $n$ eigenvalues.}

\label{fig:mco_full}
\end{figure}

\begin{figure}
    \centering
    \begin{subfigure}[b]{.49\textwidth}
        \includegraphics[width=\textwidth]{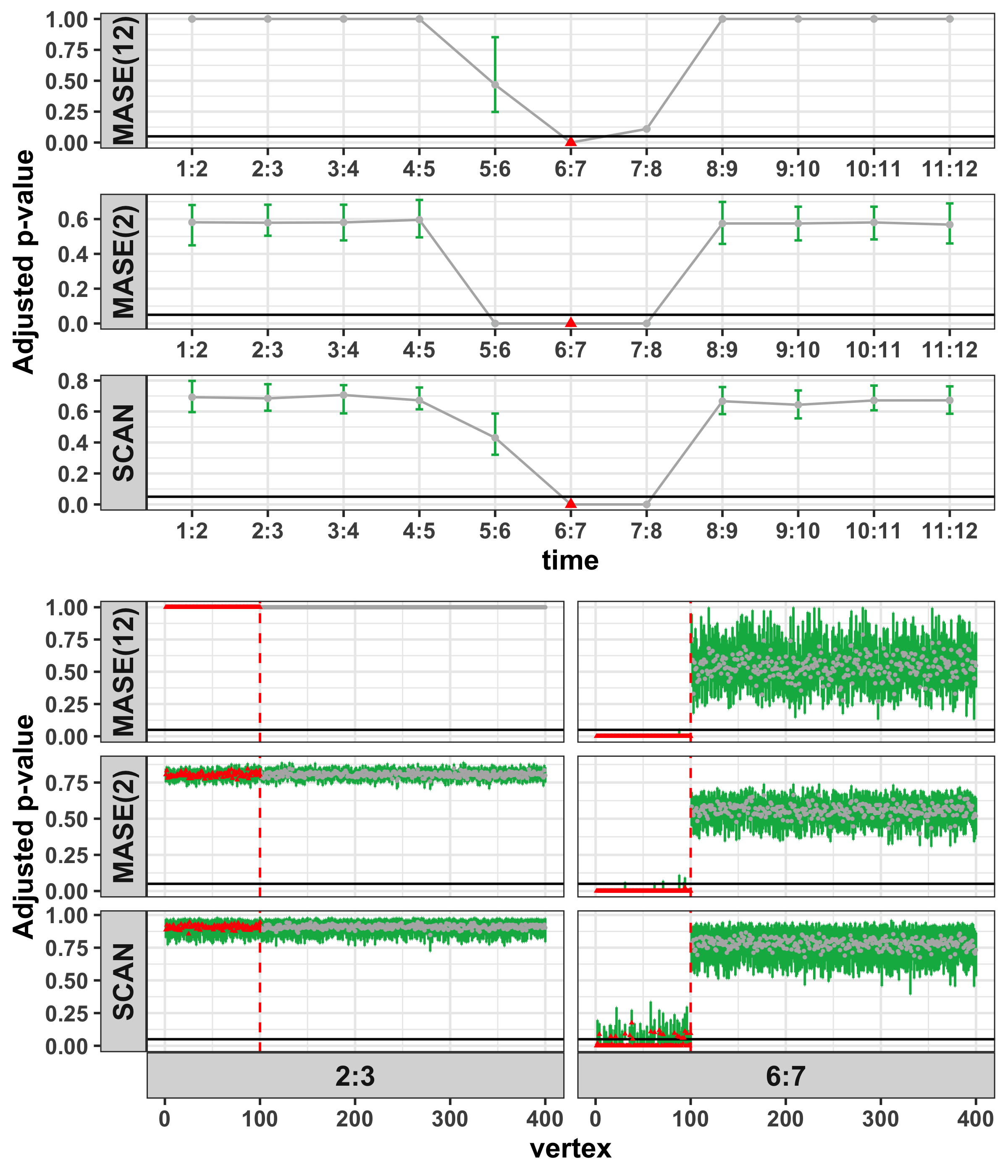}
        \caption{Scenario $2$} 
        \label{fig:sdmultid4mar09_full}
    \end{subfigure} 
    \begin{subfigure}[b]{.49\textwidth}
        \includegraphics[width=\textwidth]{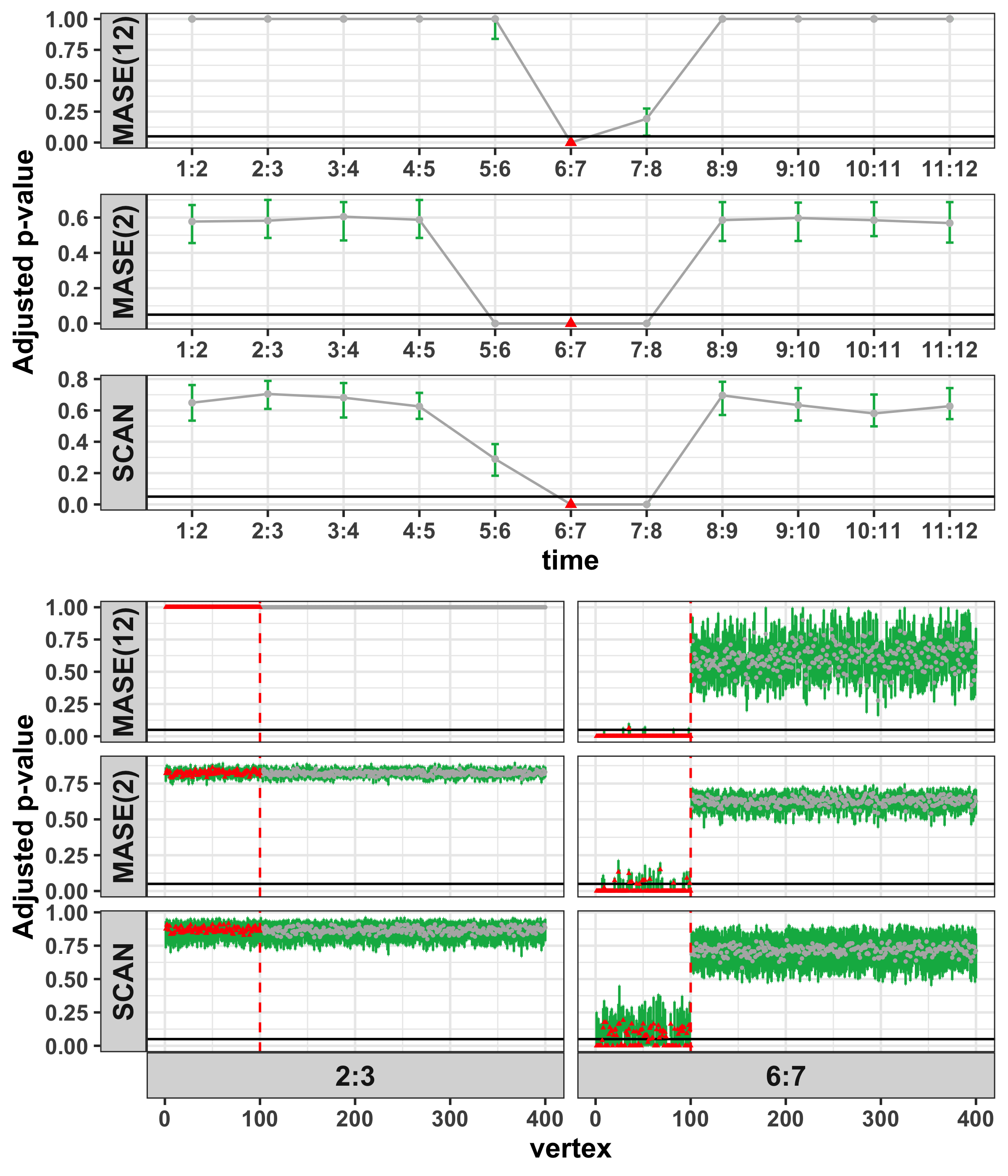}
        \caption{Scenario $3$} 
        \label{fig:ddmultid4mar09_full}
    \end{subfigure}
    \caption{Hypothesis testing for time series of graphs under scenarios $2$ and $3$. This figure is similar to Figure~\ref{fig:mmsbm4} but differs in its approach to determining the optimal number of dimensions: here, the elbow is selected based on an analysis of all $n$ eigenvalues.}
    \label{fig:mmsbm4_full}
\end{figure}

\begin{figure}
    \centering
    \begin{subfigure}[b]{.59\textwidth}
    \includegraphics[width=\textwidth]{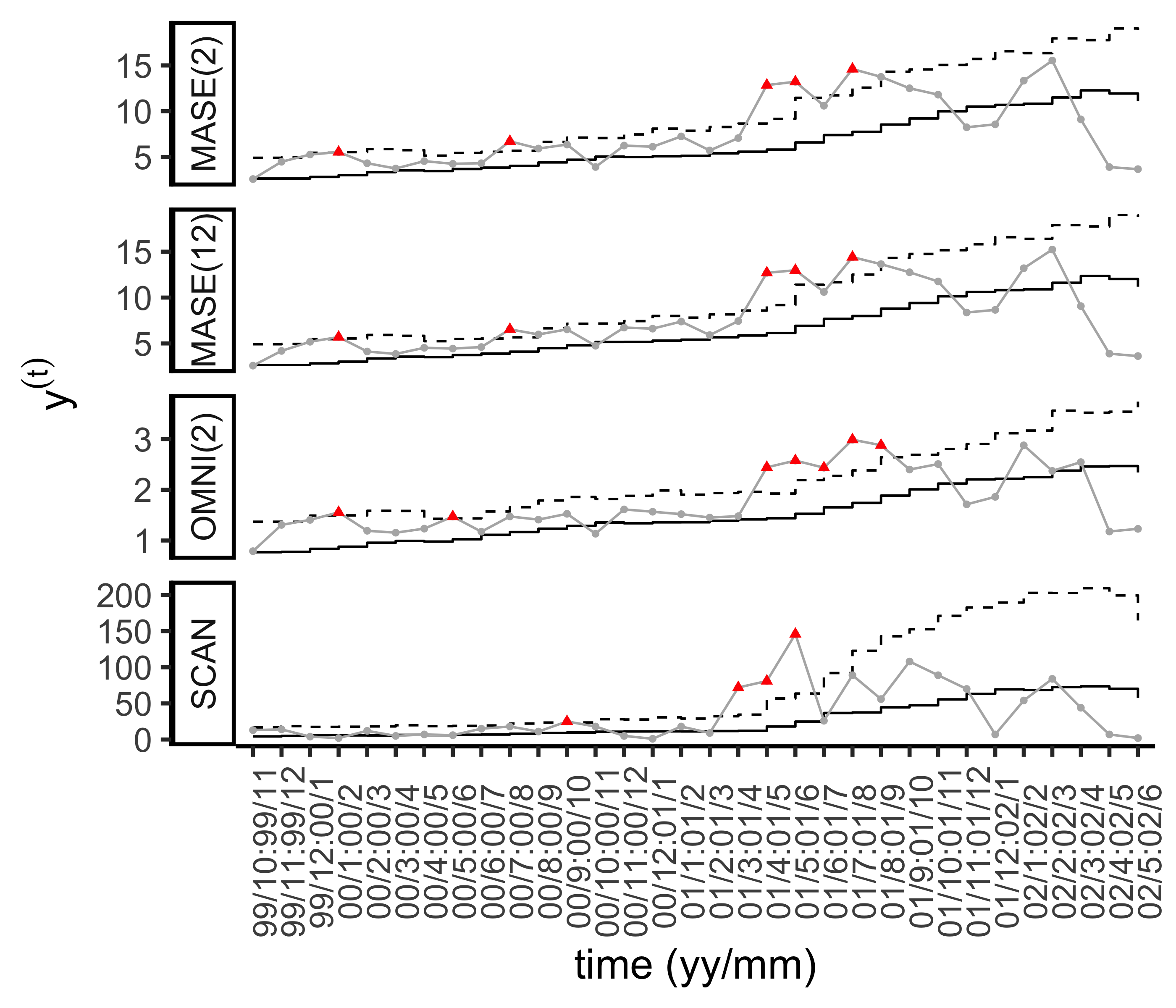}
    \caption{\label{fig:enronGraphADqcc_full}}
    \end{subfigure} 
    \begin{subfigure}[b]{.39\textwidth}
    \includegraphics[width=\textwidth]{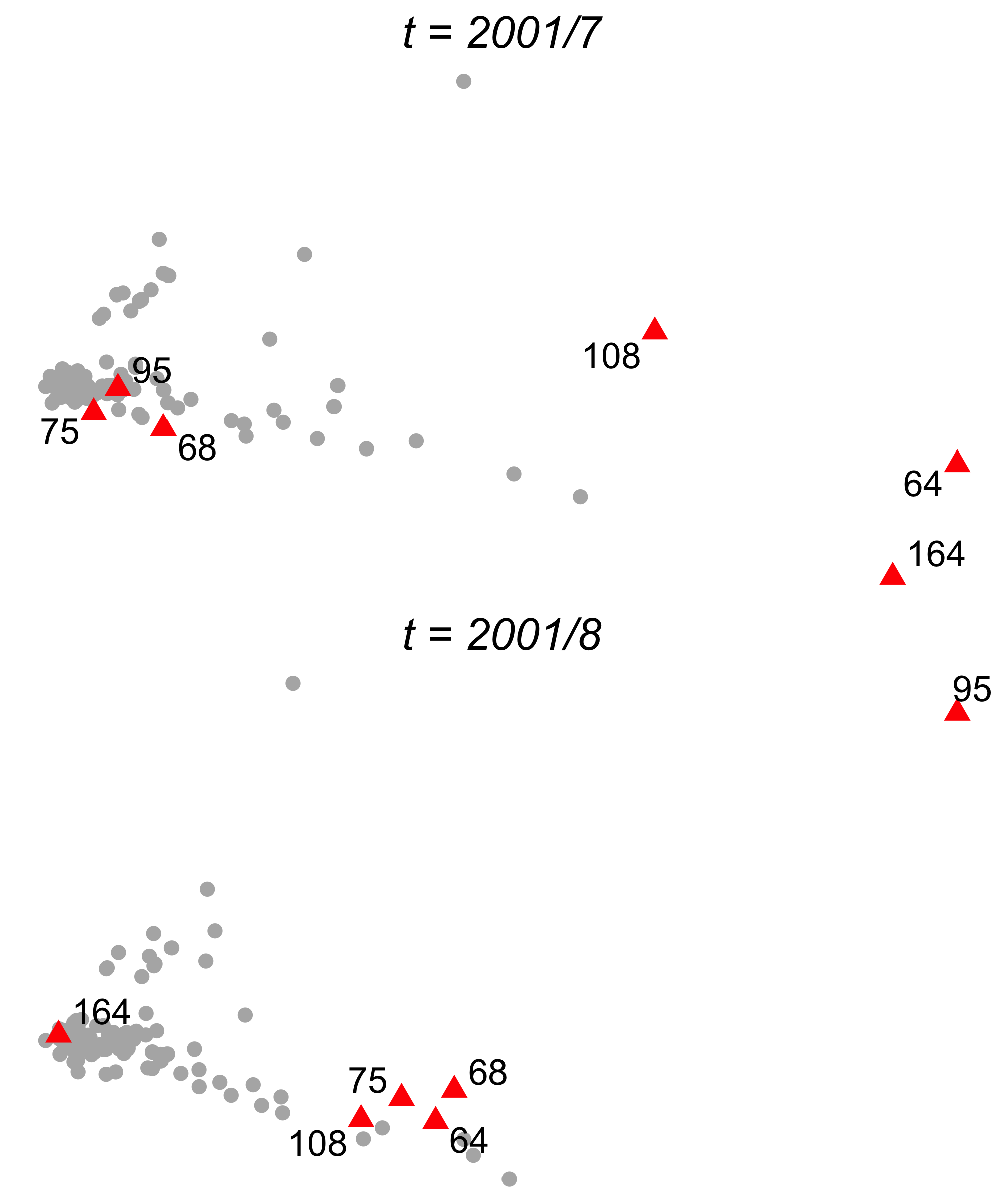}
    \caption{\label{fig:enronVertexADqcc_full}}
    \end{subfigure}
    \caption{Analysis of the Enron email graphs (Nov 1998 - Jun 2002). These figures are similar to Figures~\ref{fig:enronGraphADqcc} and~\ref{fig:enronVertexADqcc} but differ in their approach to determining the optimal number of dimensions: here, the elbow is selected based on an analysis of all $n$ eigenvalues.}

\end{figure}

\begin{figure}[h]
\centering

\includegraphics[width=\linewidth]{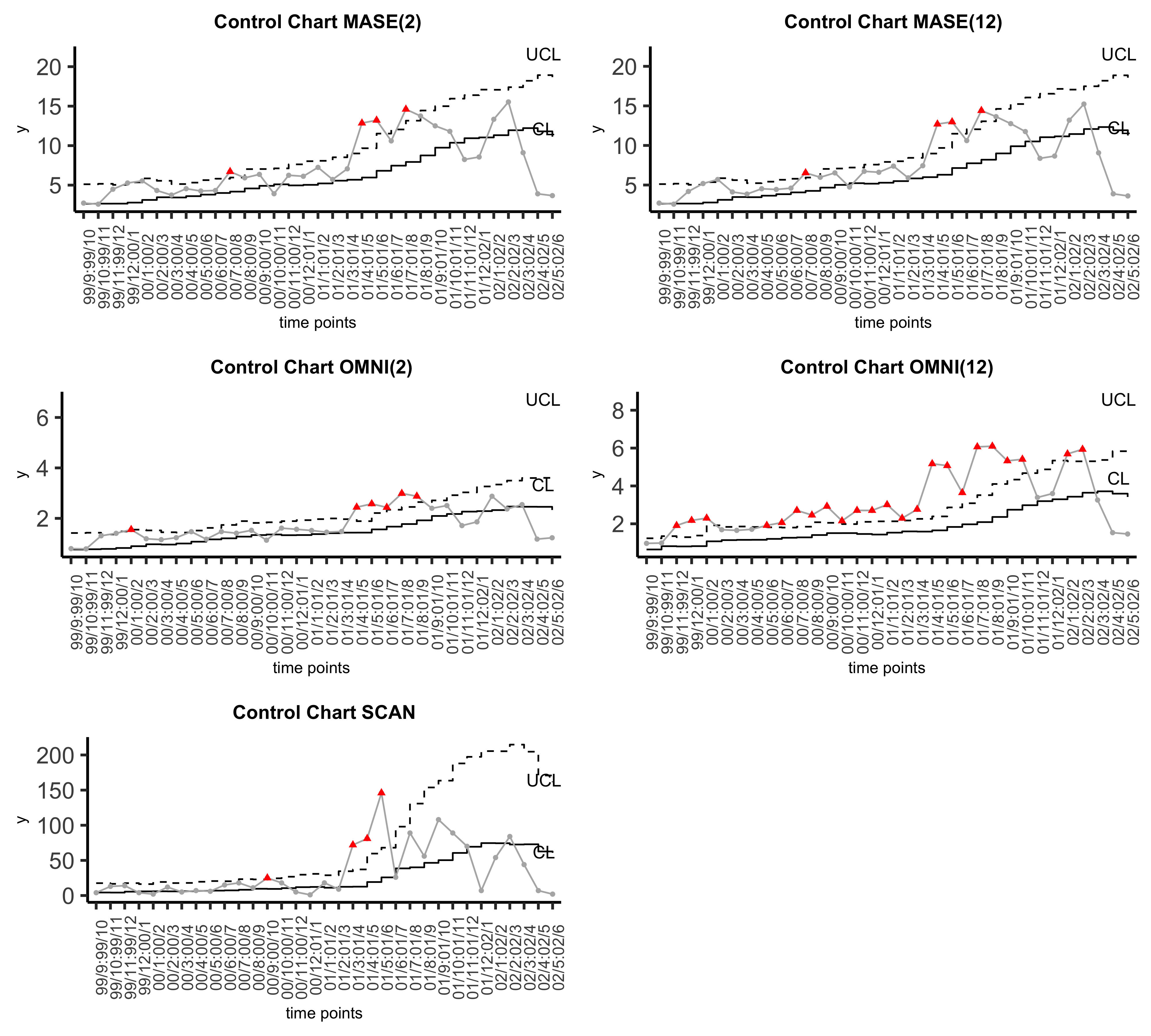}
\caption{GraphAD control charts with the moving averages (solid line), and three times the adjusted moving sample range (dashed line) with the training period $l=10$. This figure is similar to Figure~\ref{fig:enronGraphAD_cc_l10} but differs in its approach to determining the optimal number of dimensions: here, the elbow is selected based on an analysis of all $n$ eigenvalues.}
\label{fig:enronGraphAD_cc_l10_full}
\end{figure}

\begin{figure}[h]
\centering
\includegraphics[width=\linewidth]{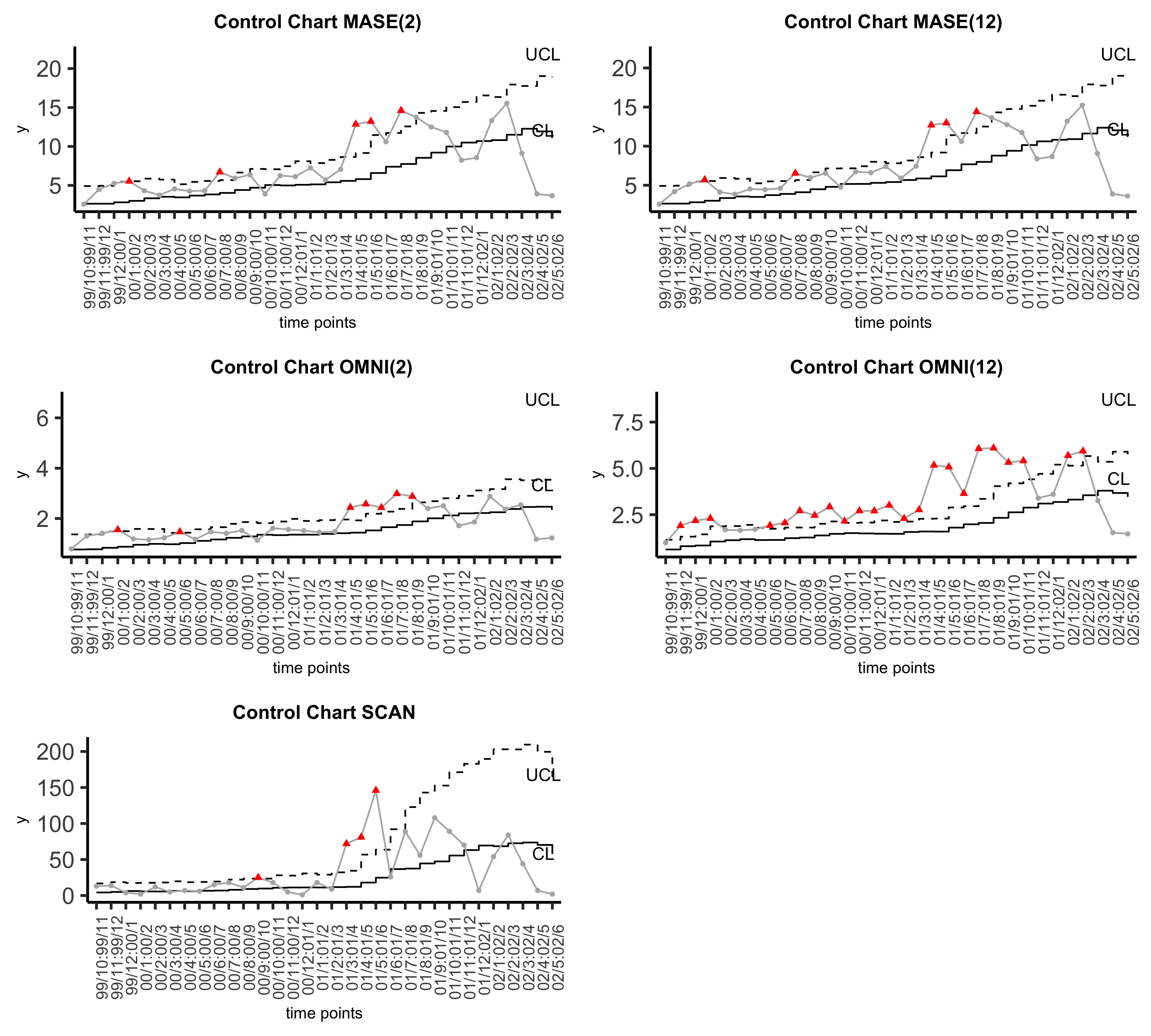}
\caption{GraphAD control charts with the moving averages (solid line), and three times the adjusted moving sample range (dashed line) with the training period $l=11$. This figure is similar to Figure~\ref{fig:enronGraphAD_cc_l11} but differs in its approach to determining the optimal number of dimensions: here, the elbow is selected based on an analysis of all $n$ eigenvalues.}
\label{fig:enronGraphAD_cc_l11_full}
\end{figure}

\begin{figure}[h]
\centering
\includegraphics[width=\linewidth]{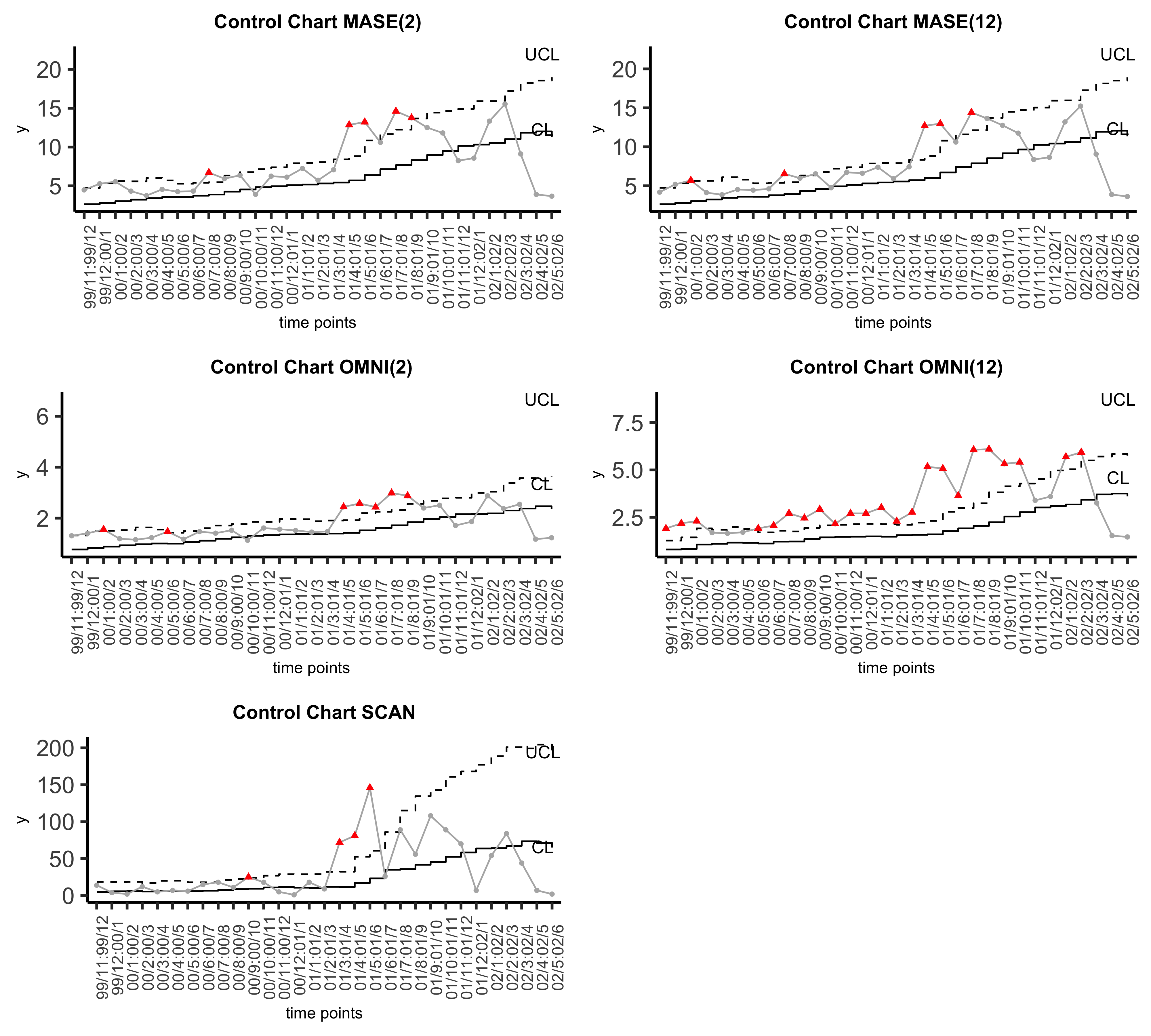}
\caption{GraphAD control charts with the moving averages (solid line), and three times the adjusted moving sample range (dashed line) with the training period $l=12$. This figure is similar to Figure~\ref{fig:enronGraphAD_cc_l12} but differs in its approach to determining the optimal number of dimensions: here, the elbow is selected based on an analysis of all $n$ eigenvalues.}
\label{fig:enronGraphAD_cc_l12_full}
\end{figure}

\begin{figure}[h]
\centering
\includegraphics[width=\linewidth]{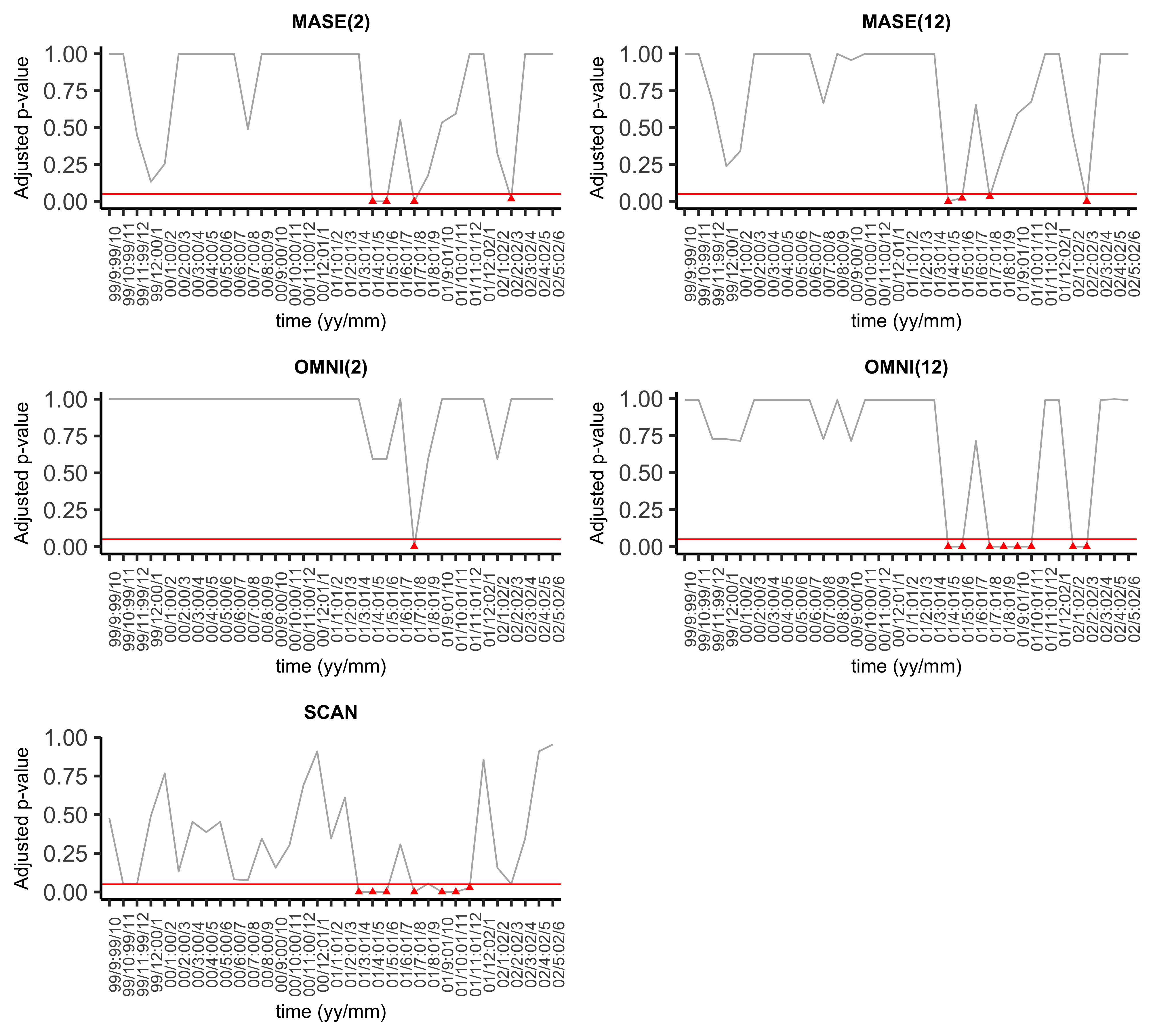}
\caption{Hypothesis testing for time series of
Enron email graphs using $l=10$ with the elbow selected based on all $n$ eigenvalues.}
\label{fig:enronGraphAD_pval_l10_full}
\end{figure}

\begin{figure}[h]
\centering
\includegraphics[width=\linewidth]{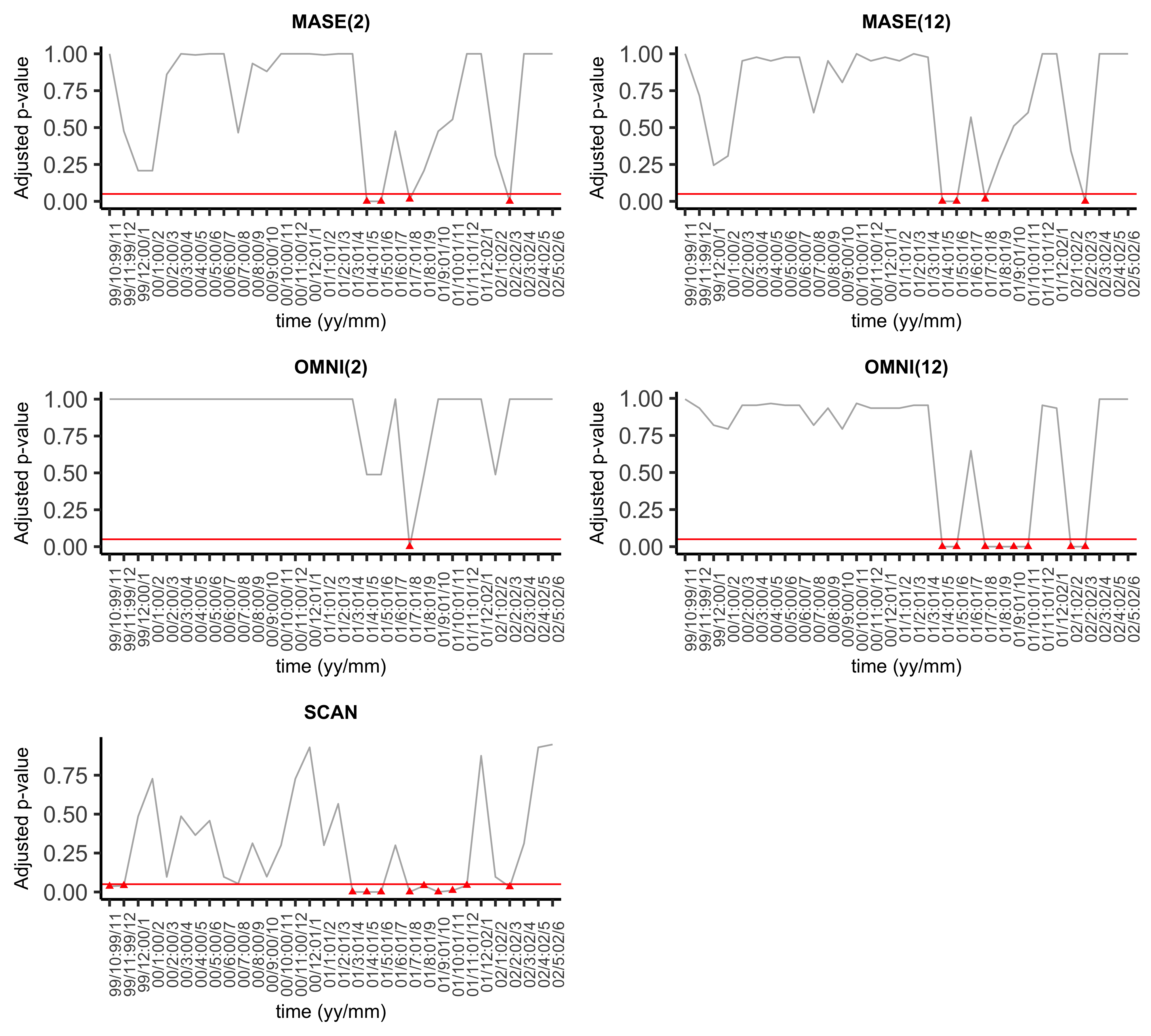}
\caption{Hypothesis testing for time series of
Enron email graphs using $l=11$ with the elbow selected based on all $n$ eigenvalues.}
\label{fig:enronGraphAD_pval_l11_full}
\end{figure}

\begin{figure}[h]
\centering

\includegraphics[width=\linewidth]{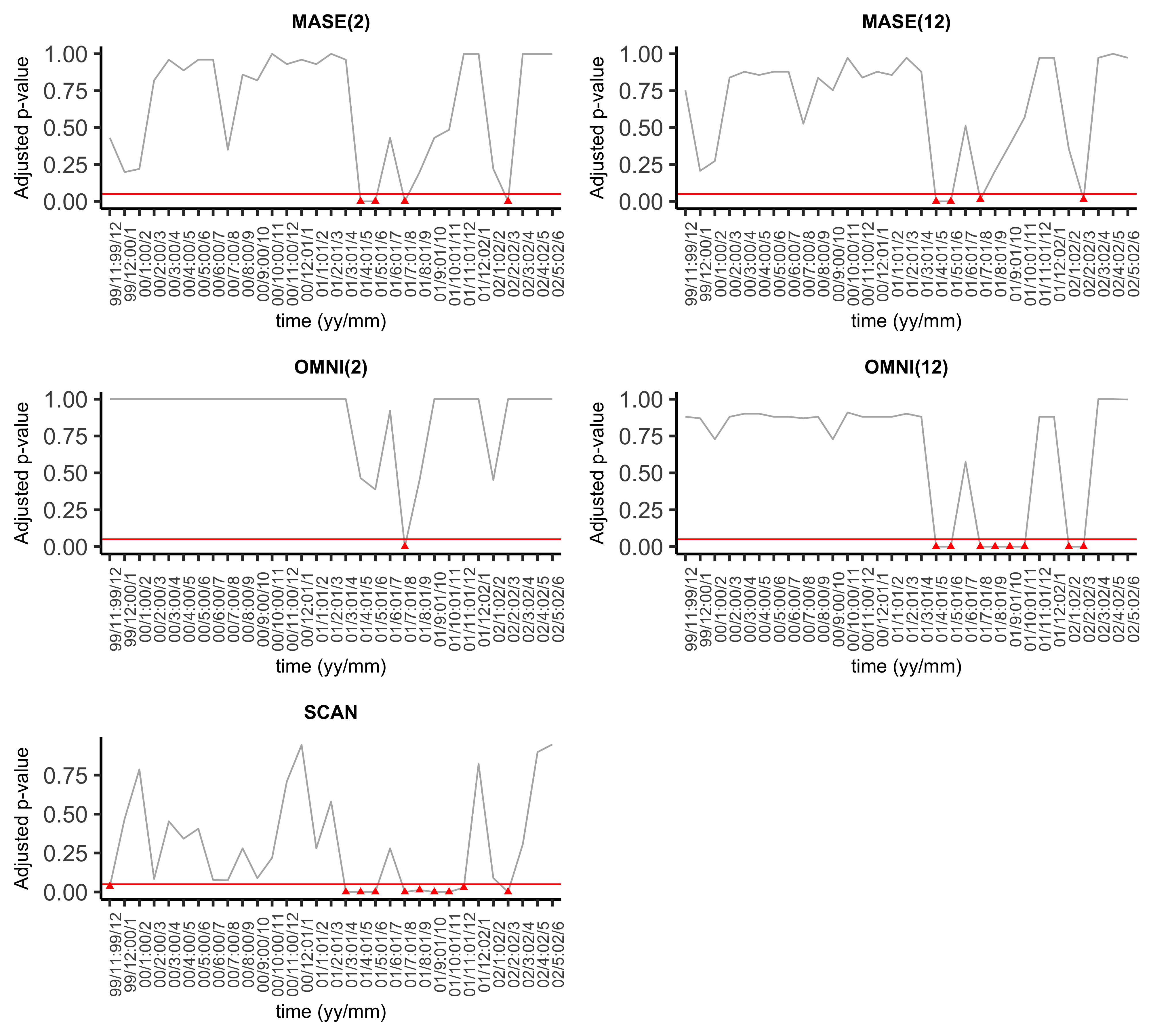}
\caption{Hypothesis testing for time series of
Enron email graphs using $l=12$ with the elbow selected based on all $n$ eigenvalues.}
\label{fig:enronGraphAD_pval_l12_full}
\end{figure}

\clearpage

\section*{Appendix F: Asymptotic Result via Simulation Study with Increasing Number of Vertices $n$}

We included the power and FPR on graphs generated according to scenario 2 from section 5.3 to demonstrate our asymptotic result via simulation study with increasing vertices  $n$.
\begin{figure}[H]
	\centering
	\includegraphics[width=\linewidth]{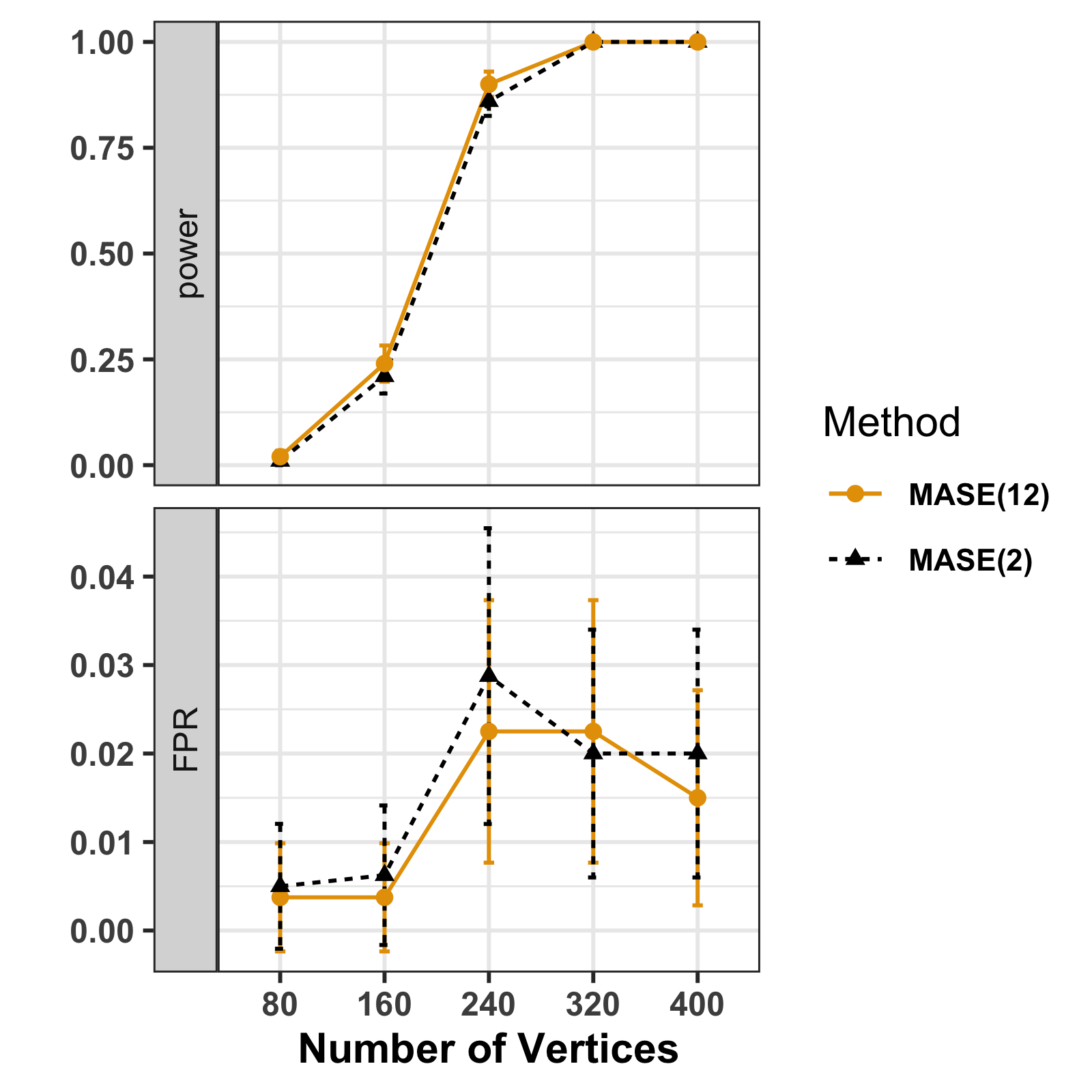}
	\caption{Asymptotic power and false positive rate as the number of vertices \( n \) increases to 80, 160, 240, 320, and 400. The top panel shows the power, while the bottom panel displays the false positive rate. The results are based on Scenario 2 from section 5.3 with \(\alpha = 0\).}
	
	\label{fig:NeuroAD}
\end{figure}

\end{document}